\documentclass[fleqn,3p,onecolumn,11pt,nofootinbib,showkeys,showpacs]{revtex4-1}

\usepackage[small,justification=centerlast]{caption}
\usepackage{graphicx}
\usepackage{bm}
\usepackage{mathrsfs}
\usepackage[latin1]{inputenc}
\usepackage{stmaryrd}
\usepackage{array}
\usepackage{psfrag}
\usepackage{dsfont}
\usepackage{epsfig}
\usepackage{titletoc}
\usepackage{float}   
\usepackage{wrapfig} 
\usepackage{amsmath}
\usepackage[latin1]{inputenc}
\usepackage{amsmath}\usepackage{amssymb,stmaryrd}
\usepackage{mathrsfs}
\usepackage{array}
\usepackage{graphicx}
\usepackage{psfrag}
\usepackage{dsfont}
\usepackage{epsfig}
\usepackage{cancel}
\usepackage[dvips]{feynmp}
\usepackage{upgreek}
\usepackage{subfig}
\usepackage{tabularx}
\usepackage{xcolor}
\usepackage{units}
\usepackage{textcomp}
\usepackage{wasysym}
\usepackage{stmaryrd}
\usepackage{dutchcal}

\usepackage{textfit} 

\usepackage{hyperref}

\renewcommand{\eqref}[1]{\mbox{Eq.~(\ref{#1})}}
\newcommand{\tabref}[1]{\mbox{Tab.~\ref{#1}}}
\newcommand{\figref}[1]{\mbox{Fig.~\ref{#1}}}
\newcommand{\secref}[1]{\mbox{Sec.~\ref{#1}}}

\newcommand{\appref}[1]{\mbox{App.~\ref{#1}}}

\begin{document}

\title{Eikonal approximation, Finsler structures, and implications for Lorentz-violating photons in weak gravitational fields}

\author{M. Schreck} \email{mschreck@indiana.edu}
\affiliation{Indiana University Center for Spacetime Symmetries, Indiana University, Bloomington, Indiana 47405-7105}

\begin{abstract}
The current article shall contribute to understanding the classical analogue of the minimal photon sector in the Lorentz-violating
Standard-Model Extension (SME). It is supposed to complement all studies performed on classical point-particle equivalents of SME
fermions. The classical analogue of a photon is not a massive particle being described by a usual
equation of motion, but a geometric ray underlying the eikonal equation. The first part of the paper will set up the necessary tools
to understand this correspondence for interesting cases of the minimal SME photon sector. In conventional optics the eikonal equation
follows from an action principle, which is demonstrated to work in most (but not all) Lorentz-violating cases as well. The integrands of
the action functional correspond to Finsler structures, which establishes the connection to Finsler geometry. The second part of the
article treats Lorentz-violating light rays in a weak gravitational background by
implementing the principle of minimal coupling. Thereby it is shown how Lorentz violation in the photon sector can be constrained
by measurements of light bending at massive bodies such as the Sun. The phenomenological studies are based on the currently
running ESA mission GAIA and the planned NASA/ESA mission LATOR. The final part of the paper discusses certain aspects
of explicit Lorentz violation in gravity based on the setting of Finsler geometry.
\end{abstract}

\maketitle

\newpage
\setcounter{equation}{0}
\setcounter{section}{0}
\renewcommand{\theequation}{\arabic{section}.\arabic{equation}}

\section{Introduction}

During the past 15 years plenty of progress has been made in understanding {\em CPT}- and Lorentz violation and its possible implications
on physics from both a theoretical and a phenomenological point of view. This was made possible by establishing the Standard-Model
Extension (SME) in 1998 \cite{Colladay:1998fq} and by the subsequent tireless work of people in our community eager to study
imprints of Planck-scale physics detectable by experiments operating at much smaller energies. The SME is a powerful framework incorporating
all Lorentz-violating operators into the Standard Model of elementary particles and General Relativity. It neither modifies the gauge structure of
the Standard Model nor does it introduce new particles. The power-counting renormalizable contributions of the SME are grouped into its
minimal part where the remaining higher-order operators comprise the nonminimal SME \cite{Kostelecky:2009zp,Kostelecky:2011gq,Kostelecky:2013rta}.
This framework allows for astounding experimental tests of Lorentz invariance where even presently some experiments reach a sensitivity
of the Planck scale square (see \cite{Kostelecky:2008ts} for a yearly updated compilation of experimental contraints on Lorentz-violating
coefficients). Since Lorentz violation implies {\em CPT}-violation according to a theorem by Greenberg \cite{Greenberg:2002uu},
the Standard-Model Extension involves all {\em CPT}-odd operators as a subset. Note that Lorentz violation has been predicted
by various prototypes of fundamental theories such as string theory \cite{Kostelecky:1988zi,Kostelecky:1991ak,Kostelecky:1994rn}, loop
quantum gravity \cite{Gambini:1998it,Bojowald:2004bb}, noncommutative spacetime \cite{AmelinoCamelia:1999pm,Carroll:2001ws},
spacetime foam \cite{Klinkhamer:2003ec,Bernadotte:2006ya,Hossenfelder:2014hha}, and models with nontrivial spacetime topology
\cite{Klinkhamer:1998fa,Klinkhamer:1999zh}.

In the recent past profound studies of modified quantum field theories based on the SME were performed at tree-level and including
quantum corrections. The result of these studies is that most sectors are free of any inconsistencies \cite{Kostelecky:2000mm,oai:arXiv.org:hep-ph/0101087,Casana-etal2009,Casana-etal2010,Klinkhamer:2010zs,Schreck:2011ai,Cambiaso:2012vb,Colladay:2014dua,Maniatis:2014xja,Schreck:2013gma,Schreck:2013kja,Cambiaso:2014eba,Schreck:2014qka,Colladay:2014dua,Albayrak:2015ewa}. Furthermore the SME was explicitly
shown to be renormalizable at one loop \cite{Kostelecky:2001jc,Colladay:2006rk,Colladay:2007aj,Colladay:2009rb} where latest computations have
demonstrated renormalizability of the modified quantum electrodynamics \cite{Santos:2015koa} and the pure Yang-Mills sector \cite{Santos:2014lfa}
at infinite-loop order using algebraic techniques. Therefore as long as the SME is restricted to Minkowski spacetime,
it seems to be a reasonable, well-behaved, and model-independent test framework for Planck-scale physics.

The gravitational sector of the SME was constructed in the seminal article \cite{Kostelecky:2003fs}. In the aftermath, studies on its theory
and phenomenology were performed in a successive series of papers \cite{Bailey:2006fd,Kostelecky:2007kx,Kostelecky:2008in,Bailey:2009me,Kostelecky:2010ze,Tasson:2010nr,Tasson:2012nx,Tasson:2012au,Bonder:2013sca,Bonder:2015maa} with recent investigations of even nonminimal operators in short-range gravity tests \cite{Bailey:2014bta,Long:2014swa}.
One of the most important theoretical results of \cite{Kostelecky:2003fs} is a no-go theorem stating that explicit Lorentz violation is incompatible with the
geometric framework of General Relativity, which is Riemannian geometry. Considering Lorentz-violating matter in a gravitational background
results in modified conservation laws of the energy-momentum tensor based on Noether's theorem. However Lorentz violation does {\em a
priori} not modify the geometrical base such as the Bianchi identities of the Riemann curvature tensor. Due to the Einstein equations the
second Bianchi identity is tightly bound to the conservation of energy-momentum, which is then incommensurate with the modified matter
sector.

A possibility of circumventing this clash is to perform phenomenological studies in theories resting on spontaneous Lorentz violation.
This means that a Lorentz-violating background field arises dynamically as the vacuum expectation value of a vector or tensor field. Such models have been studied
since the early 1990s \cite{Kostelecky:1988zi,Kostelecky:1989jp,Kostelecky:1989jw,Bluhm:2008yt,Hernaski:2014jsa,Bluhm:2014oua} (even
before the SME existed) and they can be considered as one of the motivations that lead to the construction of the SME. The
crucial point within models of spontaneous Lorentz violation is to take into account the Nambu-Goldstone modes that are linked to the
symmetry breaking. This can lead to arduous perturbative calculations within such a theory.

For these reasons it would be preferable to have a setup available that allows for incorporating explicit Lorentz violation into a curved
background without possible tensions with the underlying geometrical properties. A suggestion was already given in \cite{Kostelecky:2003fs} along the same lines
as the no-go theorem: introducing an alternative geometrical framework that can include preferred directions naturally. Such an extension
of Riemannian geometry has been known in the mathematics community for almost 100 years. It is named Finsler geometry in reference to the
famous mathematician Finsler who studied generalized path length functionals in his Ph.D. thesis \cite{Finsler:1918,Cartan:1933} (cf.~\cite{Bao:2000}
for a comprehensive mathematical overview on the subject).

Finsler geometry has been applied to various fields of physics \cite{Antonelli:1993}. In the context of the Standard-Model Extension it found its use just
a couple of years ago when it was shown that the minimal Lorentz-violating fermion sector can be mapped to classical-particle descriptions
\cite{Kostelecky:2010hs,Kostelecky:2011qz,Kostelecky:2012ac,Colladay:2012rv,Russell:2015gwa}.
The corresponding Lagrange functions are closely linked to Finsler structures, i.e., generalized path length functionals. Recently a
nonminimal case was studied \cite{Schreck:2014hga} as well as classical-particle trajectories in electromagnetic fields and modified spin
precession based on an isotropic set of minimal fermion coefficients \cite{Schreck:2014ama}. In \cite{Silva:2013xba} a particular class of
Finsler spaces known as bipartite is investigated closer from a physics point of view and \cite{Foster:2015yta} suggests classical-mechanics
systems that are linked to three-dimensional versions of Finsler $b$ space \cite{Kostelecky:2011qz}. In a very recent paper
\cite{Colladay:2015wra} $b$ space is discussed from a mathematical point of view. Its indicatrix (surface of constant value of the Finsler
structure) is a two-valued deformation from a sphere that is characterized by singularities with ambiguous derivatives. Considering the indicatrix as an
algebraic variety, the Hironaka theorem says that such singularities can be removed \cite{Hironaka:1964}. In \cite{Colladay:2015wra} a
coordinate transformation was found, which allows to remove the singular sets and to glue the remaining parts together appropriately. This
results in a well-defined mathematical description of $b$ space that can be used for future physical investigations.

The goals of the current article are threefold. First, analogous classical equivalents for the minimal {\em CPT}-even photon sector of the SME
shall be found. Second, with these equivalents at hand we intend to study phenomenological aspects of Lorentz-violating photons in weak gravitational fields. Last but not
least we would like to understand various consequences of this approach on the base of Finsler geometry. The procedures to be developed
will differ extensively from the SME fermion counterparts. The paper is organized as follows. In \secref{sec:construction-classical-lagrangians}
the Lorentz-violating framework, which all investigations are based on, is introduced. A brief review on Finsler geometry and Finsler
structures in the SME fermion sector is given in \secref{sec:finsler-structures-photon}, followed by an explanation of the method to constructing
Finsler structures in the photon sector. In that section we investigate different cases that are the most interesting ones from a
physics point of view. In the geometric-optics approximation photons are described by the eikonal equation, which forms the cornerstone of
\secref{sec:classical-equations-of-motion}. It is demonstrated how the Finsler structures obtained are linked to the eikonal equation for
the different sectors analyzed in the previous section. Since the isotropic modification of the {\em CPT}-even sector can be considered to be
the most important one, all forthcoming studies will be based on the latter. Section \ref{sec:gravitational-backgrounds} is dedicated to
investigating the isotropic eikonal equation in a weak gravitational
background. We develop a phenomenological framework to study light bending at massive bodies within such a theory. In this context
prospects are given on detecting isotropic Lorentz violation of photons propagating in a gravitational background. This is carried out for two
space-based missions: GAIA and LATOR. The final part of the paper is more theoretical. In \secref{sec:modified-energy-momentum-conservation}
the modified conservation law of the energy-momentum tensor is investigated, interpreting the results from the point of view of explicit
versus spontaneous Lorentz violation. Last but not least, in \secref{sec:properties-isotropic-finsler} we examine the properties of the isotropic
spacetime that has been subject to the studies in \secref{sec:gravitational-backgrounds} from a Finsler-geometric point of view. The most
important findings in total are concluded on and discussed in \secref{sec:conclusion}. Essential calculational details can be found in Appx.
\ref{sec:lagrangians-massive-photons} to \ref{sec:eikonal-equation-inhomogeneous-anisotropic}. Throughout the article natural units with
$\hbar=c=1$ are chosen unless otherwise stated.

\section{Construction of classical Lagrangians}
\label{sec:construction-classical-lagrangians}

The base of the current article is formed by the minimal SME photon sector whose action $S_{\upgamma}$ is comprised of {\em CPT}-even
modified Maxwell (mM) \cite{Colladay:1998fq,Kostelecky:2002hh,BaileyKostelecky2004} theory and {\em CPT}-odd Maxwell-Chern-Simons
(MCS) theory \cite{Carroll:1989vb,Colladay:1998fq,Kostelecky:2002hh,BaileyKostelecky2004}:
\begin{subequations}
\label{eq:action-modified-maxwell-theory}
\begin{align}
S_{\upgamma}&=\int_{\mathbb{R}^4}\mathrm{d}^4x\,\left[\mathcal{L}_\text{mM}(x)+\mathcal{L}_{\mathrm{MCS}}(x)+\mathcal{L}_{\mathrm{mass}}(x)\right]\,, \displaybreak[0]\\[2ex]
\label{eq:lagrange-density-modmax}
\mathcal{L}_\text{mM}(x)&=-\frac{1}{4}\,\eta^{\mu\rho}\,\eta^{\nu\sigma}\,F_{\mu\nu}(x)F_{\rho\sigma}(x)-\frac{1}{4}\,(k_F)^{\mu\nu\varrho\sigma}\,F_{\mu\nu}(x)F_{\varrho\sigma}(x)\,, \displaybreak[0]\\[2ex]
\label{eq:lagrange-density-mcs}
\mathcal{L}_{\mathrm{MCS}}(x)&=\frac{m_{\scriptscriptstyle{\mathrm{CS}}}}{2}(k_{AF})^{\kappa}\varepsilon_{\kappa\lambda\mu\nu}A^{\lambda}(x)F^{\mu\nu}(x)\,, \displaybreak[0]\\[2ex]
\mathcal{L}_{\mathrm{mass}}(x)&=m_{\upgamma}^2A_{\mu}(x)A^{\mu}(x)\,.
\end{align}
\end{subequations}
Here $F_{\mu\nu}(x)\equiv \partial_{\mu}A_{\nu}(x)-\partial_{\nu}A_{\mu}(x)$ is the electromagnetic field strength tensor that involves the \textit{U}(1)
gauge field $A_{\mu}(x)$. The fields are defined on Minkowski spacetime with metric $(\eta_{\mu\nu})=\mathrm{diag}(1,-1,-1,-1)$.
The totally antisymmetric Levi-Civita symbol in four spacetime dimensions is denoted as $\varepsilon^{\mu\nu\varrho\sigma}$
with $\varepsilon^{0123}=1$.
The controlling coefficients characteristic for the framework considered are comprised in the fourth-rank observer tensor $(k_F)^{\mu\nu\varrho\sigma}$
and the observer vector $(k_{AF})^{\kappa}$. Both have dimensionless components and they do not transform covariantly with respect to particle
Lorentz transformations, which renders this theory explicitly Lorentz-violating. The field operator of modified Maxwell theory is of dimension four,
whereas the operator of MCS theory has mass dimension three. Therefore MCS theory involves the Chern-Simons mass scale $m_{\scriptscriptstyle{\mathrm{CS}}}$
for dimensional consistency. It is well-known that a photon mass term encoded in $\mathcal{L}_{\mathrm{mass}}$ (with the photon mass
$m_{\upgamma}$) violates \textit{U}(1) gauge invariance. It has been introduced here for certain purposes that will be explained below, but for
most occasions $m_{\upgamma}$ will be set to zero. Anyhow in \cite{Colladay:2014dua} it was demonstrated that certain birefringent cases
of modified Maxwell theory require a nonvanishing photon mass (at least in intermediate calculations) to have a consistent Gupta-Bleuler
quantization. Finally, a gauge fixing term will be omitted in the action, since all considerations will be carried out at the classical level.

\subsection{Classical Lagrangians and Finsler structures}
\label{sec:classical-lagrangians-finsler-structures}

The major goal is to understand how Lorentz-violating photons can be described in the context of gravity. Since Einstein's relativity is a classical
theory, it is reasonable to obtain a classical analogue of the quantum field theory based on the action of \eqref{eq:action-modified-maxwell-theory}.
With such an analogue at hand it should be possible to study how an explicitly Lorentz-violating theory of gravity could be constructed consistently.
As an introduction to the topic the mapping procedure of the SME fermion sector to a classical point-particle description \cite{Kostelecky:2010hs} shall be reviewed.
From a quantum theoretical point of view a particle can be understood as a suitable superposition of free-field solutions with dispersion relation
\begin{equation}
\label{eq:dispersion-relation-general}
f(p_{\mu},m_{\psi},k_x)=0\,,\quad (p_{\mu})=\begin{pmatrix}
p_0 \\
\mathbf{p} \\
\end{pmatrix}\,,
\end{equation}
such that its probability density is nonzero in a localized region and drops off to zero sufficiently fast outside. Here $p_0$ is the particle energy,
$\mathbf{p}$ its three-momentum, $m_{\psi}$ the fermion mass, and $k_x$ denotes a particular set of Lorentz-violating coefficients where $x$
represents a Lorentz index structure. The physical propagation velocity of such a wave packet is the group velocity
\begin{equation}
\mathbf{v}_{\mathrm{gr}}\equiv \frac{\partial p_0}{\partial\mathbf{p}}\,.
\end{equation}
A classical, relativistic pointlike particle is assumed to propagate with four-velocity $u^{\mu}=\gamma(1,\mathbf{v})$ where $\mathbf{v}$ is
the three-velocity. To map the wave packet to such a classical particle, it makes sense to identify the group velocity components with the
appropriate spatial four-velocity components:
\begin{equation}
\label{eq:group-velocity-correspondence}
\mathbf{v}_{\mathrm{gr}}\overset{!}{=} -\frac{\mathbf{u}}{u^0}\,.
\end{equation}
The minus sign has its origin in the different position of the spatial index on both sides of the equation. Since the physics of the classical
particle rests on a Lagrange function $L=L(u^0,\mathbf{u})$, its construction is of paramount importance. If the Lagrange function is positive
homogeneous of degree one, i.e., $L(\lambda u^0,\lambda\mathbf{u})=\lambda L(u^0,\mathbf{u})$ for $\lambda>0$, the action is parameterization-invariant.
In this case the physics does not depend on the way how the particle trajectory is parameterized, which is a very reasonable property to have.
Positive homogeneity gives the following condition on the Lagrange density according to Euler's theorem \cite{Bao:2000}:
\begin{equation}
\label{eq:lagrange-function}
L=-p_{\mu}u^{\mu}\,,\quad p_{\mu}=-\frac{\partial L}{\partial u^{\mu}}\,,
\end{equation}
with the conjugate momentum $p_{\mu}$. The latter is identified with the momentum that appears in the quantum theoretical dispersion
relation of \eqref{eq:dispersion-relation-general}. The global minus sign in its definition has been introduced such that the nonrelativistic kinetic energy is
positive. Now Eqs.~(\ref{eq:dispersion-relation-general}), (\ref{eq:group-velocity-correspondence}),
and (\ref{eq:lagrange-function}) comprise a set of five conditions that shall be used to determine $p_{\mu}$ and $L$. Hence all four-momentum
components and the Lagrange function are supposed to be solely expressed in terms of four-velocity components.

The Lagrange functions corresponding to the standard fermion dispersion law $p_0^2-\mathbf{p}^2-m_{\psi}^2=0$ read $L=\pm m_{\psi}\sqrt{(u^0)^2-\mathbf{u}^2}$.
The two signs are the classical counterparts of the particle-antiparticle solutions at the level of quantum field theory. It can be checked that the five equations above are fulfilled for
this choice of $L$. The latter can also be written in the form $L=\pm m_{\psi}\sqrt{r_{\mu\nu}u^{\mu}u^{\nu}}$ with $r_{\mu\nu}$ known as the
intrinsic metric. This metric is essential to determine lengths of vectors and angles enclosed by vectors. In the particular case considered it
corresponds to the (indefinite) Minkowski metric: $r_{\mu\nu}=\eta_{\mu\nu}$. This is not surprising, since the starting point to obtaining the Lagrange
function was a field theory defined in Minkowski spacetime. By a Wick rotation the Lagrange function is related to a new function $F$ based
on a positive definite intrinsic metric:
\begin{equation}
F(y)\equiv F(\mathbf{y},y^4)\equiv \frac{\mathrm{i}}{m_{\psi}}L(\mathrm{i}y^4,\mathbf{y})=\sqrt{r_{ij}y^iy^j}\,,\quad r_{ij}=\mathrm{diag}(1,1,1,1)_{ij}\,.
\end{equation}
Promoting $r_{ij}$ to an arbitrary position-dependent metric $r_{ij}(x)$, the function $F$ becomes dependent on $x$: $F(y)\mapsto F(x,y)$.
It can then be interpreted as the integrand of a path length functional of a Riemannian manifold $M$ where $y\in T_xM$. A Finsler structure is a
generalization of that obeying the following properties:

\begin{itemize}

\item[1)] $F(x,y)>0$,
\item[2)] $F(x,y)\in C^{\infty}$ for all $y\in T_xM\setminus \{\mathrm{slits}\}$,
\item[3)] positive homogeneity in $y$, i.e., $F(x,\lambda y)=\lambda F(x,y)$ for $\lambda>0$, and
\item[4)] the derived metric (Finsler metric)
\begin{equation}
g_{ij}\equiv \frac{1}{2}\frac{\partial^2 F^2}{\partial y^i\partial y^j}\,,
\end{equation}
is positive definite.

\end{itemize}
Prominent examples for Finsler structures that are outside the scope of Riemannian geometry are Randers structures, $F(y)=\alpha+\beta$,
and Kropina structures, $F(y)=\alpha^2/\beta$, with $\alpha=\sqrt{a_{ij}y^iy^j}$ and $\beta=b_iy^i$ where $a_{ij}$ is a Riemannian
metric and $b_i$ a one-form. There are certain theorems available to classify Finsler structures using various kinds of torsions. The
most important one is the Cartan torsion $C_{ijk}$, which is given by \cite{Bao:2004}
\begin{equation}
\label{eq:cartan-torsion}
C_{ijk}\equiv \frac{1}{2}\frac{\partial g_{ij}}{\partial y^k}=\frac{1}{4}\frac{\partial^3F^2}{\partial y^i\partial y^j\partial y^k}\,.
\end{equation}
In some books $C_{ijk}$ is defined with an additional prefactor $F$ (see, e.g., \cite{Bao:2000}). The mean Cartan torsion reads as
follows:
\begin{equation}
\label{eq:mean-cartan-torsion}
\mathbf{I}\equiv I_iy^i\,,\quad I_i\equiv g^{jk}C_{ijk}\,,\quad (g^{ij})\equiv (g_{ij})^{-1}\,,
\end{equation}
with the inverse derived metric $g^{ij}$. Deicke's theorem says that a Finsler space is Riemannian if and only if $\mathbf{I}$ vanishes
\cite{Deicke:1953}. The Matsumoto torsion provides a further set of quantities that are very useful to classify Finsler structures:
\begin{equation}
\label{eq:matsumoto-torsion}
M_{ijk}\equiv C_{ijk}-\frac{1}{n+1}(I_ih_{jk}+I_jh_{ik}+I_kh_{ij})\,,\quad h_{ij}\equiv F \frac{\partial^2F}{\partial y^i\partial y^j}\,.
\end{equation}
Here $n$ is the dimension of the Finsler structure considered \cite{Bao:2004}. According to the Matsumoto-H\={o}j\={o} theorem
a Finsler structure is either of Randers or Kropina type if and only if the Matsumoto torsion is equal to zero \cite{Matsumoto:1978}.
These theorems will be used frequently throughout the paper to classify Finsler structures encountered.

According to the rules recalled at the beginning of the current section classical Lagrange functions of the SME fermion sector were
derived in \cite{Kostelecky:2010hs,Colladay:2012rv,Schreck:2014hga,Schreck:2014ama,Russell:2015gwa}. In the articles
\cite{Colladay:2012rv,Kostelecky:2011qz,Kostelecky:2012ac,Schreck:2014hga} their corresponding Finsler structures were examined.
In this paper analogous investigations shall be performed for the minimal SME photon sector based on the action of
\eqref{eq:action-modified-maxwell-theory}. It will become evident that the possible techniques used differ from the procedures adopted
for the fermion sector.

\subsection{Maxwell-Chern-Simons theory}

In the current section the {\em CPT}-even photon sector components $(k_F)^{\mu\nu\varrho\sigma}$ in \eqref{eq:lagrange-density-modmax} will be set to zero restricting our
considerations to the MCS term of \eqref{eq:lagrange-density-mcs} only. Furthermore the photon mass $m_{\upgamma}$ will be set to zero as well.
In the seminal article \cite{Carroll:1989vb} the magnitude of $m_{\scriptscriptstyle{\mathrm{CS}}}(k_{AF})^{\kappa}$ was constrained tightly due to
the absence of astrophysical birefringence. A collection of all constraints on components of $m_{\scriptscriptstyle{\mathrm{CS}}}(k_{AF})^{\kappa}$
can be found in the data tables \cite{Kostelecky:2008ts}. In spite of the tight bounds, MCS theory is very interesting from a theoretical point of view.
The structure of the quantum field theory based on MCS theory is quite involved, which was shown by extensive investigations carried out in
\cite{oai:arXiv.org:hep-ph/0101087}.\footnote{Note that the global prefactor of MCS-theory is different in the latter reference.} The smoking-gun results of
the latter reference are that MCS theory is well-behaved as long as the
preferred spacetime direction $(k_{AF})^{\kappa}$ is spacelike. For timelike $(k_{AF})^{\kappa}$ issues with either microcausality or unitarity
arise, though. Interestingly this behavior mirrors in the classical Finsler structure of MCS theory that will be derived as follows. First of all
spacelike MCS theory shall be considered. The modified field equations in momentum space read as follows \cite{Colladay:1998fq}:
\begin{subequations}
\begin{align}
M^{\alpha\delta}(p)A_{\delta}&=0\,, \\[2ex]
M^{\alpha\delta}(p)&=\eta^{\alpha\delta}k^2-k^{\alpha}k^{\delta}-2\mathrm{i}m_{\scriptscriptstyle{\mathrm{CS}}}(k_{AF})_{\beta}\varepsilon^{\alpha\beta\gamma\delta}k_{\gamma}\,,
\end{align}
\end{subequations}
where $k_{\mu}$ is the four-momentum to be distinguished from the four-momentum $p_{\mu}$ used for fermions.
Imposing Lorenz gauge $k^{\delta}A_{\delta}=0$, the condition of a vanishing determinant of $M$ results in
\begin{equation}
k^4+4m_{\scriptscriptstyle{\mathrm{CS}}}^2\left[k^2(k_{AF})^2-(k\cdot k_{AF})^2\right]=0\,,
\end{equation}
leading to the following dispersion relations:
\begin{equation}
\omega_{1,2}=\sqrt{\mathbf{k}^2+2m_{\scriptscriptstyle{\mathrm{CS}}}^2(\mathbf{k}_{AF})^2\pm 2m_{\scriptscriptstyle{\mathrm{CS}}}\sqrt{m_{\scriptscriptstyle{\mathrm{CS}}}^2(\mathbf{k}_{AF})^4+(\mathbf{k}\cdot \mathbf{k}_{AF})^2}}\,.
\end{equation}
Here the spatial momentum $\mathbf{k}$ is not to be confused with the spatial part $\mathbf{k}_{AF}$ of the MCS vector. Following
the procedure outlined in \appref{sec:lagrangians-cpt-odd-massive-photons} leads to the Lagrange function
\begin{equation}
\label{eq:lagrange-function-mcs-theory}
L|_{\mathrm{MCS}}^{\pm}=\pm m_{\scriptscriptstyle{\mathrm{CS}}}\left(\sqrt{-(k_{AF})^2u^2}\pm \sqrt{(k_{AF}\cdot u)^2-(k_{AF})^2u^2}\right)\,.
\end{equation}
First of all, this result matches the Lagrange function first obtained in \cite{McGinnis:2014}. For spacelike $k_{AF}$ it corresponds to the Lagrange
density of the minimal fermionic $b^{\mu}$ coefficient where here $(k_{AF})^{\mu}$ takes the role of $b^{\mu}$ and the Chern-Simons
mass $m_{\scriptscriptstyle{\mathrm{CS}}}$ takes the role of the fermion mass $m_{\psi}$. This is because there exists a
correspondence between MCS theory and the fermion theory involving the $b^{\mu}$ coefficient whose Lagrangian has the form
$b^{\mu}\overline{\psi}\gamma_5\gamma_{\mu}\psi$. The associated field operator is of dimension three and it is {\em CPT}-odd
\cite{Kostelecky:2009zp}, which parallels some of the properties of MCS theory. Therefore the Wick-rotated version of
\eqref{eq:lagrange-function-mcs-theory} can be interpreted as a $b$ space. The form of the Lagrangian of \eqref{eq:lagrange-function-mcs-theory}
remains the same even for MCS theory with a timelike $k_{AF}$, which can be shown by direct computation. Undoubtedly,
issues arise for timelike $k_{AF}$, since in this case the Lagrange function is not a real function any more.

A classical Lagrange function is of mass dimension one, which is why \eqref{eq:lagrange-function-mcs-theory} is directly proportional to the single mass scale
$m_{\scriptscriptstyle{\mathrm{CS}}}$ that appears in this framework. In the limit $m_{\scriptscriptstyle{\mathrm{CS}}}\mapsto 0$ the
Lagrange function vanishes, which reveals the challenge in deriving appropriate Lagrange functions corresponding to Lorentz-violating
frameworks that do not have a dimensional scale associated to them. This is especially the case for a photon theory based on modified
Maxwell theory, which will be discussed as follows.

\subsection{Modified Maxwell theory}
\label{eq:lagrange-function-massive-modmax}

In the remainder of the paper the Chern-Simons mass $m_{\scriptscriptstyle{\mathrm{CS}}}$ will be set to zero and the Lagrange density
of MCS theory, \eqref{eq:lagrange-density-mcs}, will not be taken into account any more. The observer four-tensor $(k_F)^{\mu\nu\varrho\sigma}$
in \eqref{eq:lagrange-density-modmax} will be decomposed into contributions involving the Minkowski metric and a $(4\times 4)$ matrix
$\widetilde{\kappa}^{\mu\nu}$ according to the nonbirefringent \textit{Ansatz} \cite{BaileyKostelecky2004,Altschul:2006zz}
\begin{equation}
\label{eq:nonbirefringent-ansatz}
(k_F)^{\mu\nu\varrho\sigma}=\frac{1}{2}(\eta^{\mu\varrho}\widetilde{\kappa}^{\nu\sigma}-\eta^{\mu\sigma}\widetilde{\kappa}^{\nu\varrho}-\eta^{\nu\varrho}\widetilde{\kappa}^{\mu\sigma}+\eta^{\nu\sigma}\widetilde{\kappa}^{\mu\varrho})\,.
\end{equation}
The matrix $\widetilde{\kappa}^{\mu\nu}$ is supposed to be symmetric and traceless. Its particular choice amounts to different
Lorentz-violating cases in the minimal, {\em CPT}-even photon sector characterized by nonbirefringent photon dispersion laws at first order in the Lorentz-violating
coefficients. This means that resulting dispersion relations for the two physical photon polarization states coincide with each other at first
order in Lorentz violation. The notation --- especially for the controlling coefficients --- is mainly based on \cite{Kostelecky:2002hh}.

First of all the photon mass is kept. The equations of motion for the photon field $A_{\mu}$ in momentum space then take the following
form \cite{Colladay:1998fq,McGinnis:2014}:
\begin{subequations}
\label{eq:field-equations-modmax}
\begin{align}
M^{\alpha\delta}(k)A_{\delta}&=0\,, \\[2ex]
M^{\alpha\delta}(k)&=\eta^{\alpha\delta}(k^2-m_{\upgamma}^2)-k^{\alpha}k^{\delta}-2(k_F)^{\alpha\beta\gamma\delta}k_{\beta}k_{\gamma}\,.
\end{align}
\end{subequations}
Now different interesting cases of modified Maxwell theory (including a photon mass term) will be examined. The simplest case is undoubtedly
the isotropic one, which is characterized by a single controlling coefficient $\widetilde{\kappa}_{\mathrm{tr}}$ and one preferred timelike spacetime
direction $\xi^{\mu}$. The matrix $\widetilde{\kappa}^{\mu\nu}$ is then diagonal and it is given as follows:
\begin{subequations}
\begin{align}
\label{eq:matrix-kappas-isotropic-case}
\widetilde{\kappa}^{\mu\nu}&=2\widetilde{\kappa}_{\mathrm{tr}}\left(\xi^{\mu}\xi^{\nu}-\frac{1}{4}\xi^2\eta^{\mu\nu}\right)=\frac{3}{2}\widetilde{\kappa}_{\mathrm{tr}}\,\mathrm{diag}\left(1,\frac{1}{3},\frac{1}{3},\frac{1}{3}\right)^{\mu\nu}\,, \\[2ex]
(\xi^{\mu})&=(1,0,0,0)^T\,.
\end{align}
\end{subequations}
The dispersion equation, which follows from claiming a vanishing determinant of $M^{\alpha\delta}$ in \eqref{eq:field-equations-modmax}
using Lorenz gauge $k^{\delta}A_{\delta}=0$, results in
\begin{subequations}
\begin{align}
\label{eq:dispersion-relation-isotropic-massive}
m_{\upgamma}^2&=a^{\mu\nu}k_{\mu}k_{\nu}\,, \\[2ex]
a^{\mu\nu}&=\mathrm{diag}\left(1+\widetilde{\kappa}_{\mathrm{tr}},-[1-\widetilde{\kappa}_{\mathrm{tr}}],-[1-\widetilde{\kappa}_{\mathrm{tr}}],-[1-\widetilde{\kappa}_{\mathrm{tr}}]\right)^{\mu\nu}\,.
\end{align}
\end{subequations}
The next case to be considered is a nonbirefringent, anisotropic one that is characterized by a single (parity-even) controlling coefficient $\widetilde{\kappa}_{e-}^{11}$
and one spacelike direction $\zeta^{\mu}$. Furthermore $\widetilde{\kappa}_{\mathrm{e}-}^{22}=\widetilde{\kappa}_{e-}^{11}$, $\widetilde{\kappa}_{\mathrm{e}-}^{33}=-2\widetilde{\kappa}_{e-}^{11}$ and all remaining ones vanish. The matrix $\widetilde{\kappa}^{\mu\nu}$ for the nonbirefringent \textit{Ansatz}
is given as follows:
\begin{subequations}
\label{eq:kappas-anisotropic-nonbirefringent-case}
\begin{align}
\widetilde{\kappa}^{\mu\nu}&=3\widetilde{\kappa}_{e-}^{11}\left(\zeta^{\mu}\zeta^{\nu}-\frac{1}{4}\zeta^2\eta^{\mu\nu}\right)=\frac{3}{4}\widetilde{\kappa}_{e-}^{11}\,\mathrm{diag}(1,-1,-1,3)^{\mu\nu}\,, \\[2ex]
(\zeta^{\mu})&=(0,0,0,1)^T\,.
\end{align}
\end{subequations}
The latter has a similar structure compared to \eqref{eq:matrix-kappas-isotropic-case} and it is again diagonal. However its spatial
coefficients differ from each other revealing the anisotropy. The modified photon dispersion equation can be written in the same form as
for the isotropic case:
\begin{subequations}
\begin{align}
\label{eq:dispersion-relation-anisotropic-massive}
m_{\upgamma}^2&=b^{\mu\nu}k_{\mu}k_{\nu}\,, \\[2ex]
b^{\mu\nu}&=\mathrm{diag}\left(1+\frac{3}{2}\widetilde{\kappa}_{e-}^{11},-\left[1+\frac{3}{2}\widetilde{\kappa}_{e-}^{11}\right],-\left[1+\frac{3}{2}\widetilde{\kappa}_{e-}^{11}\right],-\left[1-\frac{3}{2}\widetilde{\kappa}_{e-}^{11}\right]\right)^{\mu\nu}\,.
\end{align}
\end{subequations}
The third particular case of modified Maxwell theory to be examined in this context is characterized by three (parity-odd) controlling coefficients $\widetilde{\kappa}_{o+}^{23}$,
$\widetilde{\kappa}_{o+}^{31}$, and $\widetilde{\kappa}_{o+}^{12}$ where all remaining ones that are not related by symmetries vanish.
Furthermore there are two preferred spacetime directions: a timelike direction $\xi^{\mu}$ and a spacelike one $\zeta^{\mu}$. The matrix
$\widetilde{\kappa}^{\mu\nu}$ in the nonbirefrigent \textit{Ansatz} can be cast into
\begin{subequations}
\begin{align}
\widetilde{\kappa}^{\mu\nu}&=\frac{1}{2}(\xi^{\mu}\zeta^{\nu}+\zeta^{\mu}\xi^{\nu})-\frac{1}{4}(\xi\cdot\zeta)\eta^{\mu\nu}\,, \\[2ex]
\label{eq:four-vectors-parity-odd}
(\xi^{\mu})&=(1,0,0,0)^T\,,\quad (\zeta^{\mu})=-2(0,\boldsymbol{\zeta})^T\,,\quad \boldsymbol{\zeta}=(\widetilde{\kappa}_{o+}^{23},\widetilde{\kappa}_{o+}^{31},\widetilde{\kappa}_{o+}^{12})^T\,.
\end{align}
\end{subequations}
Due to observer Lorentz invariance the coordinate system can be set up such that $\boldsymbol{\zeta}$ points along its third axis.
The first photon dispersion equation is quadratic and reads as follows:
\begin{subequations}
\label{eq:dispersion-relation-parity-odd-massive-1}
\begin{align}
m_{\upgamma}^2&=c^{\mu\nu}k_{\mu}k_{\nu}\,, \\[2ex]
c^{\mu\nu}&=\begin{pmatrix}
1 & 0 & 0 & -\mathcal{E} \\
0 & -1 & 0 & 0 \\
0 & 0 & -1 & 0 \\
-\mathcal{E} & 0 & 0 & -1 \\
\end{pmatrix}^{\mu\nu}\,,\quad \mathcal{E}=\sqrt{(\widetilde{\kappa}_{o+}^{23})^2+(\widetilde{\kappa}_{o+}^{31})^2+(\widetilde{\kappa}_{o+}^{12})^2}\,.
\end{align}
\end{subequations}
Note that the latter has an equivalent structure to Eqs.~(\ref{eq:dispersion-relation-isotropic-massive}), (\ref{eq:dispersion-relation-anisotropic-massive}).
However the second dispersion equation is quartic and it is given by
\begin{equation}
\label{eq:dispersion-relation-parity-odd-massive-2}
0=(k^2-m_{\upgamma}^2)^2-(k\cdot\zeta)(k\cdot\xi)(k^2-m_{\upgamma}^2)+\frac{1}{4}\left\{(k\cdot\zeta)^2+\zeta^2[(k\cdot\xi)^2-k^2]\right\}k^2\,.
\end{equation}
For $m_{\upgamma}=0$ the right-hand side of the latter factorizes into $k^2$ and a quadratic dispersion relation that differs from
\eqref{eq:dispersion-relation-parity-odd-massive-1} (for $m_{\upgamma}=0$) at second order in the controlling coefficients. The nonbirefringent
\textit{Ansatz} of \eqref{eq:nonbirefringent-ansatz} prevents birefringence to occur only at leading order in Lorentz violation.

Now the classical Lagrange functions for all cases previously introduced are given as follows. The derivation for one particular
of those is shown in \appref{sec:lagrangians-cpt-even-massive-photons} and it works analogously for the remaining ones.
For the isotropic case (denoted as $\circledcirc$) the Lagrange functions read as
\begin{subequations}
\begin{align}
\label{eq:massive-lagrangian-isotropic}
L|_{\circledcirc}^{\pm}&=\pm m_{\upgamma}\sqrt{a_{\mu\nu}u^{\mu}u^{\nu}}\,, \\[2ex]
(a_{\mu\nu})&=\mathrm{diag}\left(\frac{1}{1+\kappa_{\mathrm{tr}}},-\frac{1}{1-\kappa_{\mathrm{tr}}},-\frac{1}{1-\kappa_{\mathrm{tr}}},-\frac{1}{1-\kappa_{\mathrm{tr}}}\right)=(a^{\mu\nu})^{-1}\,.
\end{align}
\end{subequations}
For the nonbirefringent, anisotropic case ($\varobar$) they are given by
\begin{subequations}
\begin{align}
\label{eq:massive-lagrangian-anisotropic}
L|_{\varobar}^{\pm}&=\pm m_{\upgamma}\sqrt{b_{\mu\nu}u^{\mu}u^{\nu}}\,, \\[2ex]
(b_{\mu\nu})&=\mathrm{diag}\left(\frac{1}{1+(3/2)\widetilde{\kappa}_{e-}^{11}},-\frac{1}{1+(3/2)\widetilde{\kappa}_{e-}^{11}},-\frac{1}{1+(3/2)\widetilde{\kappa}_{e-}^{11}},-\frac{1}{1-(3/2)\widetilde{\kappa}_{e-}^{11}}\right) \notag \\
&=(b^{\mu\nu})^{-1}\,,
\end{align}
\end{subequations}
Finally for the first dispersion relation of the parity-odd case ($\otimes$) we obtain
\begin{subequations}
\begin{align}
\label{eq:massive-lagrangian-parity-odd}
L|_{\otimes}^{\pm}&=\pm m_{\upgamma}\sqrt{c_{\mu\nu}u^{\mu}u^{\nu}}\,, \displaybreak[0]\\[2ex]
(c_{\mu\nu})&=\begin{pmatrix}
1/(1+\mathcal{E}^2) & 0 & 0 & -\mathcal{E}/(1+\mathcal{E}^2) \\
0 & -1 & 0 & 0 \\
0 & 0 & -1 & 0 \\
-\mathcal{E}/(1+\mathcal{E}^2) & 0 & 0 & -1/(1+\mathcal{E}^2) \\
\end{pmatrix}=(c^{\mu\nu})^{-1}\,.
\end{align}
\end{subequations}
Finding a classical Lagrangian that corresponds to the quartic dispersion equation of \eqref{eq:dispersion-relation-parity-odd-massive-2} is a challenging task
that we leave for the future.
The examples for Lagrange functions of Lorentz-violating photons in \eqref{eq:massive-lagrangian-isotropic}, \eqref{eq:massive-lagrangian-anisotropic},
and \eqref{eq:massive-lagrangian-parity-odd} reveal the general behavior. When the photon dispersion equation is of the form $Q^{\mu\nu}k_{\mu}k_{\nu}=m_{\upgamma}^2$
with an invertible $(4\times 4)$ matrix $Q$ the associated Lagrange function generically reads as (see \cite{Kostelecky:2010hs} for the fermion analogue):
\begin{equation}
L^{\pm}=\pm m_{\upgamma}\sqrt{Q^{-1}_{\mu\nu}u^{\mu}u^{\nu}}\,,
\end{equation}
These Lagrange functions rely on the existence of a nonzero photon mass. In general, Lagrange functions are of mass dimension one, which is why they
have to involve some dimensionful scale characteristic for the physical problem considered. For the classical fermionic point-particle analogues studied in
\cite{Kostelecky:2010hs} this scale corresponds to the particle mass. In MCS theory the Chern-Simons mass $m_{\scriptscriptstyle{\mathrm{CS}}}$
takes the role of the characteristic dimensionful scale as we saw in \eqref{eq:lagrange-function-mcs-theory}. However since modified Maxwell theory
does not involve a dimensionful scale, a photon mass $m_{\upgamma}$ had to be introduced to construct Lagrange functions for the
classical point-particle analogues.

\subsection{Classical wavefront}

A photon mass is undoubtedly not an attractive feature in a theory, since the mass term violates gauge invariance. Even if a photon mass has
to be introduced as an intermediate ingredient to regularize infrared divergences in quantum corrections or to grant
a consistent quantization of a particular Lorentz-violating framework, cf.~\cite{Colladay:2014dua}, it should be possible to consider the limit
$m_{\upgamma}\mapsto 0$ at the end of any calculation. For this reason an alternative procedure shall be developed to obtain the classical
analogue of (Lorentz-violating) photons. Classically, an electromagnetic pulse makes up a wavefront that can be interpreted as a
surface in four-dimensional spacetime: $w=w(t,\mathbf{x})=0$. In a Lorentz-invariant theory it fulfills the following equation \cite{Fock:1959}:
\begin{equation}
\left(\frac{\partial w}{\partial t}\right)^2-(\boldsymbol{\nabla}w)^2=0\,.
\end{equation}
Computing the square root and choosing one particular sign results in:
\begin{equation}
\label{eq:wave-front-equation}
\frac{\partial w}{\partial t}-\sqrt{\left(\frac{\partial w}{\partial x}\right)^2+\left(\frac{\partial w}{\partial y}\right)^2+\left(\frac{\partial w}{\partial z}\right)^2}=0\,.
\end{equation}
The latter is a Hamilton-Jacobi equation where $w$ is understood as the action $S$ and the expression on the right-hand side as the Hamilton function:
\begin{equation}
\label{eq:hamilton-jacobi-equation}
\frac{\partial S}{\partial t}+H(\mathbf{x},\boldsymbol{\nabla}S)=0\,,\quad S(t,\mathbf{x})=w(t,\mathbf{x})\,,\quad H(\mathbf{x},\mathbf{k})=-\sqrt{\mathbf{k}^2}\,,
\end{equation}
where $\mathbf{k}$ is the wave vector (momentum). Examples that obey \eqref{eq:wave-front-equation} are
\begin{subequations}
\begin{align}
w&=t-\widehat{\mathbf{a}}\cdot \mathbf{x}\,,\quad |\widehat{\mathbf{a}}|=1\,, \\[2ex]
w&=t-\sqrt{\mathbf{x}^2}\,.
\end{align}
\end{subequations}
The first describes a plane wavefront with unit normal vector $\widehat{\mathbf{a}}$ and the second a spherical wavefront. This can be seen by
equating $w$ with zero and considering a fixed value for $t$. Introducing $\lambda$ as a parameter for the trajectory of the wave, both wavefronts
can be differentiated with respect to $\lambda$, which leads to
\begin{subequations}
\begin{align}
\frac{\partial w}{\partial\lambda}&=u^0-\widehat{\mathbf{a}}\cdot\mathbf{u}\,, \\[2ex]
\frac{\partial w}{\partial\lambda}&=u^0-\sqrt{\mathbf{u}^2}\,,\quad u^0\equiv \frac{\mathrm{d}t}{\mathrm{d}\lambda}\,,\quad \mathbf{u}\equiv \frac{\mathrm{d}\mathbf{x}}{\mathrm{d}\lambda}\,.
\end{align}
\end{subequations}
At a first glance it may be assumed that the latter are suitable Lagrange functions, since they are positively homogeneous of degree one. However
computing the derived metrics $g_{\mu\nu}$ according to
\begin{equation}
g_{\mu\nu}\equiv\frac{1}{2}\frac{\partial^2L^2}{\partial u^{\mu}\partial u^{\nu}}\,,
\end{equation}
quickly reveals that their resulting determinants vanish. Therefore such a $g_{\mu\nu}$ is not invertible and definitely fails to describe a
possible (pseudo)-Finsler structure. This is a result that can be shown to hold in general. Assume that a Lagrange function $L$
exists describing the classical wave-front analogue of photons. Then the associated conjugated momentum $p_{\mu}$ must be lightlike
to obey the photon dispersion relation:
\begin{equation}
\label{eq:lightlike-canonical-momentum}
p_{\mu}=-\frac{\partial L}{\partial u^{\mu}}\,,\quad p_{\mu}=-f(u^0,u)\begin{pmatrix}
1 \\
\pm 1 \\
\end{pmatrix}_{\mu}\,.
\end{equation}
Due to rotational symmetry in the Lorentz-invariant case it is sufficient to consider a (1+1)-dimensional spacetime, which is why a lightlike
$p_{\mu}$ must be of the form stated in \eqref{eq:lightlike-canonical-momentum} with a $C^{\infty}$ function $f(u^0,u)$ where
$u\equiv |\mathbf{u}|$. The derived metric is then given by
\begin{align}
g_{\mu\nu}&=\frac{1}{2}\frac{\partial^2 L^2}{\partial u^{\mu}\partial u^{\nu}}= L\frac{\partial^2 L}{\partial u^{\mu}\partial u^{\nu}}+\frac{\partial L}{\partial u^{\mu}}\frac{\partial L}{\partial u^{\nu}} \notag \\
&= L\begin{pmatrix}
f^{(1)} & f^{(2)} \\
\pm f^{(1)} & \pm f^{(2)} \\
\end{pmatrix}_{\mu\nu}+\begin{pmatrix}
f^2 & \pm f^2 \\
\pm f^2 & f^2 \\
\end{pmatrix}_{\mu\nu}=\begin{pmatrix}
 Lf^{(1)}+f^2 &  Lf^{(2)}\pm f^2 \\
\pm( Lf^{(1)}+f^2) & \pm( Lf^{(2)}\pm f^2) \\
\end{pmatrix}_{\mu\nu}\,,
\end{align}
where $(1)$ denotes differentiation with respect to $u^0$ and $(2)$ means differentiation by $u$.
It clearly holds that $\det(g_{\mu\nu})=0$ irrespective of the unknown function $f$. Therefore a Lagrange function $ L$ with an invertible
derived metric cannot exist in the photon case. Because of this an alternative procedure has to be developed to assign a possible (pseudo)-Finsler
structure to photons, which will be examined in what follows.

\section{Finsler structures of the photon sector}
\label{sec:finsler-structures-photon}
\setcounter{equation}{0}

In the previous section it was motivated that the usual method to finding Finsler structures in the fermion sector does not seem to work in the
minimal {\em CPT}-even photon sector. The reason is the absence of a dimensionful physical scale needed for dimensional consistency of a
Lagrange function. Photons must be treated differently from fermions to obtain something like a classical description. This shall be undertaken
in the current section.

\subsection{Lorentz-invariant case}
\label{sec:lorentz-invariant-case}

To become familiar with our goals, the situation in standard electrodynamics will be described first. In a Lorentz-invariant vacuum
Maxwell's equations in momentum space read as follows:
\begin{subequations}
\begin{align}
\label{eq:maxwell-equations-1}
\mathbf{k}\times \mathbf{B}+\omega\mathbf{E}&=\mathbf{0}\,,\quad \mathbf{k}\times \mathbf{E}-\omega\mathbf{B}=\mathbf{0}\,, \\[2ex]
\label{eq:maxwell-equations-2}
\mathbf{k}\cdot \mathbf{E}&=0\,,\quad \mathbf{k}\cdot \mathbf{B}=0\,.
\end{align}
\end{subequations}
Here $\mathbf{E}$ is the electric field, $\mathbf{B}$ the magnetic flux density, $\mathbf{k}$ the wave vector, and $\omega$ the frequency.
The dispersion relation can be derived directly from the wave equation. The latter is obtained by computing the cross product of the
wave vector and, e.g., the first of \eqref{eq:maxwell-equations-1} where the second equation has to be plugged in subsequently:
\begin{equation}
\label{eq:lorentz-invariant-wave-equation}
\mathbf{k}\times (\mathbf{k}\times \mathbf{B})+\omega\,\mathbf{k}\times \mathbf{E}=\mathbf{k}(\mathbf{k}\cdot \mathbf{B})-\mathbf{k}^2\mathbf{B}+\omega^2\,\mathbf{B}=(\omega^2-\mathbf{k}^2)\mathbf{B}=\mathbf{0}\,.
\end{equation}
Here the second of \eqref{eq:maxwell-equations-2} is used as well, which says that in a Lorentz-invariant vacuum the magnetic field is
transverse. Equation (\ref{eq:lorentz-invariant-wave-equation}) has nontrivial solutions for the magnetic field only in case of $\omega^2=\mathbf{k}^2$,
which immediately leads to the dispersion relation $\omega=|\mathbf{k}|$ of electromagnetic waves. The dispersion equation
\begin{equation}
\label{eq:dispersion-equation-standard}
\omega^2-\mathbf{k}^2=0\,,
\end{equation}
is the base to determine the Finsler structure associated to standard Maxwell theory. The
method is introduced in \cite{Antonelli:1993} and will be described as follows. Let $M$ be a Finsler manifold and $F=F(x,y)$ the corresponding
Finsler structure with $x\in M$ and $y\in T_xM$ where $T_xM$ is the tangent space at $x$. The indicatrix $S_xM$ at a point $x$ of a Finsler space
is the set of all $y$ where the Finsler structure takes the constant value 1, i.e., $S_xM=\{y\in T_xM|F(x,y)=1\}$. Note that a Finsler structure
defines an indicatrix, but conversely each indicatrix determines a Finsler structure \cite{Constantinescu:2009}.

Finsler himself expressed the idea that an indicatrix might model the phase velocity of light waves in both isotropic and anisotropic materials. Hence
what is needed to associate a Finsler structure to a photon theory is an indicatrix \cite{Antonelli:1993}. The phase velocity vector is defined as
$\mathbf{v}_{\mathrm{ph}}\equiv \widehat{\mathbf{k}}v_{\mathrm{ph}}$ with $v_{\mathrm{ph}}=\omega/|\mathbf{k}|$ and the unit wave vector is
$\widehat{\mathbf{k}}\equiv \mathbf{k}/|\mathbf{k}|$. Since \eqref{eq:dispersion-equation-standard} still depends on both the energy and the
momentum components, we divide it by $|\mathbf{k}|^2$. This results in an equation that involves the phase velocity and quantities of zero mass
dimension:
\begin{equation}
\label{eq:phase-velocity-standard}
v_{\mathrm{ph}}^2-1=0\,.
\end{equation}
Now \eqref{eq:phase-velocity-standard} can be considered as the indicatrix of the associated Finsler structure that it still to be found. This is
accomplished using Okubo's technique, which is outlined in \cite{Antonelli:1993,Bao:2000}. Consider a surface within a Finsler manifold $M$ that is
described by an equation $f(x,y)=0$. A function $F(y)$ taking a constant value 1 on such a surface can be found by solving the equation $f(x,y/F(y))=0$
with respect to $F(y)$ where the solution does not necessarily have to be unique. Denoting the phase velocity by
$v_{\mathrm{ph}}\equiv |\mathbf{u}|$ with $\mathbf{u}\equiv (u^1,u^2,u^3)$ we perform the
replacement $u^i\mapsto u^i/F(\mathbf{u})$ and obtain from \eqref{eq:dispersion-equation-standard}
\begin{equation}
\frac{\mathbf{u}^2}{F(\mathbf{u})^2}-1=0\,.
\end{equation}
The latter can be solved for $F(\mathbf{u})$ immediately:
\begin{equation}
\label{eq:finsler-structure-standard}
F(\mathbf{u})|_{\text{LI}}^{\pm}=\pm\sqrt{\mathbf{u}^2}=\pm\sqrt{r_{ij}u^iu^j}\,,\quad r_{ij}=\mathrm{diag}(1,1,1)_{ij}\,.
\end{equation}
As long as the intrinsic metric $r_{ij}$ is positive definite, which is the case for the particular $r_{ij}$ given, $F(\mathbf{u})|_{\text{LI}}^{+}$ fulfills all
properties of \secref{sec:classical-lagrangians-finsler-structures}. Therefore it can be interpreted as a three-dimensional Finsler structure where
the derived metric $g_{\mathrm{LI},ij}^{\pm}$ corresponds to the intrinsic metric. Since the Cartan torsion vanishes, it must be a Riemannian structure according to
Deicke's theorem.

\subsection{Isotropic case}
\label{sec:isotropic-case}

In the Lorentz-violating case modified Maxwell's equations can be constructed by using Eqs.~(4) -- (6) of \cite{Kostelecky:2002hh}.
A Lorentz-violating vacuum behaves like an effective medium for electromagnetic waves, which is why Maxwell's equations now
involve nontrivial permeability and permittivity tensors. In momentum space they read as follows (where the spatial indices of $\mathbf{k}$
are understood to be upper ones):
\begin{subequations}
\begin{align}
\label{eq:maxwell-equations-modified-1}
\mathbf{k}\times \mathbf{H}+\omega\mathbf{D}&=\mathbf{0}\,,\quad \mathbf{k}\times \mathbf{E}-\omega\mathbf{B}=\mathbf{0}\,, \\[2ex]
\label{eq:maxwell-equations-modified-2}
\mathbf{k}\cdot \mathbf{D}&=0\,,\quad \mathbf{k}\cdot \mathbf{B}=0\,.
\end{align}
\end{subequations}
The first two of these deliver relationships between
the electric displacement $\mathbf{D}$, the magnetic field $\mathbf{H}$, the electric field $\mathbf{E}$, and the magnetic flux density
$\mathbf{B}$. The transformation between $(\mathbf{D},\mathbf{H})$ and $(\mathbf{E},\mathbf{B})$ is governed by $(3\times 3)$ matrices
$\kappa_{\scriptscriptstyle{DE}}$, $\kappa_{\scriptscriptstyle{DB}}$, $\kappa_{\scriptscriptstyle{HE}}$, and $\kappa_{\scriptscriptstyle{HB}}$
comprising the controlling coefficients and they are given by Eq. (4) in the
latter reference. In the isotropic case considered here the matrices $\kappa_{\scriptscriptstyle{DB}}$ and $\kappa_{\scriptscriptstyle{HE}}$ do not contribute. It then holds
that
\begin{subequations}
\label{eq:isotropic-case-property-tensors}
\begin{align}
\mathbf{H}&=\mu^{-1}\mathbf{B}\,,\quad \mu^{-1}=\mathds{1}_3+\kappa_{\scriptscriptstyle{HB}}=\mathds{1}_3-\kappa_{\scriptscriptstyle{DE}}\,, \\[2ex]
\mathbf{D}&=\varepsilon\mathbf{E}\,,\quad \varepsilon=\mathds{1}_3+\kappa_{\scriptscriptstyle{DE}}\,, \\[2ex]
\kappa_{\scriptscriptstyle{DE}}&=\widetilde{\kappa}_{\mathrm{tr}}\,\mathrm{diag}(1,1,1)=-\kappa_{\scriptscriptstyle{HB}}\,, \\[2ex]
\varepsilon\mu&=n^2\,\mathrm{diag}(1,1,1)\,,\quad n^{-1}=\mathcal{A}\equiv \sqrt{\frac{1-\kappa_{\mathrm{tr}}}{1+\kappa_{\mathrm{tr}}}}\,.
\end{align}
\end{subequations}
Maxwell's equations in momentum space will be needed to obtain the dispersion relations.
Each of the equations involves different fields. However to obtain the dispersion relation, a single equation is required that
contains one of the four fields only. Since according to \eqref{eq:isotropic-case-property-tensors} the different fields are related by
matrices proportional to the unit matrix, the standard procedure outlined in \secref{sec:lorentz-invariant-case} works here:
\begin{align}
\mathbf{k}\times (\mathbf{k}\times \mathbf{E})-\omega(\mathbf{k} \times \mathbf{B})&=\mathbf{k}\times (\mathbf{k}\times \mathbf{E})-\omega\mu(\mathbf{k}\times \mathbf{H}) \notag \\
&=\mathbf{k}\times (\mathbf{k}\times \mathbf{E})+\omega^2\mu\mathbf{D}=\mathbf{k}\times (\mathbf{k}\times \mathbf{E})+\omega^2\varepsilon\mu\mathbf{E}=\mathbf{0}\,.
\end{align}
Writing the equation explicitly in matrix form leads to
\begin{equation}
\begin{pmatrix}
n^2\omega^2-(k_2^2+k_3^2) & k_1k_2 & k_1k_3 \\
k_1k_2 & n^2\omega^2-(k_1^2+k_3^2) & k_2k_3 \\
k_1k_3 & k_2k_3 & n^2\omega^2-(k_1^2+k_2^2) \\
\end{pmatrix}\begin{pmatrix}
E^1 \\
E^2 \\
E^3 \\
\end{pmatrix}=\begin{pmatrix}
0 \\
0 \\
0 \\
\end{pmatrix}\,,
\end{equation}
where $\mathbf{E}\equiv (E^1,E^2,E^3)^T$ is the electric field strength vector. Lowering the indices of the components of $\mathbf{k}$ does not
lead to changes, since the components always appear in bilinear combinations.
The condition of a vanishing determinant of the coefficient matrix, which is demanded for the existence of nontrivial solutions for the electric field,
leads to the dispersion equation
\begin{equation}
\label{eq:dispersion-relation-isotropic}
0=n^2\omega^2(n^2\omega^2-\mathbf{k}^2)^2\,.
\end{equation}
From this we obtain the spurious solution $\omega=0$ associated to a nonpropagating wave and the modified dispersion relation
$\omega=\mathcal{A}|\mathbf{k}|$. Now we again need an indicatrix. A reasonable choice to start with is \eqref{eq:dispersion-relation-isotropic}.
Dividing the latter by the prefactor and computing the square root does not change the set of physical zeros for $\omega$, i.e., we can also take
\begin{equation}
n^2\omega^2-\mathbf{k}^2=0\,.
\end{equation}
A subsequent division by $|\mathbf{k}|^2$ results in the indicatrix of the related Finsler structure:
\begin{equation}
\label{eq:phase-velocity-isotropic}
v_{\mathrm{ph}}^2-\mathcal{A}^2=0\,.
\end{equation}
Using Okubo's technique we obtain $F(\mathbf{u})$ immediately:
\begin{subequations}
\begin{align}
0&=\frac{\mathbf{u}^2}{F(\mathbf{u})^2}-\mathcal{A}^2\,, \\[2ex]
\label{eq:finsler-structure-isotropic}
F(\mathbf{u})|_{\circledcirc}^{\pm}&=\pm\frac{1}{\mathcal{A}}\sqrt{\mathbf{u}^2}=\pm\frac{1}{\mathcal{A}}\sqrt{r_{ij}u^iu^j}\,,\quad r_{ij}=\mathrm{diag}(1,1,1)_{ij}\,,
\end{align}
\end{subequations}
where the symbol $\circledcirc$ denotes ``isotropic.''
Comparing \eqref{eq:finsler-structure-isotropic} to \eqref{eq:finsler-structure-standard} we see that the only difference in comparison to the Lorentz-invariant
case is the prefactor $1/\mathcal{A}$. This is not surprising, as the case considered is isotropic and the result involves the spatial velocity components only.
For a positive definite $r_{ij}$, $F(\mathbf{u})|_{\circledcirc}^{+}$ fulfills all properties of a Finsler structure where the derived metric is given by $g_{\circledcirc,ij}^{\pm}=r_{ij}/\mathcal{A}^2$.
Due to the isotropy the latter is still Riemannian, which can be explicitly checked via the Cartan torsion. In comparison to the Lorentz-invariant case it involves a global
scaling factor.

\subsection{Anisotropic, nonbirefringent case}
\label{sec:anisotropic-nonbirefringent-case}

The anisotropic case with a single modified dispersion relation reveals some peculiar properties.
The matrices relating the different electromagnetic fields with each other are given by
\begin{subequations}
\begin{align}
\label{eq:medium-tensors-anisotropic-nonbirefringent}
\kappa_{\scriptscriptstyle{DE}}&=\frac{3}{2}\widetilde{\kappa}_{e-}^{11}\,\mathrm{diag}(1,1,-1)=-\kappa_{\scriptscriptstyle{HB}}\,, \\[2ex]
\kappa_{\scriptscriptstyle{DB}}&=\kappa_{\scriptscriptstyle{HE}}=\mathbf{0}_3\,,
\end{align}
\end{subequations}
with the $(3\times 3)$ zero matrix $\mathbf{0}_3$.
The matrices $\kappa_{\scriptscriptstyle{DE}}$ and $\kappa_{\scriptscriptstyle{HB}}$ are diagonal as well, but the difference to the isotropic case is that they are
no longer proportional to the identity matrix. This is not surprising due to the preferred spacelike direction $\boldsymbol{\zeta}$ pointing along the third spatial
axis where there is a residual isotropy in the plane perpendicular to this axis. Therefore the first two components of the diagonal matrix $\varepsilon\mu$ are
equal, but the third differs from those:
\begin{equation}
\varepsilon\mu=\mathrm{diag}(n_1^2,n_2^2,n_3^2)\,,\quad n_1=n_2=\frac{1}{\mathcal{B}}\,,\quad n_3=\mathcal{B}\,,\quad \mathcal{B}\equiv \sqrt{\frac{1-(3/2)\widetilde{\kappa}_{e-}^{11}}{1+(3/2)\widetilde{\kappa}_{e-}^{11}}}\,.
\end{equation}
Now we again need an equation that can serve as a basis for the indicatrix of the associated Finsler space. Multiplying the second of
\eqref{eq:maxwell-equations-modified-1} with $\mu^{-1}$, computing the cross product with $\mathbf{k}$, and using
the first of \eqref{eq:maxwell-equations-modified-1} leads to an equation for the electric field
vector:
\begin{equation}
\mathbf{k}\times \left[(\mu^{-1}(\mathbf{k}\times \mathbf{E})\right]+\omega^2\varepsilon \mathbf{E}=\mathbf{0}\,.
\end{equation}
Multiplying the latter with an appropriate prefactor, in matrix form it reads as follows:
\begin{equation}
\begin{pmatrix}
\omega^2-k_2^2-k_3^2n_3^2 & k_1k_2 & k_1k_3n_3^2 \\
k_1k_2 & \omega^2-k_1^2-k_3^2n_3^2 & k_2k_3n_3^2 \\
k_1k_3n_3^2 & k_2k_3n_3^2 & (\omega^2-k_1^2-k_2^2)n_3^2 \\
\end{pmatrix}\begin{pmatrix}
E^1 \\
E^2 \\
E^3 \\
\end{pmatrix}=\begin{pmatrix}
0 \\
0 \\
0 \\
\end{pmatrix}\,.
\end{equation}
Lowering the components of $\mathbf{k}$ does not produce any changes.
The determinant condition for this system of equations leads to
\begin{subequations}
\begin{align}
0&=\omega^2(\omega^2-k_{\bot}^2-k_{\scalebox{0.6}{$\|$}}^2n_3^2)^2\,, \\[2ex]
k_{\scalebox{0.6}{$\|$}}&\equiv \mathbf{k}\cdot \boldsymbol{\zeta}\,,\quad k_{\bot}\equiv |\mathbf{k}-k_{\scalebox{0.6}{$\|$}}\boldsymbol{\zeta}|\,.
\end{align}
\end{subequations}
For convenience the three-momentum $\mathbf{k}$ is decomposed into a component $k_{\scalebox{0.6}{$\|$}}$ along the preferred spatial direction
$\boldsymbol{\zeta}=(0,0,1)^T$ and into a component $k_{\bot}$ perpendicular to $\boldsymbol{\zeta}$. This results in the spurious solution $\omega=0$
and a single dispersion relation for electromagnetic waves:
\begin{equation}
\label{eq:dispersion-relation-anisotropic}
\omega=\sqrt{k_{\bot}^2+\mathcal{B}^2k_{\scalebox{0.6}{$\|$}}^2}\,.
\end{equation}
Here the remaining isotropy perpendicular to the preferred direction becomes evident as well.
The photon will only be affected by Lorentz violation in case it has a momentum component pointing along the preferred direction. Note that the result of
\eqref{eq:dispersion-relation-anisotropic} is very interesting from the perspective that the underlying Lorentz-violating framework is anisotropic, but
in spite of this anisotropy there is only a single dispersion relation. In contrast, birefringence, i.e., the property of having two different dispersion laws dependent
on photon polarization seems to always occur in anisotropic media in nature. The reason that there is a single dispersion relation here only is the
extreme fine tuning of permeability and permittivity (cf.~\eqref{eq:medium-tensors-anisotropic-nonbirefringent}), which can most probably not be found
in any materials.

Now, the equation for the indicatrix follows from
\begin{equation}
\omega^2-k_{\bot}^2-k_{\scalebox{0.6}{$\|$}}^2n_3^2=0\,,
\end{equation}
in dividing it by $\mathbf{k}^2$. Introducing the angle $\vartheta$ between the wave vector $\mathbf{k}$ and the preferred spatial direction $\boldsymbol{\zeta}$ leads to
\begin{subequations}
\begin{align}
\label{eq:pre-indicatrix-anisotropic}
0&=v_{\mathrm{ph}}^2-\sin^2\vartheta-\mathcal{B}^2\cos^2\vartheta\,, \\[2ex]
\cos\vartheta&\equiv \widehat{\mathbf{k}}\cdot \boldsymbol{\zeta}\,,\quad \widehat{\mathbf{k}}\equiv \frac{\mathbf{k}}{|\mathbf{k}|}\,.
\end{align}
\end{subequations}
Thinking of $\vartheta$ as the polar angle in spherical coordinates, \eqref{eq:pre-indicatrix-anisotropic} can be reinterpreted using
\begin{equation}
v_{\mathrm{ph}}^2=\mathbf{u}^2\,,\quad \cos\vartheta=\frac{u^3}{|\mathbf{u}|}\,,\quad \sin\vartheta=\frac{\sqrt{(u^1)^2+(u^2)^2}}{|\mathbf{u}|}\,,
\end{equation}
as follows:
\begin{equation}
\label{eq:indicatrix-anisotropic}
\mathbf{u}^4-\left[(u^1)^2+(u^2)^2+\mathcal{B}^2(u^3)^2\right]=0\,.
\end{equation}
The latter is the equation that determines the indicatrix. Okubo's technique can again be used to obtain a Finsler structure directly when
$\mathbf{u}$ is replaced by $\mathbf{u}/F(\mathbf{u})$ in \eqref{eq:indicatrix-anisotropic}:
\begin{equation}
\mathbf{u}^4-F(\mathbf{u})^2\left[(u^1)^2+(u^2)^2+\mathcal{B}^2(u^3)^2\right]=0\,.
\end{equation}
This leads to the result
\begin{equation}
\label{eq:finsler-structure-anisotropic}
F(\mathbf{u})|_{\varobar}^{\pm}=\pm\frac{\mathbf{u}^2}{\sqrt{(u^1)^2+(u^2)^2+\mathcal{B}^2(u^3)^2}}\,,
\end{equation}
which can also be written in the form
\begin{equation}
\label{eq:finsler-structure-anisotropic}
F(\mathbf{u})|_{\varobar}^{\pm}=\pm\frac{r_{ij}u^iu^j}{\sqrt{s_{ij}u^iu^j}}\,,\quad r_{ij}=\mathrm{diag}(1,1,1)_{ij}\,,\quad s_{ij}=\mathrm{diag}(1,1,\mathcal{B}^2)_{ij}\,.
\end{equation}
Here $\varobar$ means ``anisotropic.'' In principle the Finsler structure can be interpreted to involve an intrinsic metric $r_{ij}$ and a second
metric $s_{ij}$. Since the background considered is flat, it is reasonable to take $r_{ij}$ as the metric that determines the lengths of vectors
and the angles between vectors. For general $r_{ij}$ and $s_{ij}$ the derived metric is given by
\begin{subequations}
\begin{align}
g_{ij}&=F(\mathbf{u})|_{\varobar}^{\pm}\frac{\partial^2F(\mathbf{u})|_{\varobar}^{\pm}}{\partial u^i\partial u^j}+\frac{\partial F(\mathbf{u})|_{\varobar}^{\pm}}{\partial u^i}\frac{\partial F(\mathbf{u})|^{\pm}_{\varobar}}{\partial u^j}\,, \displaybreak[0]\\[2ex]
\frac{\partial F(\mathbf{u})|_{\varobar}^{\pm}}{\partial u^i}&=\pm\frac{1}{(s_{ab}u^au^b)^{3/2}}Q_{iklm}u^ku^lu^m\,, \displaybreak[0]\\[2ex]
\frac{\partial^2F(\mathbf{u})|_{\varobar}^{\pm}}{\partial u^i\partial u^j}&=\mp\frac{3s_{jn}}{(s_{ab}u^au^b)^{5/2}}Q_{iklm}u^ku^lu^mu^n \notag \displaybreak[0]\\
&\phantom{{}={}}\pm\frac{1}{(s_{ab}u^au^b)^{3/2}}Q_{iklm}\left(\delta^{kj}u^lu^m+\delta^{jl}u^ku^m+\delta^{mj}u^ku^l\right)\,, \displaybreak[0]\\[2ex]
Q_{iklm}&=2s_{kl}r_{im}-r_{kl}s_{im}\,.
\end{align}
\end{subequations}
This result is not very illuminating. When contracted with appropriate velocity components it collapses to $(F(\mathbf{u})|_{\varobar}^{\pm})^2$,
which follows from its homogeneity of degree 2:
\begin{equation}
g^{\pm}_{\varobar,ij}u^iu^j=(F(\mathbf{u})|_{\varobar}^{\pm})^2\,,\quad g^{\pm}_{\varobar,ij}\equiv \frac{1}{2}\frac{\partial^2(F(\mathbf{u})|_{\varobar}^{\pm})^2}{\partial u^i\partial u^j}\,.
\end{equation}
Now the following properties of $F(\mathbf{u})|_{\varobar}^{\pm}$ can be deduced:

\begin{itemize}

\item[1)] $F(\mathbf{u})|_{\varobar}^+>0$ if $r_{ij}$ is positive definite,
\item[2)] $F(\mathbf{u})|_{\varobar}^{\pm}\in C^{\infty}$ for $\mathbf{u}\in TM\setminus \{0\}$ as well as positive definite $s_{ij}$,
\item[3)] $F(\lambda\mathbf{u})|_{\varobar}^{\pm}=\lambda F(\mathbf{u})|_{\varobar}^{\pm}$ for $\lambda>0$, i.e., positive homogeneity,
\item[4)] and the derived metric $g_{ij}$ is positive definite as long as $s_{ij}$ is positive definite.

\end{itemize}

Therefore as long as both $r_{ij}$ and $s_{ij}$ are positive definite, which in particular is the case for $r_{ij}$ and
$s_{ij}$ given in \eqref{eq:finsler-structure-anisotropic}, $F(\mathbf{u})|_{\varobar}^+$ defines a three-dimensional Finsler
structure, indeed. Furthermore both the Cartan and the Matsumoto torsion can be computed to be able to classify this Finsler structure. The
results are complicated and they do not provide further insight, which is why they will be omitted. However they are nonzero in general whereby
according to Deicke's theorem, \eqref{eq:finsler-structure-anisotropic} is not a Riemannian structure and according
to the Matsumoto-H\={o}j\={o} theorem it is neither a Randers nor
a Kropina structure. The result corresponds to Eq.~(4.2.2.6) of \cite{Antonelli:1993} where $a=\mathcal{B}$ and $b=1$ in their notation.
They denote this type of Finsler structure as a second-order Kropina structure in resemblance to a Kropina structure $F(\mathbf{u})=\alpha^2/\beta$
with $\alpha=\sqrt{a_{ij}u^iu^j}$ and $\beta=b_iu^i$. In the latter reference
\eqref{eq:finsler-structure-anisotropic} appears in the context of light propagation in uniaxial media. The numerator involves the
Euclidean intrinsic metric $r_{ij}$ only, whereas the denominator is characterized by another metric $s_{ij}$. The latter could be thought of as the metric
governing physics, since it involves the physical quantity $\mathcal{B}$.

\subsection{Anisotropic, birefringent (at second order) case}
\label{sec:anisotropic-birefringent-case}

The penultimate example provides a case of modified Maxwell theory that has not been considered in
\secref{eq:lagrange-function-massive-modmax}. It is parity-even and characterized by two preferred spacelike directions:
\begin{equation}
\label{eq:preferred-directions-anisotropic-case}
(\zeta_1^{\mu})=\begin{pmatrix}
0 \\
\boldsymbol{\zeta}_1 \\
\end{pmatrix}\,,\quad (\zeta_2^{\mu})=\begin{pmatrix}
0 \\
\boldsymbol{\zeta}_2 \\
\end{pmatrix}\,,\quad \boldsymbol{\zeta}_1=\begin{pmatrix}
\sin\eta \\
0 \\
\cos\eta \\
\end{pmatrix}\,,\quad \boldsymbol{\zeta}_2=\begin{pmatrix}
-\sin\eta \\
0 \\
\cos\eta \\
\end{pmatrix}\,.
\end{equation}
They are normalized and enclose an angle of $2\eta$. We consider an observer frame with one nonzero controlling coefficient $\mathcal{G}$.
Then the $(4\times 4)$ matrix employed in the nonbirefringent \textit{Ansatz} reads
\begin{equation}
\widetilde{\kappa}^{\mu\nu}=\mathcal{G}\left(\zeta_1^{\mu}\zeta_2^{\nu}+\zeta_1^{\nu}\zeta_2^{\mu}-\frac{1}{2}(\zeta_1\cdot \zeta_2)\eta^{\mu\nu}\right)\,.
\end{equation}
This corresponds to the following choices for the matrices that appear in Maxwell's equations:
\begin{subequations}
\begin{align}
\kappa_{\scriptscriptstyle{DE}}&=\mathcal{G}\,\mathrm{diag}(1,\cos(2\eta),-1)=-\kappa_{\scriptscriptstyle{HB}}\,, \\[2ex]
\kappa_{\scriptscriptstyle{DB}}&=\kappa_{\scriptscriptstyle{HE}}=\mathbf{0}_3\,.
\end{align}
\end{subequations}
Hence there are nontrivial permeability and permittivity tensors, but the electric and magnetic fields do still not mix.
Using these matrices, modified Maxwell's equations can be obtained according to the procedure used in \secref{sec:anisotropic-nonbirefringent-case}.
The condition of a vanishing coefficient determinant for nontrivial solutions results in an equation for the dispersion relation:
\begin{align}
\label{eq:dispersion-relation-off-shell-anisotropic-nonbirefringent}
0&=\omega^2\left[(1-\mathcal{G}^2)\omega^2-(1+\mathcal{G})[1-\mathcal{G}\cos(2\eta)]k_1^2-(1-\mathcal{G}^2)k_2^2-(1-\mathcal{G})[1-\mathcal{G}\cos(2\eta)]k_3^2\right] \notag \\
&\phantom{{}={}}\times \left\{\omega^2[1+\mathcal{G}\cos(2\eta)]-(1+\mathcal{G})k_1^2-[1+\mathcal{G}\cos(2\eta)]k_2^2-(1-\mathcal{G})k_3^2\right\}\,.
\end{align}
In contrast to the anisotropic case considered in \secref{sec:anisotropic-nonbirefringent-case} the current framework is characterized by two distinct
modified dispersion relations. They can be written in the form
\begin{subequations}
\label{eq:dispersion-relations-anisotropic-birefringent}
\begin{align}
\omega_1&=\sqrt{\mathcal{G}_1k_1^2+k_2^2+\mathcal{G}_2k_3^2}\,, \\[2ex]
\omega_2&=\sqrt{\widetilde{\mathcal{G}}_1k_1^2+k_2^2+\widetilde{\mathcal{G}}_2k_3^2}\,, \\[2ex]
\label{eq:finsler-structure-anisotropic-birefringence-1-constants}
\mathcal{G}_1&\equiv\frac{1-\mathcal{G}\cos(2\eta)}{1-\mathcal{G}}\,,\quad \mathcal{G}_2\equiv\frac{1-\mathcal{G}\cos(2\eta)}{1+\mathcal{G}}\,, \\[2ex]
\label{eq:finsler-structure-anisotropic-birefringence-2-constants}
\widetilde{\mathcal{G}}_1&\equiv\frac{1+\mathcal{G}}{1+\mathcal{G}\cos(2\eta)}\,,\quad \widetilde{\mathcal{G}}_2\equiv\frac{1-\mathcal{G}}{1+\mathcal{G}\cos(2\eta)}\,.
\end{align}
\end{subequations}
Evidently the contribution associated to the second three-momentum component stays unmodified which is reasonable, since the preferred directions
of \eqref{eq:preferred-directions-anisotropic-case} do not point along the second spatial axis. Each dispersion relation can be expanded for $\mathcal{G}\ll 1$
showing that they differ at second order in Lorentz violation. In general the nonbirefringent \textit{Ansatz} of \eqref{eq:nonbirefringent-ansatz} works at
leading order only. Besides, the dispersion relations depend on the angle $2\eta$ enclosed by the two preferred directions. With the normalized
propagation direction of the electromagnetic wave given by $\widehat{\mathbf{k}}$, the latter encloses the angles $\theta_1$, $\theta_2$ with the first
and the second preferred direction, respectively. These are given by:
\begin{subequations}
\begin{align}
\cos\theta_1&=\widehat{\mathbf{k}}\cdot \boldsymbol{\zeta}_1=\widehat{k}^1\sin\eta+\widehat{k}^3\cos\eta\,, \\[2ex]
\cos\theta_2&=\widehat{\mathbf{k}}\cdot \boldsymbol{\zeta}_2=-\widehat{k}^1\sin\eta+\widehat{k}^3\cos\eta\,.
\end{align}
\end{subequations}
The components of the propagation direction vector $\widehat{\mathbf{k}}$ can now be expressed in terms of the angles $\theta_1$, $\theta_2$, and
$\eta$. Note that $\widehat{\mathbf{k}}$ is a unit vector by construction:
\begin{equation}
\label{eq:direction-angle}
\widehat{k}^1=\frac{\cos\theta_1-\cos\theta_2}{2\sin\eta}\,,\quad \widehat{k}^3=\frac{\cos\theta_1+\cos\theta_2}{2\cos\eta}\,,\quad \widehat{k}^2=\sqrt{1-(\widehat{k}^1)^2-(\widehat{k}^3)^2}\,.
\end{equation}
Now the two individual factors of \eqref{eq:dispersion-relation-off-shell-anisotropic-nonbirefringent} are considered giving the modified dispersion relations.
Dividing each by the wave-vector magnitude $|\mathbf{k}|$, introducing the phase velocity, and expressing all propagation direction components
by the angles of \eqref{eq:direction-angle}, equations for the phase velocities are obtained as before:
\begin{subequations}
\begin{align}
\label{eq:phase-velocity-equations-anisotropic}
0&=(1-\mathcal{G}^2)v_{\mathrm{ph}}^2+\frac{\mathcal{G}}{2}\left\{4\cos(\theta_1)\cos(\theta_2)-\mathcal{G}\left[\cos(2\theta_1)+\cos(2\theta_2)\right]\right\}-1\,, \\[2ex]
0&=\left[1+\mathcal{G}\cos(2\eta)\right](1-v_{\mathrm{ph}}^2)-2\mathcal{G}\cos(\theta_1)\cos(\theta_2)\,.
\end{align}
\end{subequations}
In dividing the second equation by $-[1+\mathcal{G}\cos(2\eta)]$ and expanding both equations to linear order in $\mathcal{G}$ these results correspond to each
other as expected. Now we are in a position to interpret the latter equations geometrically, which will lead us directly to the Finsler structures associated
to this particular sector. In doing so, the velocity $\mathbf{u}$ is introduced and both the phase velocity and the angles $\theta_1$, $\theta_2$ are
expressed by the magnitude or components of $\mathbf{u}$ as follows:
\begin{subequations}
\label{eq:correspondence-angle-velocities}
\begin{align}
v_{\mathrm{ph}}&=|\mathbf{u}|\,, \\[2ex]
\cos\theta_1&=\frac{u^1}{|\mathbf{u}|}\sin\eta+\frac{u^3}{|\mathbf{u}|}\cos\eta\,, \\[2ex]
\cos\theta_2&=-\frac{u^1}{|\mathbf{u}|}\sin\eta+\frac{u^3}{|\mathbf{u}|}\cos\eta\,.
\end{align}
\end{subequations}
Inserting those into \eqref{eq:phase-velocity-equations-anisotropic} and using Okubo's technique leads to two distinct Finsler structures. The first is given
by
\begin{equation}
\label{eq:finsler-structure-anisotropic-birefringence-1}
F(\mathbf{u})|_{\varovee}^{(1)\pm}=\pm\frac{r_{ij}u^iu^j}{\sqrt{s_{ij}u^iu^j}}\,,\quad r_{ij}=\mathrm{diag}(1,1,1)_{ij}\,,\quad s_{ij}=\mathrm{diag}(\mathcal{G}_1,1,\mathcal{G}_2)_{ij}\,,
\end{equation}
and the second reads as
\begin{equation}
\label{eq:finsler-structure-anisotropic-birefringence-2}
F(\mathbf{u})|_{\varovee}^{(2)\pm}=\pm\frac{r_{ij}u^iu^j}{\sqrt{s_{ij}u^iu^j}}\,,\quad r_{ij}=\mathrm{diag}(1,1,1)_{ij}\,,\quad s_{ij}=\mathrm{diag}(\widetilde{\mathcal{G}}_1,1,\widetilde{\mathcal{G}}_2)_{ij}\,.
\end{equation}
Here $\varovee$ means ``anisotropic and birefringent (at second order).'' The four Finsler structures obtained have a form analogous to the Finsler structure
found in \eqref{eq:finsler-structure-anisotropic} of \secref{sec:anisotropic-nonbirefringent-case}. This is not surprising, since both sectors are anisotropic
but parity-even. Having birefringence at second order in Lorentz violation does obviously not affect the form of the Finsler structure. In such a case we can
obtain several distinct Finsler structures that differ from each other at second order in the controlling coefficients via the metrics $s_{ij}$.

In \eqref{eq:finsler-structure-anisotropic} the latter $s_{ij}$ differs from the standard Euclidean metric only by the component $s_{33}$. Here both $s_{11}$
and $s_{33}$ are modified by Lorentz violation where they also depend on the angle $\eta$ enclosed by the two preferred directions.
The component $s_{22}$ is standard, which again reflects the fact that the preferred directions have a vanishing second component. Since $s_{ij}$
involves the physical (dimensionless) constants $\mathcal{G}_i$ and $\widetilde{\mathcal{G}}_i$ for $i=1\dots 2$, it is reasonable to say that $s_{ij}$
seems to govern the physical properties of photon propagation in these cases.

\subsection{Parity-odd case}
\label{sec:parity-odd-case}

The final interesting sector considered involves the three parity-odd coefficients $\widetilde{\kappa}_{o+}^{12}$, $\widetilde{\kappa}_{o+}^{31}$, and
$\widetilde{\kappa}_{o+}^{23}$ and it will turn out to be the most complicated one. The preferred spacetime directions are given in
\eqref{eq:four-vectors-parity-odd} and the matrices relating the electromagnetic fields to each other read
\begin{subequations}
\begin{align}
\kappa_{\scriptscriptstyle{DE}}&=\mathbf{0}_3\,,\quad \kappa_{\scriptscriptstyle{HB}}=\mathbf{0}_3\,, \\[2ex]
\kappa_{\scriptscriptstyle{DB}}&=\begin{pmatrix}
0 & \widetilde{\kappa}_{o+}^{12} & -\widetilde{\kappa}_{o+}^{31} \\
-\widetilde{\kappa}_{o+}^{12} & 0 & \widetilde{\kappa}_{o+}^{23} \\
\widetilde{\kappa}_{o+}^{31} & -\widetilde{\kappa}_{o+}^{23} & 0 \\
\end{pmatrix}\,,\quad \kappa_{\scriptscriptstyle{HE}}=-\kappa_{\scriptscriptstyle{DB}}^T=\kappa_{\scriptscriptstyle{DB}}\,.
\end{align}
\end{subequations}
The relationships between the fields are given by
\begin{subequations}
\begin{align}
\mathbf{D}&=\mathbf{E}+\kappa_{\scriptscriptstyle{DB}}\mathbf{B}\,, \\[2ex]
\mathbf{H}&=\kappa_{\scriptscriptstyle{HE}}\mathbf{E}+\mathbf{B}=\kappa_{\scriptscriptstyle{DB}}\mathbf{E}+\mathbf{B}\,.
\end{align}
\end{subequations}
In contrast to the aforementioned cases the parity-odd case has the peculiarity that the electric fields mix with the magnetic fields.
Therefore obtaining an equation for the electric field from Maxwell's equations is more involved
here. Nevertheless it can be accomplished along the following chain of steps:
\begin{subequations}
\begin{align}
\mathbf{0}&=\kappa_{\scriptscriptstyle{DB}}(\mathbf{k}\times \mathbf{E})-\omega\kappa_{\scriptscriptstyle{DB}}\mathbf{B}\,, \\[2ex]
\mathbf{0}&=\kappa_{\scriptscriptstyle{DB}}(\mathbf{k}\times \mathbf{E})-\omega(\mathbf{D}-\mathbf{E})\,, \\[2ex]
\mathbf{0}&=\kappa_{\scriptscriptstyle{DB}}(\mathbf{k}\times \mathbf{E})+\mathbf{k}\times \mathbf{H}+\omega\mathbf{E}\,, \\[2ex]
\mathbf{0}&=\kappa_{\scriptscriptstyle{DB}}(\mathbf{k}\times \mathbf{E})+\mathbf{k}\times (\kappa_{\scriptscriptstyle{DB}}\mathbf{E}+\mathbf{B})+\omega\mathbf{E}\,, \\[2ex]
\mathbf{0}&=\kappa_{\scriptscriptstyle{DB}}(\mathbf{k}\times \mathbf{E})+\mathbf{k}\times (\kappa_{\scriptscriptstyle{DB}}\mathbf{E})+\frac{1}{\omega}\mathbf{k}\times (\mathbf{k}\times \mathbf{E})+\omega\mathbf{E}\,.
\end{align}
\end{subequations}
Inserting the explicit vectors and a subsequent multiplication with $\omega$ leads to the following system in matrix form:
\begin{subequations}
\label{eq:modified-maxwell-equations-parity-odd}
\begin{align}
\begin{pmatrix}
0 \\
0 \\
0 \\
\end{pmatrix}&=(A+B)\begin{pmatrix}
E^1 \\
E^2 \\
E^3 \\
\end{pmatrix}\,, \displaybreak[0]\\[2ex]
A&=\begin{pmatrix}
\omega^2-(k_2^2+k_3^2) & k_1k_2 & k_1k_3 \\
k_1k_2 & \omega^2-(k_1^2+k_3^2) & k_2k_3 \\
k_1k_3 & k_2k_3 & \omega^2-(k_1^2+k_2^2) \\
\end{pmatrix}\,, \displaybreak[0]\\[2ex]
B&=\omega\begin{pmatrix}
-2(\widetilde{\kappa}_{o+}^{31}k_2+\widetilde{\kappa}_{o+}^{12}k_3) & \widetilde{\kappa}_{o+}^{31}k_1+\widetilde{\kappa}_{o+}^{23}k_2 & \widetilde{\kappa}_{o+}^{12}k_1+\widetilde{\kappa}_{o+}^{23}k_3 \\
\widetilde{\kappa}_{o+}^{31}k_1+\widetilde{\kappa}_{o+}^{23}k_2 & -2(\widetilde{\kappa}_{o+}^{23}k_1+\widetilde{\kappa}_{o+}^{12}k_3) & \widetilde{\kappa}_{o+}^{12}k_2+\widetilde{\kappa}_{o+}^{31}k_3 \\
\widetilde{\kappa}_{o+}^{12}k_1+\widetilde{\kappa}_{o+}^{23}k_3 & \widetilde{\kappa}_{o+}^{12}k_2+\widetilde{\kappa}_{o+}^{31}k_3 & -2(\widetilde{\kappa}_{o+}^{23}k_1+\widetilde{\kappa}_{o+}^{31}k_2) \\
\end{pmatrix}\,.
\end{align}
\end{subequations}
The total system can be completely decomposed into the standard part of $A$ and a Lorentz-violating contribution comprised in $B$.
Note that here $B$ gets a global minus sign when lowering the indices of the $\mathbf{k}$ components. Therefore the determinant
condition results in the following equation for the photon energy:
\begin{equation}
\label{eq:dispersion-relation-parity-odd}
\omega^2(\omega^2-2\omega\,\boldsymbol{\zeta}\cdot \mathbf{k}-\mathbf{k}^2)\left[(\omega-\boldsymbol{\zeta}\cdot\mathbf{k})^2-(1+\boldsymbol{\zeta}^2)\mathbf{k}^2\right]=0\,.
\end{equation}
Here $\boldsymbol{\zeta}\equiv (\widetilde{\kappa}_{o+}^{23},\widetilde{\kappa}_{o+}^{31},\widetilde{\kappa}_{o+}^{12})^T$ is the spatial part
of the second preferred spacetime direction and $\mathbf{k}$ is understood to have lower components. The second and the third of the three
factors can be solved for the energy giving two distinct dispersion relations:
\begin{subequations}
\begin{align}
\label{eq:dispersion-relation-parity-odd-1}
\omega_1&=\boldsymbol{\zeta}\cdot\mathbf{k}+\sqrt{\mathbf{k}^2+(\boldsymbol{\zeta}\cdot\mathbf{k})^2}\,, \\[2ex]
\label{eq:dispersion-relation-parity-odd-2}
\omega_2&=\boldsymbol{\zeta}\cdot\mathbf{k}+\sqrt{1+\boldsymbol{\zeta}^2}|\mathbf{k}|\,, \\[2ex]
\cos\vartheta&=\widehat{\boldsymbol{\zeta}}\cdot \widehat{\mathbf{k}}\,,\quad \widehat{\boldsymbol{\zeta}}\equiv \frac{\boldsymbol{\zeta}}{\mathcal{E}}\,,\quad \widehat{\mathbf{k}}\equiv \frac{\mathbf{k}}{|\mathbf{k}|}\,,\quad \mathcal{E}\equiv |\boldsymbol{\zeta}|=\sqrt{(\widetilde{\kappa}_{o+}^{23})^2+(\widetilde{\kappa}_{o+}^{31})^2+(\widetilde{\kappa}_{o+}^{12})^2}\,.
\end{align}
\end{subequations}
For convenience it is again reasonable to set up the coordinate system such that $\boldsymbol{\zeta}$ points along its third axis where
$\vartheta$ is the angle between the wave vector $\mathbf{k}$ and the spatial direction. Dividing the first factor of
\eqref{eq:dispersion-relation-parity-odd} by $\mathbf{k}^2$ then leads to
\begin{equation}
v_{\mathrm{ph}}^2-2\mathcal{E}v_{\mathrm{ph}}\cos\vartheta-1=0\,.
\end{equation}
Introducing spherical polar coordinates with $v_{\mathrm{ph}}=|\mathbf{u}|$ results in
\begin{equation}
\mathbf{u}^2-2\mathcal{E}u^3-1=0\,.
\end{equation}
This is the indicatrix for the first Finsler space that can be associated to the parity-odd case. We can employ Okubo's technique to obtain
\begin{subequations}
\begin{align}
0&=\mathbf{u}^2-2\mathcal{E}F(\mathbf{u})u^3-F(\mathbf{u})^2\,, \\[2ex]
\label{eq:finsler-structure-parity-odd-1a}
F(\mathbf{u})|_{\otimes}^{(1)\pm}&=-\mathcal{E}u^3\pm \sqrt{\mathbf{u}^2+(\mathcal{E}u^3)^2}\,, \\[2ex]
\label{eq:finsler-structure-parity-odd-1b}
F(\mathbf{u})|_{\otimes}^{\boldsymbol{\zeta}(1)\pm}&=-\boldsymbol{\zeta}\cdot\mathbf{u}\pm \sqrt{\mathbf{u}^2+(\boldsymbol{\zeta}\cdot\mathbf{u})^2}\,.
\end{align}
\end{subequations}
where \eqref{eq:finsler-structure-parity-odd-1b} is the generalization of \eqref{eq:finsler-structure-parity-odd-1a} for $\boldsymbol{\zeta}$
pointing along an arbitrary direction and the symbol $\otimes$ denotes ``parity-odd.''
Without loss of generality the properties of the Finsler structure can be investigated with $\boldsymbol{\zeta}$ pointing along the third axis of the
coordinate system, which simplifies the calculations. The derived metric is again lengthy and does not seem to provide any deeper
understanding. The derived metric contracted with the spatial velocity components leads to the square of
\eqref{eq:finsler-structure-parity-odd-1b}:
\begin{equation}
g^{(1)\pm}_{\otimes,ij}u^iu^j=(F(\mathbf{u})|_{\otimes}^{(1)\pm})^2\,,\quad g^{(1)\pm}_{\otimes,ij}\equiv \frac{1}{2}\frac{\partial^2(F(\mathbf{u})|_{\otimes}^{(1)\pm})^2}{\partial u^i\partial u^j}\,.
\end{equation}
Therefore the following properties of $F(\mathbf{u})|_{\otimes}^{(1)\pm}$ in \eqref{eq:finsler-structure-parity-odd-1b} can be deduced:

\begin{itemize}

\item[1)] $F(\mathbf{u})|_{\otimes}^{(1)+}>0$ for $\mathbf{u}\in TM\setminus \{0\}$,
\item[2)] $F(\mathbf{u})|_{\otimes}^{(1)\pm}\in C^{\infty}$ for $\mathbf{u}\in TM\setminus \{0\}$,
\item[3)] $F(\lambda\mathbf{u})|_{\otimes}^{(1)\pm}=\lambda F(\mathbf{u})|_{\otimes}^{(1)\pm}$ for $\lambda>0$, and
\item[4)] the derived metric of $F(\lambda\mathbf{u})|_{\otimes}^{(1)\pm}$ is positive definite for $\mathbf{u}\in TM\setminus \{0\}$.

\end{itemize}

Due to the first item, only $F(\mathbf{u})|_{\otimes}^{(1)+}$ is a Finsler structure. Its Matsumoto torsion vanishes, whereas the Cartan torsion
does not. Furthermore when taking into account its form, $F(\mathbf{u})|_{\otimes}^{(1)+}$ must be a Randers structure. This particular type of
geometry was introduced by Randers to account for the fact that particles always move on timelike trajectories pointing forwards in time
\cite{Randers:1941}. In contrast to General Relativity his framework incorporates an additional four-vector into the metric. However this
four-vector should not be considered as a preferred spacetime direction, since it can be changed by a kind of gauge transformation without
affecting the arc length travelled by a particle. In the Lorentz-violating case considered here $\boldsymbol{\zeta}$ is a preferred direction,
indeed.

The parity-odd framework is characterized by both a preferred timelike and a spacelike direction, cf. \eqref{eq:four-vectors-parity-odd}.
For the isotropic and anisotropic cases, which are parity-even, the corresponding Finsler structures are expected to involve only bilinear
expressions such as $a_{ij}u^iu^j$, since these are invariant under $u^i\mapsto u'^i=-u^i$. Due to parity violation the Finsler structure
of the parity-odd case is expected to involve terms such as $b_iu^i$, though. The Randers structure is a very natural possibility with
this property, but it is not the only one as we shall see below.

The Finsler structure of \eqref{eq:finsler-structure-parity-odd-1b} has the same form as the corresponding dispersion relation of
\eqref{eq:dispersion-relation-parity-odd-1} not taking into account additional minus signs. Such structures could be called ``automorphic.''
They seem to appear when the dispersion equation (here \eqref{eq:dispersion-relation-parity-odd}) involves one additional
parity-odd contribution.

The parity-odd case of modified Maxwell theory has a second indicatrix, which follows from the second factor of \eqref{eq:dispersion-relation-parity-odd}
using the same procedure:
\begin{subequations}
\begin{align}
v_{\mathrm{ph}}^2-2\mathcal{E}v_{\mathrm{ph}}\cos\vartheta+\mathcal{E}^2\cos^2\vartheta-(1+\mathcal{E}^2)&=0\,, \\[2ex]
\mathbf{u}^2-2\mathcal{E}u^3+\frac{\mathcal{E}^2(u^3)^2}{\mathbf{u}^2}-(1+\mathcal{E}^2)&=0\,.
\end{align}
\end{subequations}
Okubo's technique leads to
\begin{subequations}
\begin{align}
\label{eq:finsler-structure-parity-odd-2a}
F(\mathbf{u})|_{\otimes}^{(2)\pm}&=\frac{-\mathcal{E}u^3\pm \sqrt{1+\mathcal{E}^2}|\mathbf{u}|}{1+\mathcal{E}^2-(\mathcal{E}u^3)^2/\mathbf{u}^2}\,, \\[2ex]
\label{eq:finsler-structure-parity-odd-2b}
F(\mathbf{u})|_{\otimes}^{(2)\boldsymbol{\zeta}\pm}&=\frac{-\boldsymbol{\zeta}\cdot\mathbf{u}\pm \sqrt{1+\mathcal{E}^2}|\mathbf{u}|}{1+\mathcal{E}^2-(\boldsymbol{\zeta}\cdot\mathbf{u})^2/\mathbf{u}^2}\,.
\end{align}
\end{subequations}
Let us investigate the characteristics of \eqref{eq:finsler-structure-parity-odd-2a}. We again obtain
\begin{equation}
g^{(2)\pm}_{\otimes,ij}u^iu^j=(F(\mathbf{u})|_{\otimes}^{(2)\pm})^2\,,\quad g^{(2)\pm}_{\otimes,ij}\equiv \frac{1}{2}\frac{\partial^2(F(\mathbf{u})|_{\otimes}^{(2)\pm})^2}{\partial u^i\partial u^j}\,.
\end{equation}
Hence for $F(\mathbf{u})|_{\otimes}^{(2)+}$ analogue properties hold such as for \eqref{eq:finsler-structure-parity-odd-1b}, which makes
it to a Finsler structure. Note that the latter is not automorphic, since its off-shell dispersion relation in \eqref{eq:dispersion-relation-parity-odd} does
not exclusively involve additional parity-odd terms, but also contributions like $(\boldsymbol{\zeta}\cdot \mathbf{k})^2$.
For this structure the Matsumoto torsion does not vanish, which is why it is neither a Randers nor a Kropina structure. The deviation
from a Randers structure is of second order in the controlling coefficients:
\begin{equation}
F(\mathbf{u})|_{\otimes}^{\boldsymbol{\zeta}(2)\pm}=-\boldsymbol{\zeta}\cdot\mathbf{u}\pm \sqrt{1+\mathcal{E}^2}|\mathbf{u}|+\mathcal{O}(\widetilde{\kappa}_{o+}^2)\,.
\end{equation}
Recall that the massive-photon dispersion equation of this mode, \eqref{eq:dispersion-relation-parity-odd-massive-2}, was not quadratic,
but quartic. For this reason it was challenging to derive a classical point-particle Lagrange function corresponding to the second photon
polarization.
It is also interesting to note that a large number of complications arise in the quantum field theory based on the parity-odd framework due to
the behavior of this mode \cite{Schreck:2011ai}. On the contrary the first mode is much easier to handle. The Finsler structures obtained
seem to reflect these properties. The first, given by \eqref{eq:finsler-structure-parity-odd-1b}, is a well-understood Randers structure,
whereas the second deviates from such a structure at second order in Lorentz violation, which makes its properties much more
involved to analyze.

The studies carried out in the current section will prove to be useful when describing photons in the geometric-optics approximation.
Thereby the eikonal equation will play an important role. How all these concepts are linked to each other will be clarified in the
forthcoming part of the article.

\section{Classical ray equations}
\label{sec:classical-equations-of-motion}
\setcounter{equation}{0}

Propagating electromagnetic waves can be treated in the geometric-optics approximation as long as their wave lengths can be neglected
in comparison to other physical length scales. For example this is possible for waves with low energies propagating over large
distances when physical phenomena related to the wave character (such as diffraction) do not play a role. This physical regime could be
called ``classical'' and the wave then corresponds to a geometric ray. The goal of the current section is to establish \textit{ray
equations} that describe the physical behavior of propagating rays.

Each electromagnetic pulse has a wavefront, which separates the region with nonzero electromagnetic fields from the region with
vanishing fields. At any instant of time the wavefront can be considered as a two-dimensional surface in three-dimensional space, i.e., it can be described
by an equation of the form $\psi(\mathbf{x})=t$ where $\mathbf{x}$ are spatial coordinates and $t$ is the time. The gradient $\boldsymbol{\nabla}\psi$ points
along the propagation direction and it is perpendicular to the surface. There is a relation between $\boldsymbol{\nabla}\psi$ and the refractive index $n$ of
the medium; it reads as $|\boldsymbol{\nabla}\psi|=n$. The latter is called the \textit{eikonal equation} in a subset of the literature.
In what follows, $n$ is assumed to depend on the position $\mathbf{x}$ only, but not on the velocity $\mathbf{u}$, i.e., $n=n(\mathbf{x})$.

Consider a wave propagating along a trajectory $\mathbf{x}(s)$ where $s$ is the arc length of the curve. In this parameterization the tangent vector has
magnitude 1, which is why the ray equations read as follows:
\begin{equation}
\label{eq:equations-of-motion-light}
\frac{\mathrm{d}\mathbf{x}}{\mathrm{d}s}=\frac{\boldsymbol{\nabla}\psi}{|\boldsymbol{\nabla}\psi|}\,,\quad n\frac{\mathrm{d}\mathbf{x}}{\mathrm{d}s}=\boldsymbol{\nabla}\psi\,.
\end{equation}
Computing an additional derivative of the latter with respect to $s$, its right-hand side can be expressed in terms of the refractive index as well:
\begin{equation}
\frac{\mathrm{d}}{\mathrm{d}s}\boldsymbol{\nabla}\psi=\left(\frac{\mathrm{d}\mathbf{x}}{\mathrm{d}s}\cdot \boldsymbol{\nabla}\right)\boldsymbol{\nabla}\psi=\frac{1}{n}\boldsymbol{\nabla}\psi \cdot [\boldsymbol{\nabla}(\boldsymbol{\nabla}\psi)]=\frac{1}{2n}\boldsymbol{\nabla}(\boldsymbol{\nabla}\psi)^2=\frac{1}{2n}\boldsymbol{\nabla}n^2=\boldsymbol{\nabla}n\,.
\end{equation}
Trajectories may not necessarily be parameterized by arc length. For an arbitrary parameterization with parameter $t$ we obtain
\begin{equation}
\frac{\mathrm{d}}{\mathrm{d}s}=\frac{\mathrm{d}t}{\mathrm{d}s}\frac{\mathrm{d}}{\mathrm{d}t}=\left(\frac{\mathrm{d}s}{\mathrm{d}t}\right)^{-1}\frac{\mathrm{d}}{\mathrm{d}t}=\frac{1}{|\mathbf{u}|}\frac{\mathrm{d}}{\mathrm{d}t}\,.
\end{equation}
Now the ray equations (\ref{eq:equations-of-motion-light}) can be cast into the following final form:
\begin{subequations}
\begin{align}
\label{eq:eikonal-equation-1}
\frac{\mathrm{d}}{\mathrm{d}s}\left(n\frac{\mathrm{d}\mathbf{x}}{\mathrm{d}s}\right)&=\boldsymbol{\nabla}n\,, \\[2ex]
\label{eq:equations-of-motion-light-final}
\frac{\mathrm{d}}{\mathrm{d}t}(\boldsymbol{\nabla}\psi)&=|\mathbf{u}|\boldsymbol{\nabla}n\,,\quad \boldsymbol{\nabla}\psi=n\frac{\mathbf{u}}{|\mathbf{u}|}\,.
\end{align}
\end{subequations}
The literature seems to be discordant about which equation should actually be called the eikonal equation. Some sources call the first one of
\eqref{eq:equations-of-motion-light-final} the eikonal equation, whereas others denote it as the vector magnitude of the second one. Note that the latter leads
us back to $|\boldsymbol{\nabla}\psi|=n$ (cf. the beginning of this section).
In the current paper whenever referring to the eikonal equation, we will be talking about the first one of \eqref{eq:equations-of-motion-light-final}. For
clarity, the vector magnitude of the second one will be called the \textit{wavefront equation}.
Equation (\ref{eq:equations-of-motion-light-final}) can be understood as the Euler-Lagrange equations resulting from the condition that
the following functional becomes stationary:
\begin{equation}
\label{eq:action-photon-trajectory}
L[\mathbf{x},\mathbf{u}]=\int_A^B \mathrm{d}s\,n(\mathbf{x})=\int_{T_A}^{T_B} \mathrm{d}t\,V\,,\quad V=V(\mathbf{x},\mathbf{u})=n(\mathbf{x})|\mathbf{u}|\,.
\end{equation}
The integrand of this functional is the infinitesimal optical path length and the functional itself gives the total optical path length travelled by a ray along its trajectory
between two points $A$ and $B$. Here $T_A$ is the departure time of the ray at $A$ and $T_B$ the arrival time at $B$. The optical path length is defined to
be the path length equivalent that light has to travel in vacuum to take the same time as for a given path in a medium with refractive index $n(\mathbf{x})$.
The quantity $V$ could be interpreted as the corresponding ``optical velocity.'' The functional of \eqref{eq:action-photon-trajectory} can be understood as
the base of the Fermat principle, cf. \cite{Perlick:2005hz,Torrome:2012kt}.

\subsection{Wavefront and eikonal equation in modified Maxwell theory}

The analogue of the wavefront equation in \eqref{eq:equations-of-motion-light-final} in the context of modified Maxwell theory was partially studied in \cite{Xiao:2010yx}.
The authors of the latter reference chose the coefficients contained in
$\kappa_{\scriptscriptstyle{DE}}$ and $\kappa_{\scriptscriptstyle{HB}}$ as nonvanishing where both the trace of these matrices and the matrices mixing electric and magnetic fields were
assumed to be zero. The trace components can be restored without any effort by just replacing their $\beta_E$ by $\kappa_{\scriptscriptstyle{DE}}$ and their $\beta_B$
by $\kappa_{\scriptscriptstyle{HB}}$. The wavefront equation then follows from the matrix $M_e$ in their Eq.~(38):
\begin{equation}
M_e^{ij}=\left(1-|\boldsymbol{\nabla}\psi|^2\right)\delta^{ij}+\partial^i\psi\partial^j\psi+\kappa_{\scriptscriptstyle{DE}}^{ij}-\kappa_{\scriptscriptstyle{HB}}^{kl}\varepsilon^{ink}\varepsilon^{jml}\partial^n\psi\partial^m\psi\,,
\end{equation}
where $\varepsilon^{ijk}$ is the totally antisymmetric Levi-Civita symbol in three dimensions with $\varepsilon^{123}=1$.
This matrix is multiplied with the time derivatives of the fields, which are singular on the wavefront. Therefore their Eq.~(33) can only have nontrivial solutions if the
determinant of $M_e$ vanishes. This condition directly leads to the wavefront equation within the framework considered. For the isotropic case
(cf.~\secref{sec:isotropic-case}), the anisotropic, nonbirefringent case (cf.~\secref{sec:anisotropic-nonbirefringent-case}), and the anisotropic,
birefringent sector (cf.~\secref{sec:anisotropic-birefringent-case}) we obtain
\begin{subequations}
\label{eq:eikonal-equations-isotropic-and-anisotropic}
\begin{align}
\label{eq:eikonal-equations-isotropic}
1&=\mathcal{A}^2|\boldsymbol{\nabla}\psi|^2\,, \displaybreak[0]\\[2ex]
\label{eq:eikonal-equations-anisotropic}
1&=(\partial^1\psi)^2+(\partial^2\psi)^2+\mathcal{B}^2(\partial^3\psi)^2\,, \displaybreak[0]\\[2ex]
\label{eq:eikonal-equations-anisotropic-birefringent-2}
1-\mathcal{G}^2&=|\boldsymbol{\nabla}\psi|^2+\mathcal{G}\left\{(\partial^1\psi)^2[1-\cos(2\eta)]-(\partial^3\psi)^2[1+\cos(2\eta)]\right\} \notag \\
&\phantom{{}={}}-\mathcal{G}^2\left\{(\partial^2\psi)^2+[(\partial^1\psi)^2-(\partial^3\psi)^2]\cos(2\eta)\right\}\,, \displaybreak[0]\\[2ex]
\label{eq:eikonal-equations-anisotropic-birefringent-1}
1+\mathcal{G}\cos(2\eta)&=|\boldsymbol{\nabla}\psi|^2+\mathcal{G}\left\{(\partial^1\psi)^2+(\partial^2\psi)^2\cos(2\eta)-(\partial^3\psi)^2\right\}\,.
\end{align}
\end{subequations}
These are the analogues of the wavefront equation $|\boldsymbol{\nabla}\psi|^2=n^2$ in modified Maxwell theory.
Following the lines in connection to \eqref{eq:hamilton-jacobi-equation} classical Hamilton functions can be obtained as parts of the
Hamilton-Jacobi equation describing a classical ray. For the sectors considered few lines above they read as follows:
\begin{subequations}
\begin{align}
H|_{\circledcirc}&=-\mathcal{A}\sqrt{\mathbf{k}^2}\,, \displaybreak[0]\\[2ex]
H|_{\varobar}&=-\sqrt{k_1^2+k_2^2+\mathcal{B}^2k_3^2}\,, \displaybreak[0]\\[2ex]
\sqrt{1-\mathcal{G}^2}H|_{\varovee}^{(1)}&=-\Big\{\mathbf{k}^2+\mathcal{G}\left\{k_1^2[1-\cos(2\eta)]-k_3^2[1+\cos(2\eta)]\right\}\Big. \notag \\
&\phantom{{}={}-\Big(}\Big.-\mathcal{G}^2\left[k_2^2+(k_1^2-k_3^2)\cos(2\eta)\right]\!\Big\}^{1/2}\,, \notag \\
H|_{\varovee}^{(1)}&=-\sqrt{\mathcal{G}_1k_1^2+k_2^2+\mathcal{G}_2k_3^2}\,, \displaybreak[0]\\[2ex]
\sqrt{1+\mathcal{G}\cos(2\eta)}H|_{\varovee}^{(2)}&=-\sqrt{\mathbf{k}^2+\mathcal{G}\left[k_1^2+k_2^2\cos(2\eta)-k_3^2\right]}\,, \notag \\
H|_{\varovee}^{(2)}&=-\sqrt{\widetilde{\mathcal{G}}_1k_1^2+k_2^2+\widetilde{\mathcal{G}}_2k_3^2}\,,
\end{align}
\end{subequations}
with $\mathcal{G}_1$, $\mathcal{G}_2$ of \eqref{eq:finsler-structure-anisotropic-birefringence-1-constants} and $\widetilde{\mathcal{G}}_1$,
$\widetilde{\mathcal{G}}_2$ taken from \eqref{eq:finsler-structure-anisotropic-birefringence-2-constants}. These Hamilton functions are directly
linked to the modified dispersion relations, cf.~the paragraph below \eqref{eq:dispersion-relation-isotropic} for the isotropic case,
\eqref{eq:dispersion-relation-anisotropic} for the anisotropic (nonbirefringent) sector, and \eqref{eq:dispersion-relations-anisotropic-birefringent}
for the anisotropic (birefringent) case. This nicely demonstrates that all computations are consistent with each other.

The wavefront equations (\ref{eq:eikonal-equations-isotropic-and-anisotropic}) are not suitable for our calculations, since they involve
first derivatives of the wavefront that are unclear how to be treated. Having the eikonal equations involving the refractive indices and
velocity components only would be of advantage. As a cross check with the previously obtained results the refractive indices can be derived
from \eqref{eq:eikonal-equations-isotropic-and-anisotropic}. For the isotropic case, using the second of \eqref{eq:equations-of-motion-light-final}
we obtain $|\boldsymbol{\nabla}\psi|^2=n^2$, which by inserting into \eqref{eq:eikonal-equations-isotropic} directly leads to the isotropic
result $n|_{\circledcirc}=1/\mathcal{A}$. In \eqref{eq:eikonal-equations-anisotropic} of the anisotropic (nonbirefringent) sector we can introduce
\begin{equation}
(\partial^1\psi)^2+(\partial^2\psi)^2=n^2\sin^2\vartheta\,,\quad\partial^3\psi=n\cos\vartheta\,,
\end{equation}
leading to $n|_{\varobar}=1/\sqrt{\sin^2\vartheta+\mathcal{B}^2\cos^2\vartheta}$.
The latter depends on the angle $\vartheta$ between the propagation direction and the preferred direction $\boldsymbol{\zeta}$. For the
anisotropic (birefringent) sector we insert
\begin{equation}
\partial^1\psi=n\frac{\cos\theta_1-\cos\theta_2}{2\sin\eta}\,,\quad \partial^3\psi=n\frac{\cos\theta_1+\cos\theta_2}{2\cos\eta}\,,\quad \partial^2\psi=\sqrt{n^2-(\partial^1\psi)^2-(\partial^3\psi)^3}\,,
\end{equation}
both in \eqref{eq:eikonal-equations-anisotropic-birefringent-1} and \eqref{eq:eikonal-equations-anisotropic-birefringent-2}
to obtain two refractive indices differing at second order in Lorentz violation:
\begin{subequations}
\begin{align}
n|_{\varovee}^{(1)}&=\sqrt{\frac{1-\mathcal{G}^2}{1+\mathcal{G}\left\{(\mathcal{G}/2)\left[\cos(2\theta_1)+\cos(2\theta_2)\right]-2\cos\theta_1\cos\theta_2\right\}}}\,, \displaybreak[0]\\[2ex]
n|_{\varovee}^{(2)}&=\sqrt{\frac{1+\mathcal{G}\cos(2\eta)}{1+\mathcal{G}[\cos(2\eta)-2\cos\theta_1\cos\theta_2]}}\,.
\end{align}
\end{subequations}
Based on these refractive indices the integrands of the action functional in \eqref{eq:action-photon-trajectory} can be computed.
The results are consistent with Eqs.~(\ref{eq:finsler-structure-isotropic}), (\ref{eq:finsler-structure-anisotropic}):
\begin{subequations}
\begin{align}
V(\mathbf{u})|_{\circledcirc}&=n|_{\circledcirc}|\mathbf{u}|=\frac{1}{\mathcal{A}}\sqrt{\mathbf{u}^2}=F(\mathbf{u})|_{\circledcirc}^+\,, \displaybreak[0]\\[2ex]
V(\mathbf{u})|_{\varobar}&=n|_{\varobar}|\mathbf{u}|=\frac{\sqrt{(u^1)^2+(u^2)^2+(u^3)^2}}{\sqrt{\sin^2\vartheta+\mathcal{B}^2\cos^2\vartheta}}=\frac{(u^1)^2+(u^2)^2+(u^3)^2}{\sqrt{(u^1)^2+(u^2)^2+\mathcal{B}^2(u^3)^2}} \notag \\
&=F(\mathbf{u})|_{\varobar}^+\,, \displaybreak[0]\\[2ex]
V(\mathbf{u})|_{\varovee}^{(1)}&=n|_{\varovee}^{(1)}|\mathbf{u}|=F(\mathbf{u})|_{\varovee}^{(1)+}\,,\quad V(\mathbf{u})|_{\varovee}^{(2)}=n|_{\varovee}^{(2)}|\mathbf{u}|=F(\mathbf{u})|_{\varovee}^{(2)+}\,,
\end{align}
\end{subequations}
where for the latter two \eqref{eq:correspondence-angle-velocities} has to be employed. The refractive indices obtained from the wavefront
equations correspond to the refractive indices computed directly from their definitions via the inverse phase velocity:
$n\equiv v_{\mathrm{ph}}^{-1}=|\mathbf{k}|/\omega$.
\begin{subequations}
\begin{align}
\label{eq:refraction-index-isotropic}
n|_{\circledcirc}&=v_{\mathrm{ph}}|_{\circledcirc}^{-1}=\frac{|\mathbf{k}|}{\omega|_{\circledcirc}}=\frac{1}{\mathcal{A}}\,, \\[2ex]
\label{eq:refraction-index-anisotropic}
n|_{\varobar}&=v_{\mathrm{ph}}|_{\varobar}^{-1}=\frac{|\mathbf{k}|}{\omega|_{\varobar}}=\sqrt{\frac{\mathbf{k}^2}{k_{\bot}^2+\mathcal{B}^2k_{\scalebox{0.6}{$\|$}}^2}}=\frac{1}{\sqrt{\sin^2\vartheta+\mathcal{B}^2\cos^2\vartheta}}\,, \\[2ex]
\label{eq:refraction-index-anisotropic-birefringent}
n|_{\varovee}^{(1)}&=(v_{\mathrm{ph}}|_{\varovee}^{(1)})^{-1}=\frac{|\mathbf{k}|}{\omega_1|_{\varovee}}\,,\quad n|_{\varovee}^{(2)}=(v_{\mathrm{ph}}|_{\varovee}^{(2)})^{-1}=\frac{|\mathbf{k}|}{\omega_2|_{\varovee}}\,.
\end{align}
\end{subequations}
These previously performed studies do not reveal any inconsistencies. The essential conclusion is that it should be warranted to describe the
isotropic, anisotropic (nonbirefringent), and anisotropic (birefringent) sectors of modified Maxwell theory (in the geometric-optics approximation)
with an adapted version of the eikonal equation, \eqref{eq:equations-of-motion-light-final}.

Last but not least the parity-odd sector of \secref{sec:parity-odd-case} shall be elaborated on.
The wavefront equations for the parity-odd case were not derived in \cite{Xiao:2010yx}, since in the latter reference all controlling coefficients
mixing electric and magnetic fields were set to zero. Adapting the procedure used allows to derive them nevertheless.
The authors of \cite{Xiao:2010yx} consider the values of the fields directly on the wavefront, e.g., for the electric field:
$\mathbf{E}_0(\mathbf{x})=\mathbf{E}(t,\mathbf{x})|_{t=\psi(\mathbf{x})}$. In what follows, all fields evaluated on the wavefront will be
denoted by an additional ``0'' as an index. The spatial derivative on the wavefront is then given by:
\begin{equation}
\frac{\partial \mathbf{E}_0}{\partial x^j}=\frac{\partial \mathbf{E}}{\partial x^j}+\dot{\mathbf{E}}\frac{\partial\psi}{\partial x^j}\,.
\end{equation}
Based on this procedure, from Maxwell's equations four equations can be derived that involve field components on the wavefront and
field derivatives only:
\begin{subequations}
\begin{align}
\boldsymbol{\nabla}\times \mathbf{E}_0&=-\dot{\mathbf{B}}+\boldsymbol{\nabla}\psi \times \dot{\mathbf{E}}\,,\quad \boldsymbol{\nabla}\times \mathbf{H}_0=\dot{\mathbf{D}}+\boldsymbol{\nabla}\psi\times \dot{\mathbf{H}}\,, \\[2ex]
\boldsymbol{\nabla}\cdot \mathbf{D}_0&=\boldsymbol{\nabla}\psi\cdot \dot{\mathbf{D}}\,,\quad \boldsymbol{\nabla}\cdot \mathbf{B}_0=\boldsymbol{\nabla}\psi\cdot \dot{\mathbf{B}}\,, \\[2ex]
\mathbf{D}&=\mathbf{E}+\kappa_{\scriptscriptstyle{DB}}\mathbf{B}\,,\quad \mathbf{H}=\kappa_{\scriptscriptstyle{DB}}\mathbf{E}+\mathbf{B}\,.
\end{align}
\end{subequations}
These must be combined to obtain an equation that involves the time derivatives of only a single field, e.g., the electric field
and field values on the wavefront that may not necessarily include only a single field. This can be carried out via the following chain of
steps:
\begin{subequations}
\begin{align}
\boldsymbol{\nabla}\times \mathbf{E}_0&=-\dot{\mathbf{H}}+\kappa_{\scriptscriptstyle{DB}}\dot{\mathbf{E}}+\boldsymbol{\nabla}\psi \times \dot{\mathbf{E}}\,, \displaybreak[0]\\[2ex]
\boldsymbol{\nabla}\psi \times (\boldsymbol{\nabla}\times \mathbf{E}_0)&=-\boldsymbol{\nabla}\psi\times \dot{\mathbf{H}}+\boldsymbol{\nabla}\psi \times \kappa_{\scriptscriptstyle{DB}}\dot{\mathbf{E}}+\boldsymbol{\nabla}\psi\times(\boldsymbol{\nabla}\psi\times \dot{\mathbf{E}})\,, \displaybreak[0]\\[2ex]
\boldsymbol{\nabla}\psi\times (\boldsymbol{\nabla}\times \mathbf{E}_0)&=\dot{\mathbf{D}}-\boldsymbol{\nabla}\times \mathbf{H}_0+\boldsymbol{\nabla}\psi\times \kappa_{\scriptscriptstyle{DB}}\dot{\mathbf{E}}+\boldsymbol{\nabla}\psi\times (\boldsymbol{\nabla}\psi\times \dot{\mathbf{E}})\,, \displaybreak[0]\\[2ex]
\boldsymbol{\nabla}\psi\times (\boldsymbol{\nabla}\times \mathbf{E}_0)&=\dot{\mathbf{E}}+\kappa_{\scriptscriptstyle{DB}}\dot{\mathbf{B}}-\boldsymbol{\nabla}\times \mathbf{H}_0+\boldsymbol{\nabla}\psi\times \kappa_{\scriptscriptstyle{DB}}\dot{\mathbf{E}}+\boldsymbol{\nabla}\psi\times (\boldsymbol{\nabla}\psi\times \dot{\mathbf{E}})\,, \displaybreak[0]\\[2ex]
\boldsymbol{\nabla}\psi\times (\boldsymbol{\nabla}\times \mathbf{E}_0)&=\dot{\mathbf{E}}+\kappa_{\scriptscriptstyle{DB}}(\boldsymbol{\nabla}\psi\times \dot{\mathbf{E}}-\boldsymbol{\nabla}\times \mathbf{E}_0)-\boldsymbol{\nabla}\times \mathbf{H}_0 \notag \\
&\phantom{{}={}}+\boldsymbol{\nabla}\psi\times \kappa_{\scriptscriptstyle{DB}}\dot{\mathbf{E}}+\boldsymbol{\nabla}\psi\times (\boldsymbol{\nabla}\psi\times \dot{\mathbf{E}})\,.
\end{align}
\end{subequations}
The resulting equation then reads
\begin{align}
\dot{\mathbf{E}}+\kappa_{\scriptscriptstyle{DB}}\boldsymbol{\nabla}\psi\times \dot{\mathbf{E}}+\boldsymbol{\nabla}\psi \times \kappa_{\scriptscriptstyle{DB}}\dot{\mathbf{E}}+\boldsymbol{\nabla}\psi\times (\boldsymbol{\nabla}\psi\times \dot{\mathbf{E}})&=\boldsymbol{\nabla}\psi\times (\boldsymbol{\nabla}\times \mathbf{E}_0)+\boldsymbol{\nabla}\times \mathbf{H}_0 \notag \\
&\phantom{{}={}}+\kappa_{\scriptscriptstyle{DB}}\boldsymbol{\nabla}\times \mathbf{E}_0\,.
\end{align}
The condition for a vanishing determinant of the matrix on the left-hand side for the existence of nontrivial solutions leads to
the wavefront equation for the parity-odd case. For consistency we pull the index of $\boldsymbol{\nabla}\psi$ down:
\begin{equation}
\label{eq:eikonal-equation-parity-odd}
\left(1-2\boldsymbol{\zeta}\cdot \boldsymbol{\nabla}\psi-|\boldsymbol{\nabla}\psi|^2\right)\left[1-(1+\boldsymbol{\zeta}^2)|\boldsymbol{\nabla}\psi|^2-2(\boldsymbol{\zeta}\cdot\boldsymbol{\nabla}\psi)+(\boldsymbol{\zeta}\cdot\boldsymbol{\nabla}\psi)^2\right]=0\,.
\end{equation}
Inserting the second of \eqref{eq:equations-of-motion-light-final} in the first factor of \eqref{eq:eikonal-equation-parity-odd} results in
\begin{equation}
1-2\boldsymbol{\zeta}\cdot \boldsymbol{\nabla}\psi-|\boldsymbol{\nabla}\psi|^2=1-2n\boldsymbol{\zeta}\cdot \widehat{\mathbf{u}}-n^2 \overset{!}{=} 0\,.
\end{equation}
The latter can be solved with respect to the refractive index $n$ to give
\begin{equation}
\label{eq:refraction-index-parity-odd-1}
n|_{\otimes}^{\boldsymbol{\zeta}(1)}=-\boldsymbol{\zeta}\cdot \widehat{\mathbf{u}}+\sqrt{1+(\boldsymbol{\zeta}\cdot \widehat{\mathbf{u}})^2}=-\mathcal{E}\cos\vartheta+\sqrt{1+\mathcal{E}^2\cos^2\vartheta}\,,
\end{equation}
where only the positive-sign solution delivers a physically meaningful refractive index.
Hence the result obtained from the eikonal equation is consistent with \eqref{eq:finsler-structure-parity-odd-1b}, which can be seen upon close
inspection:
\begin{equation}
V(\mathbf{u})|_{\otimes}^{\boldsymbol{\zeta}(1)}=n|_{\otimes}^{\boldsymbol{\zeta}(1)}|\mathbf{u}|=-\boldsymbol{\zeta}\cdot\mathbf{u}+\sqrt{\mathbf{u}^2+(\boldsymbol{\zeta}\cdot\mathbf{u})^2}=F(\mathbf{u})|_{\otimes}^{\boldsymbol{\zeta}(1)+}\,.
\end{equation}
The same procedure applied to the second factor of \eqref{eq:eikonal-equation-parity-odd} leads to:
\begin{equation}
1-(1+\mathcal{E}^2)n^2-2n(\boldsymbol{\zeta}\cdot \widehat{\mathbf{u}})+n^2(\boldsymbol{\zeta}\cdot \widehat{\mathbf{u}})^2=0\,.
\end{equation}
Therefore the refractive index reads
\begin{equation}
\label{eq:refraction-index-parity-odd-2}
n|_{\otimes}^{\boldsymbol{\zeta}(2)}=\frac{-\boldsymbol{\zeta}\cdot \widehat{\mathbf{u}}+\sqrt{1+\mathcal{E}^2}}{1+\mathcal{E}^2-(\boldsymbol{\zeta}\cdot\widehat{\mathbf{u}})^2}=\frac{-\mathcal{E}\cos\vartheta+\sqrt{1+\mathcal{E}^2}}{1+\mathcal{E}^2\sin^2\vartheta}\,,
\end{equation}
which is consistent with \eqref{eq:finsler-structure-parity-odd-2b}
\begin{equation}
V(\mathbf{u})|_{\otimes}^{\boldsymbol{\zeta}(2)}=n|_{\otimes}^{\boldsymbol{\zeta}(2)}|\mathbf{u}|=\frac{-\boldsymbol{\zeta}\cdot\mathbf{u}+\sqrt{1+\mathcal{E}^2}|\mathbf{u}|}{1+\mathcal{E}^2-(\boldsymbol{\zeta}\cdot\mathbf{u})^2/\mathbf{u}^2}=F(\mathbf{u})|_{\otimes}^{\boldsymbol{\zeta}(2)+}\,.
\end{equation}
The refractive indices obtained from the wavefront equations for the isotropic and anisotropic cases, Eqs.~(\ref{eq:eikonal-equations-isotropic}) --
(\ref{eq:eikonal-equations-anisotropic-birefringent-1}), respectively, are consistent with the usual definition of the refractive index via the inverse phase
velocity (cf. Eqs~(\ref{eq:refraction-index-isotropic}) -- (\ref{eq:refraction-index-anisotropic-birefringent}). However this does not seem to be
the case for the parity-odd sector. Inspecting
Eqs.~(\ref{eq:dispersion-relation-parity-odd-1}), (\ref{eq:dispersion-relation-parity-odd-2}) and the latter results for the refractive indices
of Eqs.~(\ref{eq:refraction-index-parity-odd-1}), (\ref{eq:refraction-index-parity-odd-2}) reveals the inconsistency:
\begin{subequations}
\begin{align}
\frac{|\mathbf{k}|}{\omega_1|_{\otimes}}&=\frac{1}{\mathcal{E}\cos\vartheta+\sqrt{1+\mathcal{E}^2\cos^2\vartheta}}\neq n|_{\otimes}^{\boldsymbol{\zeta}(1)}\,, \\[2ex]
\frac{|\mathbf{k}|}{\omega_2|_{\otimes}}&=\frac{1}{\mathcal{E}\cos\vartheta+\sqrt{1+\mathcal{E}^2}}\neq n|_{\otimes}^{\boldsymbol{\zeta}(2)}\,.
\end{align}
\end{subequations}
The definition of the refractive index via the inverse of the phase velocity rests on the existence of a nonzero permeability and permittivity.
However for the parity-odd case they both vanish and the electric fields even mix with the magnetic fields, which is why the ordinary
definition of the refractive index does not seem to be reasonable. A further origin of the issue may be that Okubo's method does
not produce Finsler structures in a unique manner. We conclude that it may be problematic to treat the parity-odd
case of modified Maxwell theory with the eikonal equation. Finding a solution to this clash is an interesting open problem.

\section{Gravitational backgrounds}
\label{sec:gravitational-backgrounds}
\setcounter{equation}{0}

The physics for a classical point-particle equivalent to a massive fermion rests on its Lagrangian. The procedure of deriving those within
the framework of the SME works for massive particles only where in the limit of a vanishing particle mass the Lagrangian goes to zero. So far we have demonstrated that the
important quantity to describe the physics of electromagnetic waves in the geometric-optics approximation is the refractive index. The reason is that the motion of photons
is much more restricted than the motion of a massive particle. After all, for a particle with mass moving in a potential the initial position,
direction, and velocity can be chosen freely. On the contrary, for a photon the initial position and direction only are not fixed, whereas its
initial speed is determined by the refractive index at its starting point.

In the previous sections it was shown how to establish connections between various cases of the minimal SME photon sector and
certain Finsler geometries. The Finsler geometries found were discovered to be closely related to the various refractive indices where
only for the parity-odd case of the {\em CPT}-even sector such a connection is not manifest. The refractive indices found are independent
of the spacetime position such as the controlling coefficients, which corresponds to the analogue of a homogeneous medium in optics.
However the refractive index can depend on the three spatial velocity components. In other words, in such cases the refractive index depends
on angles enclosed between the propagation direction and preferred directions. This situation is reminiscent of anisotropic media in optics.
Hence Finsler structures related to the {\em CPT}-even photon sector are three-dimensional in contrast to the Finsler
structures obtained from Wick-rotating classical Lagrangians of massive particles. Besides, note that in the photon case no Wick
rotation is necessary, since the intrinsic metric involved is already of Euclidean signature.

These results shall serve as a base to study light rays in the geometric-optics approximation in the presence of Lorentz violation.
As we saw, for most cases these can be described by the eikonal equation, cf.~\eqref{eq:equations-of-motion-light-final}:
\begin{equation}
\frac{\mathrm{d}}{\mathrm{d}s}\left[n\frac{\mathrm{d}\mathbf{x}}{\mathrm{d}s}\right]=\boldsymbol{\nabla}_{\mathbf{x}}n\,,
\end{equation}
where $n$ is the refractive index of the medium considered. On the right-hand side the gradient is understood to be computed
with respect to the position vector $\mathbf{x}$. The photon trajectory is given by $\mathbf{x}=\mathbf{x}(s)$ and it is
parameterized by the arc length $s$. For an isotropic and homogeneous medium the refractive index is a mere constant.
In this case one immediately sees that the resulting ray equation is
\begin{equation}
\label{eq:eikonal-equation-homogeneous-medium}
\frac{\mathrm{d}^2\mathbf{x}}{\mathrm{d}s^2}=\mathbf{0}\,,
\end{equation}
whose solution is a straight line as expected. For homogeneous, but anisotropic media the refractive index depends on at least
one angle, $n=n(\vartheta)$, where further angles are suppressed for brevity. For a straight ray trajectory the angle $\vartheta$
is fixed by the initial direction and it does not change during propagation, i.e., it is not a function of $s$. Furthermore due to
homogeneity the refractive index does not change along the trajectory as well, which is why $\boldsymbol{\nabla}_{\mathbf{x}}n(\vartheta)=0$
for points on the trajectory. Therefore in this case we again end up with \eqref{eq:eikonal-equation-homogeneous-medium}.
For inhomogeneous media with $n=n(\mathbf{x})$ the eikonal equation cannot have straight-line solutions, though.

In what follows the formalism and knowledge attained shall be applied to propagating light rays in curved spacetimes with metric
$g_{\mu\nu}=g_{\mu\nu}(x)$. The trajectory of a ray in a spacetime is described by a four-vector $x^{\mu}=x^{\mu}(s)$ and it
propagates with the four-velocity $u^{\mu}\equiv \mathrm{d}x^{\mu}/\mathrm{d}s$. Propagation occurs along geodesics combined
with the nullcone condition $g_{\mu\nu}u^{\mu}u^{\nu}=0$ that has to hold locally at each spacetime point. For practical reasons,
which will become clear in the course of the current section, all forthcoming investigations will be performed in a spacetime
characterized by a line interval of the form
\begin{equation}
\label{eq:line-interval}
\mathrm{d}\tau^2=\frac{1}{A(r,\theta,\phi)}\mathrm{d}t^2-A(r,\theta,\phi)(\mathrm{d}r^2+r^2\mathrm{d}\theta^2+r^2\sin^2\theta\mathrm{d}\phi^2)\,.
\end{equation}
Here $t$ is the time, $(r,\theta,\phi)$ are spherical coordinates, and $A$ is a time-independent function. Such metrics were
proposed in \cite{Wu:1988} and they are denoted as ``generally isotropic'' where metrics with $A=A(r)$ are called ``spherically symmetric.''
The parentheses in the spatial part of \eqref{eq:line-interval} give the volume element of a three-dimensional ball and it is
multiplied by $A(r,\theta,\phi)$. The choice $A(r,\theta,\phi)=1$ in \eqref{eq:line-interval} describes Minkowski spacetime in three-dimensional
spherical coordinates. In this case the spatial coordinate surfaces with constant $r$ are two-spheres. For arbitrary $A(r,\theta,\phi)$
these surfaces are still two-spheres topologically, but their local geometry depends on $r$, $\theta$, and $\phi$. Note that the
metric describing a weak gravitational field can be brought into the generally isotropic form:
\begin{align}
\label{eq:metric-weak-gravitational-field}
(g_{\mu\nu})&=\mathrm{diag}\Big((1+2\Phi),-(1-2\Phi),-(1-2\Phi),-(1-2\Phi)\Big) \notag \\
&=\mathrm{diag}\Big(\frac{1}{1-2\Phi},-(1-2\Phi),-(1-2\Phi),-(1-2\Phi)\Big)+\mathcal{O}(\Phi^2)\,.
\end{align}
Here $\Phi=\Phi(r)=-GM/r\ll 1$ is the Newtonian potential.

In the latter paper \cite{Wu:1988} it was shown that there is a link between the eikonal equation of the geometric-optics approximation
and the null geodesic equations of a spacetime based on a line interval of \eqref{eq:line-interval}. A suitable combination of the geodesic
equations leads to
\begin{equation}
\frac{\mathrm{d}}{\mathrm{d}s}\left[A(r,\theta,\phi)\frac{\mathrm{d}\mathbf{x}}{\mathrm{d}s}\right]=\boldsymbol{\nabla}A(r,\theta,\phi)\,,
\end{equation}
i.e., $A(r,\theta,\phi)$ of \eqref{eq:line-interval} can be understood as an inhomogeneous and anisotropic refractive index. Therefore
as long as weak gravitational fields are considered, light behaves according to the geometric-optics approximation. The approximation
is expected to break down as soon as strong gravitational forces appear such as in the direct vicinity of a black hole. In this case the
original geodesic equations have to be studied instead of the eikonal approach. Note that the converse is true as well. If the eikonal
equation is known to be valid (also in flat spacetime) this corresponds to a propagating ray in a generally isotropic spacetime of
\eqref{eq:line-interval}.

\subsection{Isotropic case}
\label{sec:gravitational-backgrounds-isotropic-case}

The eikonal approach has a great potential to be applied to the propagation of light rays in a weak gravitational field permeated
by a Lorentz-violating background field. It is reasonable to start with the simplest case, which is the isotropic one investigated in
\secref{sec:isotropic-case}. With the constant refractive index $n=1/\mathcal{A}$ (in Minkowski spacetime) given by \eqref{eq:isotropic-case-property-tensors}
or \eqref{eq:refraction-index-isotropic} the eikonal equation and the corresponding spacetime, \eqref{eq:line-interval}, read as follows:
\begin{subequations}
\begin{align}
\label{eq:line-interval-isotropic-case}
\frac{\mathrm{d}}{\mathrm{d}s}\left[\frac{1}{\mathcal{A}}\frac{\mathrm{d}\mathbf{x}}{\mathrm{d}s}\right]&=\boldsymbol{\nabla}\left(\frac{1}{\mathcal{A}}\right)\,, \\[2ex]
\mathrm{d}\tau^2&=\mathcal{A}\mathrm{d}t^2-\frac{1}{\mathcal{A}}(\mathrm{d}r^2+r^2\mathrm{d}\theta^2+r^2\sin^2\theta\mathrm{d}\phi^2)\,.
\end{align}
\end{subequations}
The coordinate surfaces of the associated spacetime are spheres whose radii are scaled by $1/\sqrt{\mathcal{A}}$.
This intermediate result can now be used to introduce a gravitational background. Via the principle of minimal coupling the flat Minkowski
metric is replaced by a curved spacetime metric, $\eta_{\mu\nu}\mapsto g_{\mu\nu}(x)$, and the constant refractive index $n$ is
promoted to a spacetime-position dependent function: $n\mapsto n(r,\theta,\phi)$. The curved spacetime metric is
taken to be \eqref{eq:metric-weak-gravitational-field} for a weak gravitational field. Since the latter is spherically symmetric, it is reasonable
to assume spherical symmetry for the position-dependent refractive index, i.e., $n(r)=1/\mathcal{A}(r)$. The corresponding eikonal
equation and the line interval then read as follows:
\begin{subequations}
\begin{align}
\label{eq:refractive-index-isotropic-gravitational}
\frac{\mathrm{d}}{\mathrm{d}s}\left[n(r)\frac{\mathrm{d}\mathbf{x}}{\mathrm{d}s}\right]&=\boldsymbol{\nabla}n(r)\,,\quad n(r)\equiv \frac{1-2\Phi(r)}{\mathcal{A}(r)}\,, \\[2ex]
\label{eq:spherical-metric-isotropic-modmax}
\mathrm{d}\tau^2&=\frac{1}{n(r)}\mathrm{d}t^2-n(r)(\mathrm{d}r^2+r^2\mathrm{d}\theta^2+r^2\sin^2\theta\mathrm{d}\phi^2)\,.
\end{align}
\end{subequations}
\begin{figure}
\centering
\includegraphics{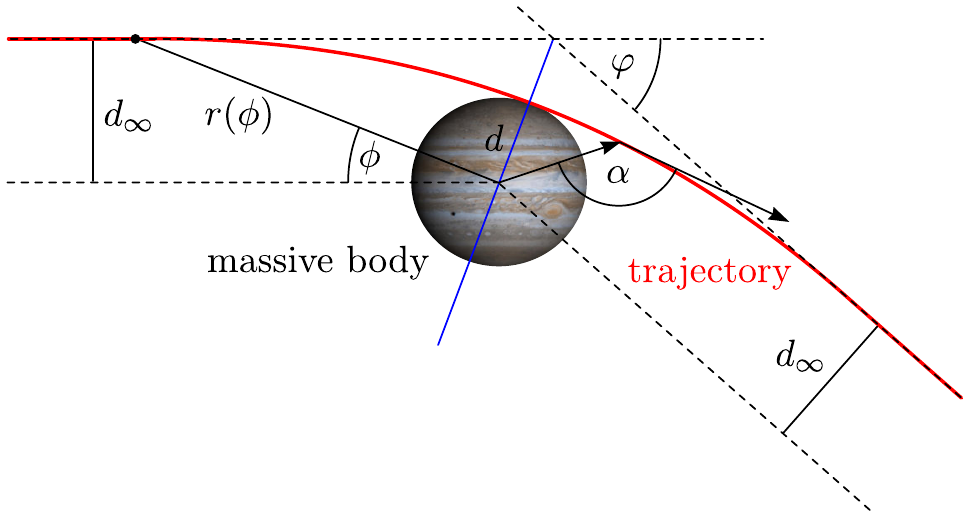}
\caption{Deflection of light near a massive body, e.g., the planet Jupiter. (The picture of Jupiter was taken by the Cassini spacecraft,
cf.~\url{http://solarsystem.nasa.gov/planets/profile.cfm?Object=Jupiter}.)}
\label{fig:light-deflection}
\end{figure}
Hence the minimal-coupling principle amounts to a refractive index that is the product of a spatial component of the weak
gravitational field metric and the spacetime-position dependent refractive index $1/\mathcal{A}(r)$ associated to the isotropic Lorentz-violating
framework considered.

The approach introduced has a paramount advantage. The physics of a Lorentz-violating photon in a (weak) gravity field can be studied
without field theory and the geodesic equations in a curved spacetime. Instead, a classical method is used replacing photons by light
rays and working in the geometric-optics approximation with the eikonal equation. In this context Lorentz symmetry violation is treated
as explicit, which is known to clash with the existence of gravitational backgrounds \cite{Kostelecky:2003fs}. The latter sections
\ref{sec:modified-energy-momentum-conservation} and \ref{sec:properties-isotropic-finsler} will be dedicated to this issue where for
now we will delve into phenomenology.

One possible application of the used approach lies in the (modified) deflection of light in the vicinity of a massive body
(cf.~\figref{fig:light-deflection}), which is an important test of gravitational theories. From a technical point of view the eikonal equation is
nonlinear, which makes it challenging to solve analytically in general. However for the isotropic case, i.e., a refractive index only depending
on the radial coordinate $r$ the formula of Bouguer follows from the eikonal equation (see, e.g., Sec.~3.2.1 of~\cite{Born:1999}):
\begin{equation}
\label{eq:bouguer-formula-nonintegrated}
n(r)r\sin\alpha=C\,.
\end{equation}
Here $C$ is a constant and $\alpha$ the angle between the tangent vector of the trajectory and the radial vector pointing from the coordinate
origin to a particular point on the trajectory. The latter relationship is the equivalent of energy and angular momentum conservation for a massive particle
in classical mechanics. Since both the distance $r$ of a particular point from the origin and the angle $\alpha$ associated
to this point does not depend on the parameterization of the trajectory, we choose to parameterize it by spherical coordinates. Thereby the
problem is restricted to the $x$-$z$-plane with $\theta=\pi/2$. The trajectory then reads $\mathbf{x}=r\widehat{\mathbf{e}}_r$
where $r=r(\phi)$ and $\widehat{\mathbf{e}}_r=\widehat{\mathbf{e}}_r(\phi)$ is the unit vector pointing in radial direction. The angle $\alpha$ is
given as follows:
\begin{equation}
\sin\alpha=\frac{r(\phi)}{\sqrt{r^2(\phi)+\dot{r}^2(\phi)}}\,,\quad \dot{r}\equiv \frac{\mathrm{d}r}{\mathrm{d}\phi}\,.
\end{equation}
Now the formula of Bouguer delivers a differential equation for $\phi(r)$. Its solution is obtained by solving the latter with respect to $\mathrm{d}\phi/\mathrm{d}r$
and by performing a subsequent integration:
\begin{equation}
\label{eq:integration-deflection-angle}
\phi(r)=C\int_d^r \frac{\mathrm{d}r}{r\sqrt{n(r)^2r^2-C^2}}\,,
\end{equation}
where $d$ is the distance of minimal proximity and the condition $\phi(d)=0$ has been set. By doing so, solving the eikonal equation has been
reduced to computing a one-dimensional integral. Now
consider a classical light ray approaching a massive body with impact parameter $d_{\infty}$, which is the distance between the particle
propagation direction in the asymptotically flat region and the parallel going through the center of mass of the body (at the coordinate origin). The
photon will travel
such that its distance to the body steadily decreases until reaching a minimum where it increases again afterwards. At the minimum distance $d$
we have that $\dot{r}=0$ and therefore $\alpha=\pi/2$. The minimum distance corresponds to the impact parameter to a very good approximation:
$d\approx d_{\infty}$. This is why \eqref{eq:bouguer-formula-nonintegrated} immediately tells us that
\begin{equation}
C=n(d)d\approx n(d_{\infty})d_{\infty}\,.
\end{equation}
Without the massive body the change $\Delta\phi$ in the angle would be equal to $\pi$ for a photon coming from an asymptotically flat region,
passing near the coordinate origin, and propagating back to infinity. Due to the body there is a deflection, which changes $\Delta\phi$ to an
angle that is slightly larger than $\pi$. Performing the integration in \eqref{eq:integration-deflection-angle} from $r=d$ to infinity gives half of
this contribution, since it only takes into account the second half of the trajectory. Therefore the deflection angle $\varphi$ is given by
\begin{equation}
\label{eq:deflection-angle}
\varphi=\Delta\phi-\pi=2C\int_d^{\infty} \frac{\mathrm{d}r}{r\sqrt{n(r)^2r^2-C^2}}-\pi\,.
\end{equation}
It can be checked that \eqref{eq:deflection-angle} gives $\varphi=0$ for $n(r)=1$ as expected. For a constant refractive index $n$ it holds that
$C=nd$. By inspecting \eqref{eq:deflection-angle} it follows immediately that a constant $n$ does not lead to any deflection. This is in contrast
to \cite{Betschart:2008yi} where for certain Lorentz-violating frameworks with {\em constant} Lorentz-violating coefficient it was
shown that there is a change in the deflection angle caused by Lorentz violation, indeed. However note that in the latter reference a Schwarzschild
black hole was considered whose line interval had not been cast into generally isotropic form, cf. \eqref{eq:line-interval}. A discussion of this difference
leading to more insight into Bouguer's formula is relegated to \appref{sec:light-deflection-schwarzschild}, since it is quite technical and probably
not of relevance for all readers.

\subsection{Phenomenology for the isotropic framework}

With the technique further developed, we are ready to carry out phenomenological calculations. The goal is to obtain predictions for the
change of the light deflection angle caused by particular Lorentz-violating frameworks. These predictions will be compared to experiment to
obtain sensitivities on controlling coefficients in the minimal SME photon sector. As the most important example light deflection
at the Sun will be discussed first. However light can be deflected at any other massive bodies such as planets.

First of all we intend to recapitulate the standard result. For vanishing Lorentz violation the deflection angle of \eqref{eq:deflection-angle} can
be computed analytically. Thereby the integral 2.266 of \cite{Gradshteyn:2007} is helpful:
\begin{equation}
\label{eq:gradshteyn-integral}
\int \frac{\mathrm{d}x}{x\sqrt{\alpha+\beta x+\gamma x^2}}=\frac{1}{\sqrt{-\alpha}}\arcsin\left(\frac{2\alpha+\beta x}{x\sqrt{\beta^2-4\alpha\gamma}}\right)\,,\quad \alpha<0\,,\quad \beta^2-4\alpha\gamma>0\,.
\end{equation}
For the Lorentz-invariant case we have
\begin{equation}
\alpha=-\frac{d}{R_S}\left(\frac{d}{R_S}+2\right)\,,\quad \beta=2\,,\quad \gamma=1\,,
\end{equation}
with the Schwarzschild radius $R_S=2GM/c^2$ of the massive body. Here $G$ is the gravitational constant, $M$ the mass of the body, and $c$
the speed of light. The conditions for $\alpha$, $\beta$, and $\gamma$ stated in \eqref{eq:gradshteyn-integral}
are fulfilled and the full analytical result for the deflection angle is given as follows:
\begin{equation}
\label{eq:standard-light-deflection}
\varphi=\frac{1+2\xi}{\sqrt{1+4\xi}}\left[\pi+2\arcsin\left(\frac{2\xi}{1+2\xi}\right)\right]-\pi=4\xi+\mathcal{O}(\xi^2)\,,\quad \xi=\frac{R_S}{2d}\,,
\end{equation}
where the latter is the first-order expansion in the dimensionless parameter $\xi\ll 1$. Now considering a light ray directly passing the
surface of the Sun (scraping incidence), $d$ is given by the radius $r_{\scriptscriptstyle{\bigodot}}$ of the Sun. Using the values of \tabref{tab:parameters-sun}
and multiplying the previous equation with $180\cdot 60^2/\pi$ leads to the well-known result $\varphi\approx 1.75''$, which lies within
few standard deviations from the mean value observed during the total eclipse in 1919 \cite{Dyson:1919,Eddington:1923}.
\begin{table}[b]
\centering
\begin{tabular}{ccc}
\toprule
Quantity & Unit & Value \\
\colrule
$G$ & $\mathrm{m^3/(kg\cdot s^2)}$ & $6.67384\cdot 10^{-11}$ \\
$M_{\astrosun}$ & kg & $1.98910\cdot 10^{30}$ \\ 
$M_{\jupiter}$ & kg & $1.89813\cdot 10^{27}$ \\ 
$M_{\saturn}$ & kg & $5.68319\cdot 10^{26}$ \\ 
$r_{\scriptscriptstyle{\astrosun}}$ & m & $6.95508\cdot 10^8$ \\ 
$r_{\scriptscriptstyle{\jupiter}}$ & m & $6.99110\cdot 10^7$ \\ 
$r_{\scriptscriptstyle{\saturn}}$ & m & $5.82320\cdot 10^7$ \\ 
\botrule
\end{tabular}
\caption{Gravitational constant, masses, and radii of the Sun (\astrosun), Jupiter (\jupiter), and Saturn (\saturn). For Jupiter and Saturn
the average of the pole and equatorial radii is used.}
\label{tab:parameters-sun}
\end{table}
\begin{figure}
\centering
\includegraphics{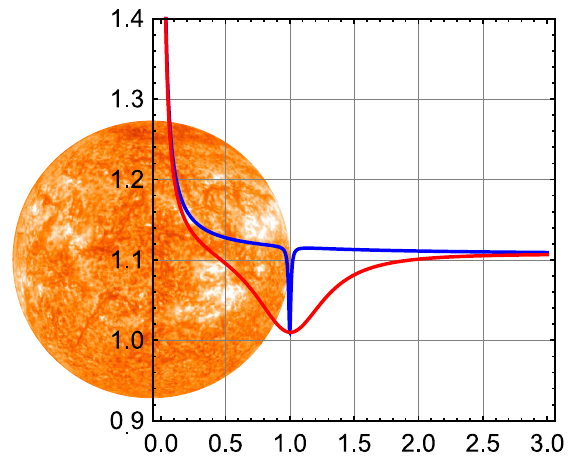}
\caption{Spacetime-position dependent refractive index (and controlling coefficient) as a function of the dimensionless parameter
$r/d$ where $d$ corresponds to the Sun radius $r_{\scriptscriptstyle{\bigodot}}$ in this example. (The picture of the Sun was taken
by SOHO -- EIT Consortium, ESA, NASA, cf.~\url{http://science.nasa.gov/science-news/science-at-nasa/2003/22apr_currentsheet}.)}
\label{fig:position-dependent-controlling-coefficient}
\end{figure}

Now the refractive index is modified due to Lorentz violation according to \eqref{eq:refractive-index-isotropic-gravitational}. Therefore
the isotropic Lorentz-violating coefficient $\widetilde{\kappa}_{\mathrm{tr}}$ is promoted to a spacetime-position dependent function
(cf.~\figref{fig:position-dependent-controlling-coefficient}).
It is assumed to only depend on the radial coordinate $r$ to keep the framework isotropic:
\begin{equation}
\widetilde{\kappa}_{\mathrm{tr}}\mapsto \widetilde{\kappa}_{\mathrm{tr}}(r)=\widetilde{\kappa}_{\mathrm{tr}}\left[1-f(r)\right]\,,
\end{equation}
with a function $f$ having special properties. The latter shall be constructed such that $1-f\geq 0$ for $r/d\geq 1$. This means that the sign of
$\widetilde{\kappa}_{\mathrm{tr}}(r)$ is fixed by the sign of the constant prefactor $\widetilde{\kappa}_{\mathrm{tr}}$. Furthermore
$\lim_{r\mapsto\infty} \widetilde{\kappa}_{\mathrm{tr}}(r)=\widetilde{\kappa}_{\mathrm{tr}}$, whereby in the asymptotically
flat region the position-dependent controlling coefficient is identified with the corresponding SME photon coefficient $\widetilde{\kappa}_{\mathrm{tr}}$
in Minkowski spacetime. The refractive index then reads as
\begin{equation}
\label{eq:refractive-index-modelling}
n(r)=\sqrt{\frac{1+\widetilde{\kappa}_{\mathrm{tr}}(r)}{1-\widetilde{\kappa}_{\mathrm{tr}}(r)}}\left(1+\frac{R_S}{r}\right)\,.
\end{equation}
From \eqref{eq:spherical-metric-isotropic-modmax} the coordinate velocity of light in this framework is given by
\begin{equation}
c=\frac{|\mathrm{d}\mathbf{r}|}{\mathrm{d}t}=\frac{1}{n(r)}\,,
\end{equation}
i.e., for $\widetilde{\kappa}_{\mathrm{tr}}>0$ it is reduced in comparison to the Lorentz-invariant case. The position dependence shall reflect
the properties of the gravitational background. The curvature radius $R_S$ is the physical scale of the background, i.e., it is reasonable to
associate it with $\widetilde{\kappa}_{\mathrm{tr}}(r)$ as well. Note that we are only interested in the behavior of the function outside of the
massive body, which means $r\geq d$. Generic functions with these properties are
\begin{subequations}
\label{eq:sample-functions}
\begin{align}
\label{eq:sample-functions-f}
f(r)&\equiv \left[1+a\left(\frac{r-d}{R_S}\right)^2\right]^{-1}\,, \\[2ex]
\label{eq:sample-functions-g}
g(r)&\equiv \frac{2\arctan\left\{a\left[1-(r-d)^2/R_S^2\right]\right\}+\pi}{2\arctan(a)+\pi}\,,
\end{align}
\end{subequations}
where $a\leq 1$ is a free, dimensionless parameter. Therefore for these particular sample functions it holds that $f(d)=1$
and $\lim_{r\mapsto\infty} f(r)=0$. Whatever the underlying theory for a possible violation of Lorentz invariance looks like, it is reasonable to assume
that the amount of Lorentz violation is influenced by a gravitational background field. Referring to a small-scale structure of spacetime where
simple models were shown to produce Lorentz-violating particle dispersion relations \cite{Klinkhamer:2003ec,Bernadotte:2006ya} the argument could be along the following lines. A
gravitational field has an energy density associated to it, cf.~\cite{Lynden-Bell:1985} for the case of spheres and black holes. Since a spacetime
foam is caused by energy fluctuations, an additional contribution of energy density associated to a gravitational field may have some influence
on it. This would render the effective controlling coefficients for Lorentz violation spacetime-position dependent. Hence for the isotropic framework
considered the refractive index directly at the surface of the Sun may have a dip for $\widetilde{\kappa}_{\mathrm{tr}}>0$ or a peak for
$\widetilde{\kappa}_{\mathrm{tr}}<0$ in its position dependence  (cf.~\figref{fig:position-dependent-controlling-coefficient}). As long as the
underlying description is not available, it is challenging to deliver a more rigorous argumentation. Hence a $\widetilde{\kappa}_{\mathrm{tr}}(r)$
including \eqref{eq:sample-functions} with the parameter $a$ controlling the width of the dip/peak must be interpreted as a phenomenological description
of such effects.

Now the modified deflection angle can be calculated in two different ways. The first is to compute the integral according to Bouguer's formula
of \eqref{eq:deflection-angle}. The second is to solve the eikonal equation directly. In \appref{sec:eikonal-equation-inhomogeneous-anisotropic}
the eikonal equation is brought into a form that is suitable for solving it. For a refractive index that has a radial dependence only,
\eqref{eq:eikonal-equation-main-result} results in
\begin{equation}
\label{eq:eikonal-equation-isotropic-numerical-basis}
0=(r^2+\dot{r}^2)r\frac{\partial n}{\partial r}+n(r^2+2\dot{r}^2-r\ddot{r})\,.
\end{equation}
Bouguer's formula is a first integral of the eikonal equation that follows from angular momentum conservation. Therefore using it allows us
to avoid the computation of one integral. Nevertheless as a cross check it is reasonable to carry out the computation with the two techniques.
Both the integral of \eqref{eq:deflection-angle} and the eikonal equation are challenging to be solved analytically for a refractive index that is
modified by Lorentz violation. Therefore we attempt to treat both cases numerically with \texttt{Mathematica}.

\begin{figure}
\centering
\subfloat[]{\label{fig:light-deflection-positive-lorentz-violation}\includegraphics{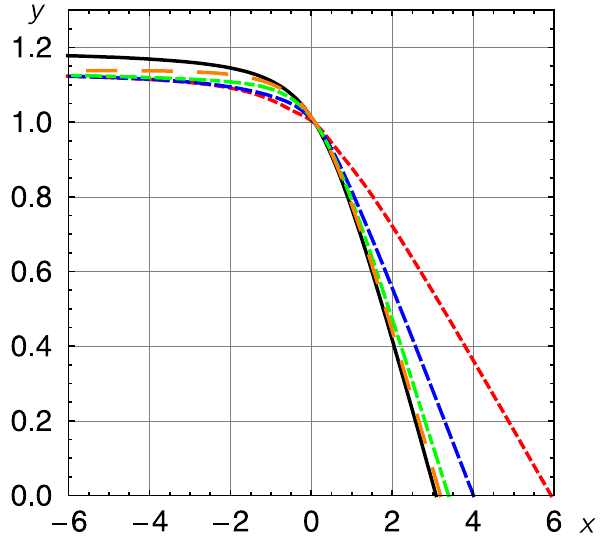}}\hspace{2cm}
\subfloat[]{\label{fig:light-deflection-negative-lorentz-violation}\includegraphics{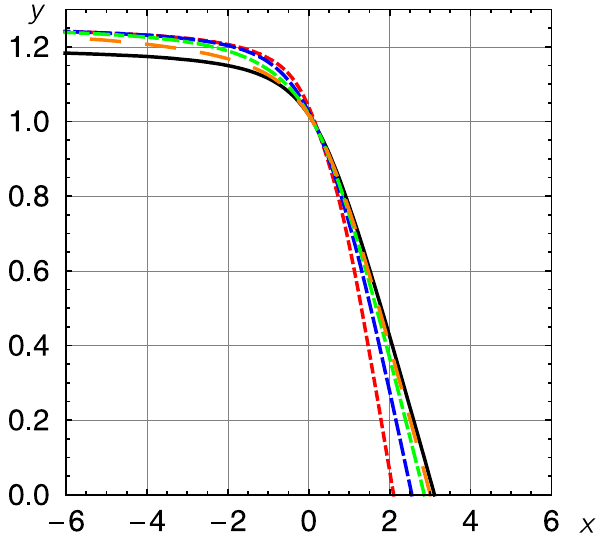}}
\caption{Solutions of the eikonal equation in the $x$-$y$-plane (in dimensions of $d$) for $\widetilde{\kappa}_{\mathrm{tr}}=1/20$
\protect\subref{fig:light-deflection-positive-lorentz-violation}
and for $\widetilde{\kappa}_{\mathrm{tr}}=-1/20$ \protect\subref{fig:light-deflection-negative-lorentz-violation}. The massive body resides
in the origin and the hypothetical value $d/R_S=5$ has been chosen. The black (plain) curve shows standard light deflection where for the
remaining $\{$red (dotted), blue (dashed), green (dashed-dotted), orange (dashed with large spaces)$\}$ curve $a=\{1,1/10,1/10^2,1/10^3\}$
has been taken successively.}
\label{fig:solution-eikonal-equation-isotropic-case}
\end{figure}

To gain some physical understanding, the eikonal equation is solved numerically for hypothetical values of $R_S$ and $\widetilde{\kappa}_{\mathrm{tr}}$
first. At the distance of minimal proximity $d$ the sample functions of \eqref{eq:sample-functions} vanish by construction. Therefore $n(d)=1+R_S/d$,
which is why $C=n(d)d$ and the impact parameter is given by
\begin{equation}
d_{\infty}=\frac{C}{n(r=\infty)}=\sqrt{\frac{1-\widetilde{\kappa}_{\mathrm{tr}}}{1+\widetilde{\kappa}_{\mathrm{tr}}}}\,(d+R_S)\,.
\end{equation}
For realistic situations, i.e., light bending at stars the Schwarzschild radius is much smaller than the distance of minimal proximity. Note that
for scraping incidence, $d$ corresponds to the radius of the star. Since bounds on the isotropic coefficient $\widetilde{\kappa}_{\mathrm{tr}}$
in flat, asymptotic spacetime are strict, it holds that $d\approx d_{\infty}$ to a good approximation. For the hypothetical values that we choose for
illustration purposes this is not necessarily warranted. Taking $d/R_S=5$ and $\widetilde{\kappa}_{\mathrm{tr}}=1/20$ we obtain the results
depicted in \figref{fig:light-deflection-positive-lorentz-violation}. The curves show the solutions of the eikonal equation where the refractive index has been modeled according to
\eqref{eq:refractive-index-modelling} using the sample function $f(r)$ of \eqref{eq:sample-functions-f} for different choices of the parameter
$a$.  Recall that the latter characterizes the width of the dip/peak in the refractive index directly at the surface of the massive body, which is
caused by Lorentz violation. Since for comparison all curves should meet at a single point, the impact parameters $d_{\infty}$ have to be
adapted properly, which is why they differ from each other.

The observation is that for increasing $a$ and $\widetilde{\kappa}_{\mathrm{tr}}>0$ the deflection angle is reduced. As long as the light
ray is far away from the massive body it experiences a refractive index that increases when the distance to the body decreases. This is
the standard behavior of the refractive index whose origin lies in nonvanishing Riemann curvature components. Upon approaching the
massive body the light ray suddenly experiences the dip where the refractive index becomes smaller for decreasing distance. The
ray then behaves contrary to the standard case and tends to be bent away from the body, which can be clearly seen in
\figref{fig:light-deflection-positive-lorentz-violation}. Note that for $\widetilde{\kappa}_{\mathrm{tr}}<0$ the dip in the refractive index turns
into a peak. Hence the behavior is opposite and the ray is bent towards the body even stronger, cf. \figref{fig:light-deflection-negative-lorentz-violation}.

From a technical point of view to solve the eikonal equation, proper initial conditions have to be considered. Since the ray is assumed to arrive
from an asymptotically flat region, the initial angle is $\phi_0=\pi$. In practice an angle lying close to $\pi$ must be chosen where the
direction of the ray initially is assumed to point along the positive horizontal axis. We express the solution of the eikonal equation as $r(\phi)=d\xi(\phi)$
with the dimensionless function $\xi(\phi)$. The initial conditions are then fixed to be $\xi(\phi_0)=\Delta$ and $\xi'(\phi_0)=-\cot(\phi_0)\Delta$
where $\phi_0=\pi-\arcsin(d_{\infty}/\Delta)$. Here $\Delta$ is a length scale with the property $\Delta \gg d$, which is tuned to increase the
precision of the numerical result. Theoretically $\Delta$ should approach infinity, which is not a possible value to choose in practice, though.
Setting the final angle in the numerical integration to $\phi_1\leq 0$ leads to numerical instabilities, which is why $\phi_1$ is taken to be slightly
larger than zero. This is supposed to be sufficient for small bending angles that appear in realistic scenarios. It is reasonable to set both the working
precision to a large number and the maximum number of steps to infinity.

There are at least two space-based missions available that could test gravity based on light deflection. Two of the most promising ones are GAIA
and LATOR. In what follows we will discuss the perspective of these missions in obtaining constraints on Lorentz violation in the (isotropic) photon
sector by performing measurements of light deflection at massive bodies. Thereby the theoretical tools developed so far will be of great use.

\subsubsection{Sensitivity of GAIA}

GAIA\footnote{The acronym originally meant ``Global Astrometric Interferometer for Astrophysics.'' Although the foreseen measurement technique
was changed upon construction of the apparatus, the acronym was kept.} \cite{Perryman:2001sp} is a space probe
that was launched in December 2013 by ESA. The mission goal is to perform measurements of positions and radial velocities of about 1\%
of the galactic stellar population, which shall generate a three-dimensional map of our galaxy. This is supposed to give information on the
galactic history, dark matter as well as extra-solar planetary systems. GAIA can measure angles with a sensitivity of around \unit[10]{\textmu arcs},
which is why it can test deflection of light at massive bodies to a high precision. However the mission parameters do not allow light to be
measured grazing the surface of the Sun. Such measurements will be possible for Jupiter and Saturn only (see
Table III in \cite{Perryman:2001sp}).

Now we intend to perform phenomenology of light bending in an isotropic Lorentz-violating framework based on the possibilities of GAIA.
Thereby sample functions are taken according to \eqref{eq:sample-functions} with different values for the parameter $a=1/10^i$ and the
range $i=0\dots 15$. Choosing a particular controlling coefficient $\widetilde{\kappa}_{\mathrm{tr}}$, the deflection angle of light in the
vicinity of Jupiter is computed with two methods. The first uses the formula of Bouguer, \eqref{eq:deflection-angle}. The second
solves the eikonal equation (\ref{eq:eikonal-equation-isotropic-numerical-basis}) numerically in analogy to what was described above.
The bending angle is then computed via the scalar product of the initial and final normalized tangent vectors. This gives an excellent
cross check for the results, since the two methods are independent from each other.
\begin{table}[b]
\centering
\begin{tabular}{cc|c|c}
\toprule
 & & $f(r)$ & $g(r)$ \\
$-\log_{10}(a)$ & $-\log_{10}(\widetilde{\kappa}_{\mathrm{tr}})$ & $\varphi_{\tiny{\jupiter}}^{*}-\varphi_{\tiny{\jupiter}}$ [10\,\textmu{}arcs] & $\varphi_{\tiny{\jupiter}}^{*}-\varphi_{\tiny{\jupiter}}$ [10\,\textmu{}arcs] \\
\colrule
0 & 14 & 1.61 & 1.40 \\
$1\dots 4$ & 13 & 9.02; 5.07; 2.85; 1.60 & 8.80; 4.97; 2.80; 1.57 \\
$5\dots 8$ & 12 & 9.02; 5.07; 2.85; 1.60 & 8.84; 4.97; 2.80; 1.57 \\
$9\dots 12$ & 11 & 9.02; 5.07; 2.84; 1.59 & 8.84; 4.97; 2.79; 1.56 \\
$13\dots 15$ & 10 & 8.76; 4.67; 2.32 & 8.58; 4.57; 2.26 \\
\botrule
\end{tabular}
\caption{Numerical results for modified light deflection angles at Jupiter (scraping incidence) in the isotropic framework. The first column states (ranges of) the parameter $a$ used in
\eqref{eq:sample-functions}. The second column gives the value of the isotropic coefficient. In the third and fourth columns differences between the standard
deflection angle $\varphi_{\tiny{\jupiter}}^{*}\approx \unit[16.8]{marcs}$ and the modified angle $\varphi_{\tiny{\jupiter}}$
are shown in suitable units. For the third column the modeling function $f(r)$ of \eqref{eq:sample-functions-f} is used and for the fourth column we employ
$g(r)$ of \eqref{eq:sample-functions-g}. Each number is understood to be associated to the proper parameter $a$ in the first column where both lists are in
order. The results for a negative $\widetilde{\kappa}_{\mathrm{tr}}$ are equal to the corresponding numbers for a positive $\widetilde{\kappa}_{\mathrm{tr}}$
where the global sign of $\varphi_{\tiny{\jupiter}}^{*}-\varphi_{\tiny{\jupiter}}$ is reversed. The deviation in the absolute values shows up in the third digit
for almost all differences.}
\label{tab:results-for-gaia}
\end{table}

The bending angle obtained is then compared to the standard result. This procedure is repeated for a decreasing isotropic coefficient
$\widetilde{\kappa}_{\mathrm{tr}}$ until the difference between the modified and the standard result approximately matches the precision that
GAIA can measure angles with. This sets the sensitivity of the experiment with respect to $\widetilde{\kappa}_{\mathrm{tr}}$ in a curved
background. However it is challenging to compute the integral or to solve the eikonal equation with a high precision. We use the difference
of the results obtained from the two methods as a measure for how meaningful they are. For a conservative estimate of the sensitivity one
should keep results only if this theoretical uncertainty is much smaller than the difference between the modified and the standard bending angle.

First of all for $\widetilde{\kappa}_{\mathrm{tr}}>0$ the difference between the standard bending angle $\varphi_{\tiny{\jupiter}}^{*}$ and the modified
bending angle is positive, which shows that the bending angle is reduced by a positive Lorentz-violating coefficient $\widetilde{\kappa}_{\mathrm{tr}}$
(see the third column of \tabref{tab:results-for-gaia}). For $\widetilde{\kappa}_{\mathrm{tr}}<0$ the behavior is vice versa and the absolute numbers
mainly deviate in the third digit, which is why they are omitted in the table. We stated all differences $\varphi_{\tiny{\jupiter}}^{*}-\varphi_{\tiny{\jupiter}}$ that are
larger than and lie in the vicinity of the experimental precision of GAIA, i.e., \unit[10]{\textmu{}arcs}. Such modifications can be expected to
be detectable by this mission. From the results it becomes
clear that the sensitivity of the isotropic coefficient reduces when the width of the dip, which is controlled by the parameter $a$, decreases.
If the width lies in the order of magnitude of Jupiter's radius the sensitivity for $|\widetilde{\kappa}_{\mathrm{tr}}|$ is $10^{-14}$. In case the width
lies 15 orders of magnitude below that the sensitivity of $|\widetilde{\kappa}_{\mathrm{tr}}|$ is still $10^{-10}$. Hence the sensitivity does not decrease
as quickly as the parameter $a$. The numbers are meaningful, since the difference of the results obtained with Bouguer's formula and by solving the
eikonal equation directly is around \unit[$4\times 10^{-10}$]{\textmu{}arcs} at the maximum. The latter is interpreted as the theoretical uncertainty and it
is much smaller than $|\varphi_{\tiny{\jupiter}}^{*}-\varphi_{\tiny{\jupiter}}|$.

Obtaining the modified deflection angles for Saturn works completely analogously. The sensitivity on $\widetilde{\kappa}_{\mathrm{tr}}$ lies in the
same order of magnitude. The only difference is that even smaller $a$ could be probed based on a modeling according to \eqref{eq:sample-functions}.
The reason is that
\begin{equation}
\frac{d_{\tiny\saturn}}{R_{S,\tiny\saturn}}\approx 2.78\frac{d_{\tiny\jupiter}}{R_{S,\tiny\jupiter}}\,,
\end{equation}
whereby the additional dimensionless factor increases the contribution of $a$.

\subsubsection{Sensitivity of LATOR}

LATOR (Laser Astrometric Test of Relativity) \cite{Turyshev:2004ga,Turyshev:2009zz} is a mission that is being planned by a collaboration of NASA and ESA. It is a Michelson-Morley-type
experiment that shall perform curvature measurements in our solar system to determine the Eddington post-Newtonian parameter $\gamma$
with a precision of 1 part in $10^8$. It is considered to be a test mission for General Relativity and it is supposed to detect the frame-dragging effect
and to determine the solar quadrupole moment. The primary objective will be to measure the gravitational deflection of light by the Sun to an
accuracy of \unit[0.02]{\textmu arcs}. Such an astounding precision shall be made possible by an improved laser ranging and a long-baseline
optical interferometry system.
\begin{table}[b]
\centering
\begin{tabular}{cc|c|c}
\toprule
 & & $f(r)$ & $g(r)$ \\
$-\log_{10}(a)$ & $-\log_{10}(\widetilde{\kappa}_{\mathrm{tr}})$ & $\varphi_{\tiny{\astrosun}}^{*}-\varphi_{\tiny{\astrosun}}$ [$10^{-2}$\,\textmu{}arcs] & $\varphi_{\tiny{\astrosun}}^{*}-\varphi_{\tiny{\astrosun}}$ [$10^{-2}$\,\textmu{}arcs] \\
\colrule
0 & 16 & 1.57 & --- \\
$(0)1\dots 4$ & 15 & 8.84; 4.97; 2.80; 1.57 & 13.6; 8.58; 4.84; 2.72; 1.53 \\
$5\dots 8$ & 14 & 8.84; 4.97; 2.79; 1.56 & 8.61; 4.84; 2.72; 1.52 \\
$9\dots 11$ & 13 & 8.57; 4.56; 2.26 & 8.35; 4.43; 2.18 \\
$12\dots 13$ & 12 & 9.99; 3.91 & 9.49; 3.64 \\
$14\dots 15$ & 11 & 13.9; 4.64 & 12.7; 4.21 \\
\botrule
\end{tabular}
\caption{Numerical results for modified light deflection at the Sun in the isotropic framework (see \tabref{tab:results-for-gaia} for the meaning
of each column). The standard deflection angle for scraping incidence at the Sun is $\varphi_{\tiny{\astrosun}}^{*}\approx \unit[1.75]{arcs}$ (see
\eqref{eq:standard-light-deflection} and the subsequent paragraph). In the second line $a=0$ is associated to the first value in the fourth column.}
\label{tab:results-for-lator}
\end{table}

We carry out phenomenology as we did before by choosing different parameters $a$ for the sample functions of \eqref{eq:sample-functions}. The calculations are completely
analogous to before where the only difference is that they are carried out for the Sun using the appropriate parameters of \tabref{tab:parameters-sun}.
The essential numerical results are stated in \tabref{tab:results-for-lator}. The bending angle behaves similarly to before, i.e., it is reduced
for $\widetilde{\kappa}_{\mathrm{tr}}>0$ and it increases for $\widetilde{\kappa}_{\mathrm{tr}}<0$. The differences $\varphi_{\tiny{\astrosun}}^{*}-\varphi_{\tiny{\astrosun}}$
are listed that lie in the vicinity of the experimental precision expected for LATOR, i.e., \unit[0.02]{\textmu{}arcs}. If the width of the dip/peak
in the refractive index of the Lorentz-violating vacuum lies in the order of magnitude of Sun's radius the sensitivity for the isotropic coefficient
$|\widetilde{\kappa}_{\mathrm{tr}}|$ is $10^{-16}$. The lowest sensitivity in case of a very narrow dip/peak is $10^{-11}$. Comparing the
results determined from Bouguer's formula to the results from the numerical solution of the eikonal equation reveals differences of ca.
\unit[$2\times 10^{-7}$]{\textmu{}arcs}. Therefore the theoretical uncertainty is still much smaller than $|\varphi_{\tiny{\astrosun}}^{*}-\varphi_{\tiny{\astrosun}}|$.
Note that for the model function $g(r)$ the modification of the deflection angle for $|\widetilde{\kappa}_{\mathrm{tr}}|=10^{-16}$ is smaller than
$1.50\times 10^{-2}$\,\textmu{}arcs. Therefore assuming this model function, the sensitivity of LATOR will not be sufficient to detect a
$|\widetilde{\kappa}_{\mathrm{tr}}|$ lying in the order of magnitude of $10^{-16}$.

\subsubsection{Discussion}

According to the current (2015) version of the data tables \cite{Kostelecky:2008ts} the strictest lower bounds on $\widetilde{\kappa}_{\mathrm{tr}}$ lie
in the order of magnitude of $-10^{-16}$ where the best upper bounds are around $10^{-20}$. The isotropic coefficient of modified Maxwell theory
is challenging to be constrained in laboratory experiments, which is why these bounds are related to ultra-high energy cosmic rays. With the
precision of LATOR there would be a space-based experiment performed under controlled conditions that could have a sensitivity comparable
to the best current constraints on a negative $\widetilde{\kappa}_{\mathrm{tr}}$. This is astonishing taking into account that the precision of a
man-made experiment may match the sensitivity reached by the most energetic particles propagating through interstellar space for distances of
many lightyears. It illustrates the versatility of the technique presented to constrain Lorentz violation in the photon sector by precise measurements
of light bending at massive bodies. Note that the sensitivity does not largely depend on the model function used. This independence could be checked
for further model functions, which can be regarded as an interesting future project.

\subsection{Anisotropic (nonbirefringent) case}

\begin{figure}
\centering
\includegraphics{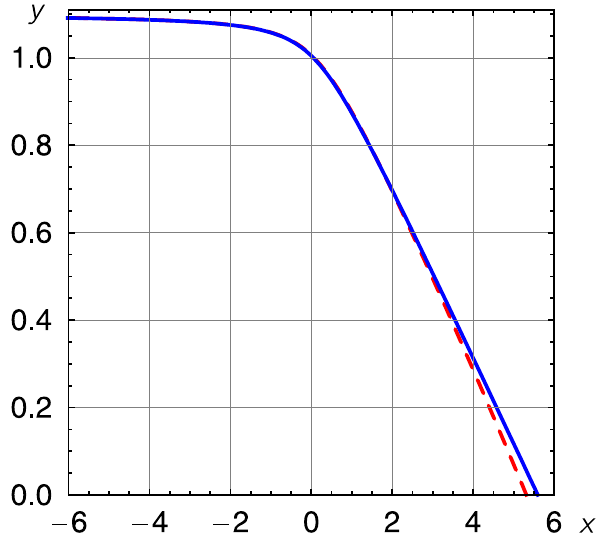}
\caption{Solution of the eikonal equation in the $x$-$y$-plane (in dimensions of $d$) with $d/R_S=5$. The blue (plain) curve shows the solution for the
Lorentz-invariant case, whereas the red (dashed) curve depicts the solution for the anisotropic case with $\widetilde{\kappa}_{e-}^{11}=0.65$.
The massive body resides in the coordinate center.}
\label{fig:solution-eikonal-both-standard-and-anisotropic}
\end{figure}

The anisotropic case of modified Maxwell theory exhibiting a single modified dispersion relation was discussed in
\secref{sec:anisotropic-nonbirefringent-case}. This particular case is characterized by a preferred spacelike direction (chosen to point along the
positive $z$-axis) and one controlling coefficient. The refractive index was found in \eqref{eq:refraction-index-anisotropic} and it was expressed
in terms of the angle $\vartheta$ enclosed between the propagation direction and the preferred axis. The possible trajectory of a light ray is
parameterized by $\mathbf{r}(\phi)=r(\phi)\widehat{\mathbf{e}}_{\phi}$ such as for the isotropic case. The angle $\vartheta$ in the refractive
index is then given by the scalar product of the tangent vector $\mathbf{t}$ and the preferred direction $\boldsymbol{\zeta}$ where it is sufficient
to work in two spatial dimensions:
\begin{equation}
\label{eq:anisotropic-angle}
\cos\vartheta=\frac{\mathbf{t}\cdot \boldsymbol{\zeta}}{|\mathbf{t}|}=\frac{r(\phi)\cos\phi+\dot{r}(\phi)\sin\phi}{\sqrt{r(\phi)^2+\dot{r}(\phi)^2}}\,.
\end{equation}
Note that for the anisotropic case angular momentum is not conserved and Bouguer's formula loses its meaning. Hence there does not seem to
be an alternative to solving the eikonal equation directly, which is carried out numerically for hypothetical values of $R_S$ and the controlling
coefficient $\widetilde{\kappa}_{e-}^{11}$. The results are shown in \figref{fig:solution-eikonal-both-standard-and-anisotropic}. In contrast
to the isotropic case, cf.~\figref{fig:solution-eikonal-equation-isotropic-case}, where the trajectory is not modified for a spacetime position independent
$\widetilde{\kappa}_{\mathrm{tr}}$ this is not the case here. For the anisotropic sector the shape of the trajectory gets distorted where the final
impact parameter decreases for $\widetilde{\kappa}_{e-}^{11}>0$. Physically this means that the ray loses angular momentum. An
interesting future research project would be to perform a similar kind of phenomenological analysis as we did for the isotropic case.

\section{Modified energy-momentum conservation}
\label{sec:modified-energy-momentum-conservation}

The phenomenology in the previous section was carried out in an explicitly Lorentz-violating framework, which is known to cause tensions in a
gravitational background \cite{Kostelecky:2003fs}. The purpose of the current section is to investigate where exactly these problems occur in our
classical description and how they can be interpreted from the point of view of an inhomogeneous medium. Therefore the energy-momentum
tensor and its conservation law will be derived for the isotropic case. The (Belinfante-Rosenfeld)
energy-momentum tensor follows from varying the corresponding Lagrangian with respect to the metric. The Finsler structure $F(\mathbf{u})|_{\circledcirc}^+$
of \eqref{eq:finsler-structure-isotropic} is the equivalent to a Lagrangian, since it appears as the integrand of the path length functional that
is stationary for the trajectory travelled by the light ray. Instating an auxiliary metric tensor $\psi_{\mu\nu}$ leads to the following result:
\begin{equation}
\label{eq:lagrangian-classical-light-ray-isotropic}
F=n|\mathbf{u}|=\sqrt{\frac{1+\widetilde{\kappa}_{\mathrm{tr}}\,\psi_{\mu\nu}\xi^{\mu}\xi^{\nu}}{1-\widetilde{\kappa}_{\mathrm{tr}}\,\psi_{\rho\sigma}\xi^{\rho}\xi^{\sigma}}}\sqrt{-\psi_{ij}u^iu^j}\,.
\end{equation}
Note that in Minkowski spacetime it holds that $\eta_{\mu\nu}\xi^{\mu}\xi^{\nu}=\xi^2=1$ and $-\eta_{ij}u^iu^j=\mathbf{u}^2$ where the
minus sign in the latter term is due to the signature of the metric chosen. Variation has to be carried out for all independent degrees
of freedom. A useful formula is
\begin{align}
\delta(A_{\mu}A^{\mu})&=\delta(\psi_{\mu\nu}A^{\mu}A^{\nu})=\psi_{\mu\nu}\delta A^{\mu}A^{\nu}+\psi_{\mu\nu}A^{\mu}\delta A^{\nu}+\delta \psi_{\mu\nu} A^{\mu}A^{\nu} \notag \\
&=2A_{\nu}\delta A^{\nu}+\delta \psi_{\mu\nu}A^{\mu}A^{\nu}\,,
\end{align}
which states the variation of a scalar product of fields. Employing this rule, the variation of $F$ can then be computed as follows:
\begin{align}
\delta F&=(\delta n)\sqrt{-\psi_{ij}u^iu^j}+n\;\!\delta\left(\sqrt{-\psi_{ij}u^iu^j}\right) \notag \displaybreak[0]\\
&=\frac{\widetilde{\kappa}_{\mathrm{tr}}}{n(1-\widetilde{\kappa}_{\mathrm{tr}})^2}\delta\left(\psi_{\mu\nu}\xi^{\mu}\xi^{\nu}\right)\sqrt{-\psi_{ij}u^iu^j}+\frac{n}{2\sqrt{-\psi_{ij}u^iu^j}}\delta\left(-\psi_{ij}u^iu^j\right) \notag \displaybreak[0]\\
&=\frac{\widetilde{\kappa}_{\mathrm{tr}}}{n(1-\widetilde{\kappa}_{\mathrm{tr}})^2}\sqrt{-\psi_{ij}u^iu^j}\left(2\xi_{\nu}\delta\xi^{\nu}+\delta \psi_{\mu\nu}\xi^{\mu}\xi^{\nu}\right) \notag \displaybreak[0]\\
&\phantom{{}={}}-\frac{n}{2\sqrt{-\psi_{ij}u^iu^j}}\left(2u_j\delta u^j+\delta \psi_{ij}u^iu^j\right)\,.
\end{align}
Now everything is available to obtain the energy-momentum tensor from $\delta F$ by considering all terms comprising a variation
of the metric. An additional prefactor containing the metric has to be taken into account in the definition. However we are interested
in the covariant conservation law of $T^{\mu\nu}$ for Minkowski spacetime, i.e., for a spacetime-position dependent refractive index
without an additional gravitational field. In this case $\psi_{\mu\nu}=\eta_{\mu\nu}$ whereby
\begin{align}
T^{\mu\nu}&\equiv\left.\frac{2}{\sqrt{|\psi|}}\frac{\delta (\sqrt{|\psi|}F)}{\delta \psi_{\mu\nu}}\right|_{\psi_{\mu\nu}=\eta_{\mu\nu}}=\frac{2\widetilde{\kappa}_{\mathrm{tr}}}{n(1-\widetilde{\kappa}_{\mathrm{tr}})^2}\sqrt{u_iu^i}\xi^{\mu}\xi^{\nu}-\frac{n}{\sqrt{u_iu^i}}\widetilde{u}^{\mu}\widetilde{u}^{\nu} \notag \\
&=n\sqrt{u_iu^i}\left[\frac{1}{2}\left(n^2-\frac{1}{n^2}\right)\xi^{\mu}\xi^{\nu}-\frac{\widetilde{u}^{\mu}\widetilde{u}^{\nu}}{u_iu^i}\right]\,.
\end{align}
Here $\psi\equiv\det(\psi_{\mu\nu})$ and $(\widetilde{u}^{\mu})\equiv (0,\mathbf{u})^T$, i.e., $\widetilde{u}^{\mu}$ involves the spatial velocity and its zeroth component
vanishes. Upon inspection of the latter result we see that the 00-component of $T^{\mu\nu}$ is made up by the preferred timelike
spacetime direction $\xi^{\mu}$ and it vanishes for $n=1$, i.e., in a Lorentz-invariant vacuum. The spatial part solely comprises
products of three-velocity components and the mixed components vanish. Now the partial derivative of the energy-momentum tensor
in Minkowski spacetime leads to:
\begin{align}
\label{eq:conservation-energy-momentum}
\partial_{\mu}T^{\mu\nu}&=(\partial_{\mu}n)\sqrt{\mathbf{u}^2}\left[\frac{1}{2}\left(n^2-\frac{1}{n^2}\right)\xi^{\mu}\xi^{\nu}-\frac{\widetilde{u}^{\mu}\widetilde{u}^{\nu}}{\mathbf{u}^2}\right]+n\sqrt{\mathbf{u}^2}\partial_{\mu}\left[\frac{1}{2}\left(n^2-\frac{1}{n^2}\right)\right]\xi^{\mu}\xi^{\nu} \notag \\
&=\frac{T^{\mu\nu}}{n}\partial_{\mu}n+\left(n^2+\frac{1}{n^2}\right)\sqrt{\mathbf{u}^2}\,\xi^{\mu}\xi^{\nu}(\partial_{\mu}n) \notag \\
&=\sqrt{\mathbf{u}^2}\left[\frac{1}{2}\left(3n^2+\frac{1}{n^2}\right)\xi^{\mu}\xi^{\nu}-\frac{\widetilde{u}^{\mu}\widetilde{u}^{\nu}}{\mathbf{u}^2}\right](\partial_{\mu}n)\,.
\end{align}
An interesting observation is that the timelike contribution can be expressed in terms of the metric $\widetilde{g}_{\mu\nu}$ appearing
in \eqref{eq:spherical-metric-isotropic-modmax}:
\begin{subequations}
\begin{align}
(\widetilde{g}^2)^{\mu}_{\phantom{\mu}\mu}&=3n^2+\frac{1}{n^2}\,, \\[2ex]
\label{eq:four-dimensional-isotropic-metric}
\widetilde{g}_{\mu\nu}(r)&\equiv \mathrm{diag}\left(\frac{1}{n(r)},-n(r),-n(r),-n(r)\right)_{\mu\nu}\,.
\end{align}
\end{subequations}
Note that $\widetilde{g}_{\mu\nu}$ is not associated to a gravity field but only to a nonconstant refractive index. The result obtained in
\eqref{eq:conservation-energy-momentum} describes the conservation of energy and momentum of a light ray. Its properties are in order. First, it vanishes for a constant
refractive index, i.e., energy and momentum of the ray are conserved in a homogeneous medium, in the Lorentz-invariant vacuum, and
a Lorentz-violating vacuum with a constant controlling coefficient. Second, in an inhomogeneous medium or in a Lorentz-violating
vacuum with spacetime-dependent controlling coefficient the energy-momentum tensor is not conserved, since the partial derivative
of the refractive index does not vanish in this case. As long as the refractive index is not time-dependent, $\partial_0n=0$, which is why
$\partial_{\mu}T^{\mu 0}=0$. Since the case under consideration is isotropic, the controlling coefficient and the refractive index,
respectively, can only depend on the radial coordinate: $\widetilde{\kappa}_{\mathrm{tr}}=\widetilde{\kappa}_{\mathrm{tr}}(r)$, $n=n(r)$.
Hence $\partial_{\mu}n$ has a nonvanishing component along the radial basis vector only, i.e., $\partial_rn\neq 0$ and
$\partial_{\theta}n=\partial_{\phi}n=0$. Decomposing the spatial velocity into a radial part $u^r$ and transverse components $u^{\theta}$,
$u^{\phi}$,
\begin{equation}
\mathbf{u}=u^r\mathbf{e}_r+u^{\theta}\mathbf{e}_{\theta}+u^{\phi}\mathbf{e}_{\phi}\,,
\end{equation}
the spatial part of the conservation law reads as
\begin{equation}
\label{eq:modified-energy-momentum-conservation-isotropic}
\partial_{\mu}T^{\mu i}=-\frac{1}{\sqrt{\mathbf{u}^2}}(\boldsymbol{\nabla}n\cdot\mathbf{u})u^i=-\sqrt{\mathbf{u}^2}(\boldsymbol{\nabla}n\cdot\widehat{\mathbf{u}})\widehat{u}^i
=-\sqrt{\mathbf{u}^2}(\partial_rn)\widehat{u}^r\widehat{u}^i\,,\quad \widehat{\mathbf{u}}=\frac{\mathbf{u}}{|\mathbf{u}|}\,.
\end{equation}
Several observations can be made upon inspecting the result. For a constant refractive index the right-hand side of the latter equation is
zero, which means that the spatial part of the energy-momentum conservation law is valid as well in this case. For $\partial_rn\neq 0$ it
even holds when the radial velocity component vanishes: $u^r=0$. This is a special situation that can occur for a light ray in an inhomogeneous,
isotropic medium whose refractive index has a particular $r$-dependence and when the ray is emitted tangentially to a circle with its center
lying in the coordinate origin (cf.~\cite{Evans:1985} for a beautiful paper on geometric-ray optics and its implications for certain optical
systems). The trajectory of the ray is then a circle where the refractive index is constant. The magnitude
of the three-momentum vector does not change, but only its direction. So momentum is not exchanged between the light ray and the medium,
because any momentum transfer would change the magnitude of the momentum vector. For any other case with nonzero $\partial_rn$
momentum has to be exchanged, which is why $T^{\mu\nu}$ of the ray
cannot be conserved. The net term obtained above points in the direction $\widehat{\mathbf{u}}$ of the ray at the point considered. However
the total energy-momentum tensor with $T^{\mu\nu}_{\mathrm{med}}$ of the medium included is expected to be conserved, because any
momentum change of the light ray will cause a momentum change of the medium itself.

In General Relativity local diffeomorphism invariance is tightly linked to energy-momentum conservation. In \cite{Kostelecky:2003fs} it was
shown that explicit Lorentz violation in gravity leads to a loss of diffeomorphism invariance, which then causes the energy-momentum
tensor to be no longer covariantly conserved. Note that in the latter reference the energy-momentum tensor $T_e^{\mu\nu}$ is
considered that follows by varying the Lagrangian with respect to the vierbein instead of with respect to the metric tensor. It is different
from the Belinfante-Rosenfeld energy-momentum tensor considered here even in case there is no Lorentz violation \cite{Belinfante:1940}. The covariant derivative of
$T_e^{\mu\nu}$ in \cite{Kostelecky:2003fs} involves the covariant derivative of the Lorentz-violating controlling coefficients, i.e., a
term of the structure $J^xD_{\nu}k_x$. Here $k_x$ is a generic controlling coefficient with a particular Lorentz index structure $x$
contracted with an appropriate operator $J^x$. For general curved manifolds there is no spacetime-position dependent function
satisfying $D_{\nu}k_x=0$, but only for parallelizable manifolds such as the circle $S^1$ or the two-torus $T^2=S^1\times S^1$.
In four dimensions such manifolds are rare, though, and they do not seem to be of particular interest in the context of General Relativity.

Note that the conservation law without gravitational fields is given by $\partial_{\mu}(\Theta_c)^{\mu\nu}=J^x\partial^{\nu}k_x$ with
the canonical energy-momentum tensor $(\Theta_c)^{\mu\nu}$ \cite{Kostelecky:2003fs}. Therefore if the controlling coefficient $k_x$ is dependent on spacetime
position the conservation law is modified even in flat spacetime. From a physical perspective this is not surprising, since such a
controlling coefficient implies that the vacuum behaves like an effective, inhomogeneous medium. In general the magnitude of the
three-momentum of a light ray is not conserved as was argued above. Therefore momentum has to be exchanged between the ray
and the medium. When considering explicit Lorentz violation the effective medium is considered to be nondynamical, which is why it
can neither absorb nor deliver momentum to the light ray.

Interestingly the situation is different when spontaneous violations of diffeomorphism invariance and local Lorentz symmetry are considered.
In these cases the ground state violates these symmetries dynamically by an emergent vacuum expectation value of a vector or tensor
field in a potential
\cite{Kostelecky:2003fs,Kostelecky:1988zi,Kostelecky:1989jp,Kostelecky:1989jw,Bailey:2006fd,Bluhm:2008yt,Hernaski:2014jsa,Bluhm:2014oua}.
Such models have in common that they involve massless (Nambu-Goldstone) modes where the latter appear when any global,
continuous symmetry is broken spontaneously. When the symmetry is local there can be an additional Higgs-type mechanism
absorbing the massless modes to produce massive gauge fields.
Since for spontaneous Lorentz violation the dynamics of the Lorentz-violating background field is taken into account, the energy-momentum
conservation law is restored in these theories. In the corresponding equation there is no contribution $J^x\partial^{\nu}k_x$.

From the perspective of an inhomogeneous medium translational and rotational symmetry are violated spontaneously by the atomic lattice.
The Nambu-Goldstone
modes (gapless excitations) linked to the spontaneous violation of these symmetries are the two transverse and the longitudinal types of
phonons.\footnote{The number of spontaneously broken translational and rotational generators is six. However as they are not
independent from each other, the total number of Nambu-Goldstone bosons is reduced by these constraints to be three. Relations between
broken symmetries and Nambu-Goldstone bosons in certain nonrelativistic systems such as crystals, ferromagnets, and superfluids are
nicely described in \cite{Watanabe:2013iia}.} Since the medium is now dynamical, it can absorb momentum from the ray upon producing
phonons. Hence the conservation law for the light ray remains valid in this case.

\section{Properties of the isotropic Finsler space}
\label{sec:properties-isotropic-finsler}
\setcounter{equation}{0}

In the previous section the modified conservation law for the Belinfante-Rosenfeld energy-momentum tensor for a light ray in an
isotropic, inhomogeneous medium was considered and discussed. It was found that the energy-momentum tensor is not conserved,
since the nontrivial medium is nondynamical corresponding to a background violating Lorentz symmetry explicitly. This issue persists
even in Minkowski spacetime, since in inhomogeneous media light rays behave similarly to when they propagate in gravitational
backgrounds. The most prominent example for a common effect is light bending.

The conservation law of the energy-momentum tensor for a medium with spherically symmetric refractive index,
\eqref{eq:modified-energy-momentum-conservation-isotropic}, involves both the first derivative of the refractive index and the
propagation direction of the ray at a given point. Recall that in gravitational theories with explicit Lorentz violation nonconservation
of the energy-momentum tensor clashes with the Bianchi identities of Riemannian geometry \cite{Kostelecky:2003fs}. Although we
work with an effective Lorentz-violating theory for light rays based on a nontrivial refractive index, this issue can be encountered here
as well.

The purpose of the current section is to figure out whether explicit (isotropic) Lorentz violation can be considered in a weak gravity
field in the framework of Finsler geometry such that no inconsistencies arise. As a basis we use the spacetime metric of
\eqref{eq:spherical-metric-isotropic-modmax}, which was shown to be closely linked to the isotropic case. The properties of the
latter metric shall be studied from a Finslerian point of view where we use the conventions of \cite{Bao:2000} for all geometrical
quantities. The latter are treated based on the indefinite signature of the metric in \eqref{eq:spherical-metric-isotropic-modmax}.
As a starting point an appropriate Finsler structure has to be constructed whose derived metric should correspond to
\eqref{eq:spherical-metric-isotropic-modmax}. This works for the following choice:
\begin{equation}
\label{eq:finsler-structure-isotropic-refractive-index}
F(r,\mathbf{y})=\sqrt{\mathbf{y}^2}\,,\quad \mathbf{y}^2=\frac{1}{n}(y^t)^2-n(y^r)^2-n\left[(y^{\theta})^2+(y^{\phi})^2\sin^2\theta\right]r^2\,,
\end{equation}
where the refractive index solely has a radial dependence. In what follows we write $n(r)=n$ for brevity, i.e., the argument of the
refractive index will be omitted. The vector $\mathbf{y}\in TM$ is expressed in spherical polar coordinates as
$\mathbf{y}=y^t\mathbf{e}_t+y^r\mathbf{e}_r+y^{\theta}\mathbf{e}_{\theta}+y^{\phi}\mathbf{e}_{\phi}$ with suitable basis vectors.
The spatial part of the Finsler structure is written in terms of spherical polar coordinates. The corresponding Finsler metric is
then computed according to the usual definition and it corresponds to the result of \eqref{eq:spherical-metric-isotropic-modmax}
(with the spatial part transformed to spherical coordinates):
\begin{subequations}
\begin{equation}
\label{eq:finsler-metric-isotropic}
g_{\mu\nu}\equiv\frac{1}{2}\frac{\partial^2F^2}{\partial y^{\mu}\partial y^{\nu}}=\mathrm{diag}\left(\frac{1}{n},-n,-nr^2,-nr^2\sin^2\theta\right)_{\mu\nu}\,.
\end{equation}
The inverse metric simply reads as
\begin{equation}
\label{eq:finsler-metric-isotropic-inverse}
g^{\mu\nu}=\mathrm{diag}\left(n,-\frac{1}{n},-\frac{1}{nr^2},-\frac{1}{nr^2\sin^2\theta}\right)^{\mu\nu}\,.
\end{equation}
\end{subequations}
Since the metric does not depend on $\mathbf{y}$, the Cartan connection \cite{Bao:2000} vanishes:
\begin{equation}
\label{eq:cartan-torsion-isotropic}
A_{\mu\nu\varrho}\equiv \frac{F}{2}\frac{\partial g_{\mu\nu}}{\partial y^{\varrho}}=\frac{F}{4}\frac{\partial^3F^2}{\partial y^{\mu}\partial y^{\nu}\partial y^{\varrho}}\,,
\end{equation}
Therefore according to Deicke's theorem \cite{Deicke:1953} the Finsler space considered is Riemannian. Now the base has been set up to study the geometry of
the space defined by \eqref{eq:finsler-structure-isotropic-refractive-index}. The first step is the obtain the coefficients of the affine connection
(Christoffel symbols of second kind) that are defined in analogy to the Christoffel symbols in Riemannian geometry:
\begin{equation}
\label{eq:christoffel-symbols}
\gamma^{\mu}_{\phantom{\mu}\nu\rho}=\frac{1}{2}g^{\;\!\mu\alpha}\left(\frac{\partial g_{\alpha\nu}}{\partial x^{\rho}}-\frac{\partial g_{\nu\rho}}{\partial x^{\alpha}}+\frac{\partial g_{\rho\alpha}}{\partial x^{\nu}}\right)\,.
\end{equation}
Note that summation over equal indices is understood based on Einstein's convention. The nonzero contributions read as follows:
\begin{subequations}
\begin{align}
\gamma^t_{\phantom{t}tr}&=-\frac{n'}{2n}\,,\quad \gamma^r_{\phantom{r}tt}=-\frac{n'}{2n^3}\,,\quad \gamma^r_{\phantom{r}rr}=\frac{n'}{2n}\,,\quad \gamma^r_{\phantom{r}\theta\theta}=-\frac{r(2n+rn')}{2n}\,, \\[2ex]
\gamma^r_{\phantom{r}\phi\phi}&=-\frac{r(2n+rn')}{2n}\sin^2\theta\,,\quad \gamma^{\theta}_{\phantom{\theta}r\theta}=\frac{1}{r}+\frac{n'}{2n}\,,\quad \gamma^{\theta}_{\phantom{\theta}\phi\phi}=-\sin\theta\cos\theta\,, \\[2ex]
\gamma^{\phi}_{\phantom{\phi}r\phi}&=\frac{1}{r}+\frac{n'}{2n}\,,\quad \gamma^{\phi}_{\phantom{\phi}\theta\phi}=\cot\theta\,.
\end{align}
\end{subequations}
Since torsion is assumed to vanish, the Christoffel symbols are symmetric in the latter two indices. The connection coefficients with at
least one index equal to the radial coordinate $r$ involve the first derivative of the refractive index. Furthermore they do not involve the
angle $\phi$ as expected for spherically symmetric metrics. As a next step the geodesic spray coefficients are needed:
$G^{\mu}\equiv\gamma^{\mu}_{\phantom{\mu}\nu\varrho}y^{\nu}y^{\varrho}$. The latter appear in the geodesic equations in Finsler geometry:
\begin{subequations}
\begin{align}
G^t&=-y^ty^r\frac{n'}{n}\,, \\[2ex]
G^r&=-(y^t)^2\frac{n'}{2n^3}+\frac{1}{2n}\left\{(y^r)^2n'-r(2n+rn')[(y^{\theta})^2+(y^{\phi})^2\sin^2\theta]\right\}\,, \\[2ex]
G^{\theta}&=y^ry^{\theta}\left(\frac{2}{r}+\frac{n'}{n}\right)-(y^{\phi})^2\sin\theta\cos\theta\,,\quad G^{\phi}=y^{\phi}\left[2y^{\theta}\cot\theta+y^r\left(\frac{2}{r}+\frac{n'}{n}\right)\right]\,.
\end{align}
\end{subequations}
The geodesic spray coefficients can be used to define the nonlinear connection \cite{Bao:2000} on $TM\setminus \{0\}$:
\begin{equation}
N^{\mu}_{\phantom{\mu}\nu}\equiv\frac{1}{2}\frac{\partial G^{\mu}}{\partial y^{\nu}}\,.
\end{equation}
The reasons for introducing these connection coefficients is as follows. On the one hand the basis vectors $\partial/\partial x^{\nu}$ and
$\partial/\partial y^{\nu}$ are unsuitable to be chosen as a local basis of $TTM$, since the $\partial/\partial x^{\nu}$ transform in a complicated way.
On the other hand if $\{\mathrm{d}x^{\mu},\mathrm{d}y^{\mu}\}$ is chosen as a local basis of the cotangent bundle $T^{*}TM$ the transformation
properties of $\mathrm{d}y^{\mu}$ are involved. To have the desired transformation properties for the basis of the tangent and the cotangent
bundle of $TM\setminus \{0\}$ the following basis vectors can be introduced using the nonlinear connection:
\begin{subequations}
\begin{align}
\left\{\frac{\delta}{\delta x^{\nu}},F\frac{\partial}{\partial y^{\nu}}\right\}\,,&\quad \frac{\delta}{\delta x^{\nu}}\equiv \frac{\partial}{\partial x^{\nu}}-N^{\mu}_{\phantom{\mu}\nu}\frac{\partial}{\partial y^{\mu}}\,, \\[2ex]
\left\{\mathrm{d}x^{\mu},\frac{\delta y^{\mu}}{F}\right\}\,,&\quad \delta y^{\mu}\equiv \mathrm{d}y^{\mu}+N^{\mu}_{\phantom{\mu}\nu}\mathrm{d}x^{\nu}\,.
\end{align}
\end{subequations}
For the particular case studied here the nonlinear connection coefficients can be comprised in a $(3\times 3)$ matrix that reads as
\begin{equation}
(N^{\mu}_{\phantom{\mu}\nu})=\frac{1}{2}\begin{pmatrix}
-y^rn'/n & -y^tn'/n & 0 & 0 \\
-y^tn'/n^3 & y^rn'/n & -y^{\theta}r(2n+rn')/n & -y^{\phi}r\sin^2\theta(2n+rn')/n \\
0 & y^{\theta}(2/r+n'/n) & y^r(2/r+n'/n) & -y^{\phi}\sin(2\theta) \\
0 & y^{\phi}(2/r+n'/n) & 2y^{\phi}\cot\theta & y^r(2/r+n'/n)+2y^{\theta}\cot\theta \\
\end{pmatrix}\,.
\end{equation}
To compute directional derivatives of tensor fields on Finsler manifolds, a further connection has to be found to define a covariant derivative.
It was shown that the pulled-back bundle $\pi^{*}TM$ has a linear connection associated to it, which is called the Chern connection
$\Gamma^{\mu}_{\phantom{\mu}\nu\varrho}$. Explicitly it can be obtained from the Finsler metric using the nonlinear connection $N^{\mu}_{\phantom{\mu}\nu}$:
\begin{align}
\Gamma^{\mu}_{\phantom{\mu}\nu\varrho}&=\frac{1}{2}g^{\mu\alpha}\left(\frac{\delta g_{\alpha\nu}}{\delta x^{\varrho}}-\frac{\delta g_{\nu\varrho}}{\delta x^{\alpha}}+\frac{\delta g_{\varrho\alpha}}{\delta x^{\nu}}\right) \notag \\
&=\frac{1}{2}g^{\mu\alpha}\left(\frac{\partial g_{\alpha\nu}}{\partial x^{\varrho}}-N^{\beta}_{\phantom{\beta}\varrho}\frac{\partial g_{\alpha\nu}}{\partial y^{\beta}}-\left[\frac{\partial g_{\nu\varrho}}{\partial x^{\alpha}}-N^{\beta}_{\phantom{\beta}\alpha}\frac{\partial g_{\nu\varrho}}{\partial y^{\beta}}\right]+\frac{\partial g_{\varrho\alpha}}{\partial x^{\nu}}-N^{\beta}_{\phantom{\beta}\nu}\frac{\partial g_{\varrho\alpha}}{\partial y^{\beta}}\right)\,.
\end{align}
The Chern connection is unique and formally it has the same index structure as the formal Christoffel symbols. The difference to the latter
is that the derivative $\delta/\delta x^{\mu}$ is used instead of the ordinary partial derivative $\partial/\partial x^{\mu}$. However in the particular
case studied here, $\Gamma^{\mu}_{\phantom{\mu}\nu\varrho}=\gamma^{\mu}_{\phantom{\mu}\nu\varrho}$, since the Finsler metric $g_{\mu\nu}$ does not
depend on the components of $\mathbf{y}$. Finally the Chern connection is needed to define a Finslerian version of the Riemann curvature tensor:
\begin{align}
\label{eq:finsler-curvature-tensor}
R_{\nu\phantom{\mu}\varrho\sigma}^{\phantom{\nu}\mu}&=\frac{\delta\Gamma^{\mu}_{\phantom{\mu}\nu\sigma}}{\delta x^{\varrho}}-\frac{\delta\Gamma^{\mu}_{\phantom{\mu}\nu\varrho}}{\delta x^{\sigma}}+\Gamma^{\mu}_{\phantom{\mu}\alpha\varrho}\Gamma^{\alpha}_{\phantom{\alpha}\nu\sigma}-\Gamma^{\mu}_{\phantom{\mu}\alpha\sigma}\Gamma^{\alpha}_{\phantom{\alpha}\nu\varrho} \notag \\
&=\frac{\partial\Gamma^{\mu}_{\phantom{\mu}\nu\sigma}}{\partial x^{\varrho}}-N^{\beta}_{\phantom{\beta}\varrho}\frac{\partial\Gamma^{\mu}_{\phantom{\mu}\nu\sigma}}{\partial y^{\beta}}-\left(\frac{\partial\Gamma^{\mu}_{\phantom{\mu}\nu\varrho}}{\partial x^{\sigma}}-N^{\beta}_{\phantom{\beta}\sigma}\frac{\partial\Gamma^{\mu}_{\phantom{\mu}\nu\varrho}}{\partial y^{\beta}}\right)+\Gamma^{\mu}_{\phantom{\mu}\alpha\varrho}\Gamma^{\alpha}_{\phantom{\alpha}\nu\sigma}-\Gamma^{\mu}_{\phantom{\mu}\alpha\sigma}\Gamma^{\alpha}_{\phantom{\alpha}\nu\varrho}\,.
\end{align}
Since the Chern connection coefficients correspond to the formal Christoffel symbols and the latter are independent of $y^{\mu}$, the
curvature components correspond to the Riemannian ones. They involve an additional derivative of the Christoffel symbols, which is why
they comprise second derivatives of the refractive index. Explicitly the independent curvature tensor components are stated
as follows:
\begin{subequations}
\begin{align}
R_{t\phantom{r}tr}^{\phantom{t}r}&=\frac{nn''-2n'^2}{2n^4}\,,\quad R_{t\phantom{\theta}t\theta}^{\phantom{t}\theta}=\frac{n'(2n+rn')}{4rn^4}=R_{t\phantom{\phi}t\phi}^{\phantom{t}\phi}\,, \\[2ex]
R_{r\phantom{t}tr}^{\phantom{r}t}&=\frac{nn''-2n'^2}{2n^2}\,,\quad R_{\theta\phantom{t}t\theta}^{\phantom{\theta}t}=\frac{rn'(2n+rn')}{4n^2}\,,\quad R_{\theta\phantom{\phi}\theta\phi}^{\phantom{\theta}\phi}=\frac{rn'(4n+rn')}{4n^2}\,, \\[2ex]
R_{\phi\phantom{t}t\phi}^{\phantom{\phi}t}&=\frac{rn'(2n+rn')}{4n^2}\sin^2\theta\,,\quad R_{\phi\phantom{\theta}\theta\phi}^{\phantom{\phi}\theta}=-\frac{rn'(4n+rn')}{4n^2}\sin^2\theta\,.
\end{align}
\end{subequations}
The components related by symmetries are omitted. Since the Finsler structure of \eqref{eq:finsler-structure-isotropic-refractive-index}
is Riemannian according to Deicke's theorem, we will first use the Riemannian definitions
of the Ricci tensor $\mathcal{Ric}_{\mu\nu}\equiv R_{\mu\phantom{\alpha}\alpha\nu}^{\phantom{\mu}\alpha}$ and the curvature scalar
(Ricci scalar) $\mathcal{Ric}$. These are denoted by calligraphic letters and they follow from suitable contractions of the Riemann
curvature tensor. The Ricci tensor components with equal indices deliver nonzero contributions only:
\begin{subequations}
\begin{align}
\mathcal{Ric}_{tt}&=\frac{1}{2rn^4}\left[rn'^2-n(2n'+rn'')\right]\,,\quad \mathcal{Ric}_{rr}=-\frac{1}{2rn}(2n'+rn'')\,, \\[2ex]
\mathcal{Ric}_{\theta\theta}&=\frac{r}{2n^2}\left[rn'^2-n(2n'+rn'')\right]\,,\quad \mathcal{Ric}_{\phi\phi}=\frac{r}{2n^2}\left[rn'^2-n(2n'+rn'')\right]\sin^2\theta\,, \\[2ex]
\label{eq:ricci-scalar-conventional}
\mathcal{Ric}&\equiv \mathcal{Ric}^{\mu}_{\phantom{\mu}\mu}=g^{\mu\nu}\mathcal{Ric}_{\mu\nu}=\frac{1}{2rn^3}\left[2n(2n'+rn'')-rn'^2\right]\,.
\end{align}
\end{subequations}
In Riemannian geometry the curvature tensor obeys the first and the second Bianchi identities. Especially the second one,
\begin{subequations}
\begin{align}
0&\equiv D_{\eta}R_{\lambda\phantom{\mu}\nu\kappa}^{\phantom{\lambda}\mu}+D_{\kappa}R_{\lambda\phantom{\mu}\eta\nu}^{\phantom{\lambda}\mu}+D_{\nu}R_{\lambda\phantom{\mu}\kappa\eta}^{\phantom{\lambda}\mu}\,, \\[2ex]
D_{\lambda}R_{\mu\phantom{\nu}\rho\sigma}^{\phantom{\mu}\nu}&=\partial_{\lambda}R_{\mu\phantom{\nu}\rho\sigma}^{\phantom{\mu}\nu}-\Gamma^{\alpha}_{\phantom{\alpha}\mu\lambda}R_{\alpha\phantom{\nu}\rho\sigma}^{\phantom{\alpha}\nu}+\Gamma^{\nu}_{\phantom{\nu}\alpha\lambda}R_{\mu\phantom{\alpha}\rho\sigma}^{\phantom{\mu}\alpha}-\Gamma^{\alpha}_{\phantom{\alpha}\rho\lambda}R_{\mu\phantom{\nu}\alpha\sigma}^{\phantom{\mu}\nu}-\Gamma^{\alpha}_{\phantom{\alpha}\sigma\lambda}R_{\mu\phantom{\nu}\rho\alpha}^{\phantom{\mu}\nu}\,,
\end{align}
\end{subequations}
is important in the context of General Relativity, because it leads to the statement that the Einstein tensor $G^{\mu\nu}$ is covariantly
constant:
\begin{subequations}
\begin{align}
D_{\mu}G^{\mu}_{\phantom{\mu}\nu}=\partial_{\mu}G^{\mu}_{\phantom{\mu}\nu}+\Gamma^{\mu}_{\phantom{\mu}\alpha\mu}G^{\alpha}_{\phantom{\alpha}\nu}-\Gamma^{\alpha}_{\phantom{\alpha}\nu\mu}G^{\mu}_{\phantom{\mu}\alpha}\equiv 0\,,\quad G^{\mu\nu}\equiv \mathcal{Ric}^{\mu\nu}-\frac{\mathcal{Ric}}{2}g^{\mu\nu}\,,
\end{align}
\end{subequations}
which was checked to be valid for the particular metric $g_{\mu\nu}$ of \eqref{eq:finsler-metric-isotropic}. This identity is the reason
why explicit Lorentz violation is incompatible with Riemannian geometry. Due to the Einstein equations it forces the energy-momentum
tensor to be covariantly conserved as well, which does not necessarily hold when there is a spacetime-dependent background. At this
point it is reasonable to wonder how Finsler geometry can help us to solve that problem. For the isotropic metric considered
the identity $D_{\mu}G^{\mu}_{\phantom{\mu}\nu}\equiv 0$ is inherited from the Riemannian to the Finslerian framework, since the
Finsler metric of \eqref{eq:finsler-metric-isotropic} does not comprise any dependence on $y^{\mu}$. Assuming that Finsler
geometry provides the necessary tools to circumvent the no-go theorem of \cite{Kostelecky:2003fs} in a general explicitly Lorentz-violating
setting, then it should also work for the special isotropic case studied here.

One possible approach (there may be others) might be to consider a suitable equivalent of the Einstein equations in Finsler geometry.
Such an equivalent can be based on an alternative definition of the Einstein tensor $G^{\mu\nu}$ constructed from curvature-related tensors
in the Finsler framework. These objects will be introduced in what follows. The first is obtained from the curvature tensor by contracting
the latter with two vectors $y^{\mu}/F$ according to
\begin{equation}
\label{eq:predecessor-flag-curvature}
R^{\mu}_{\phantom{\mu}\varrho}\equiv\frac{y^{\nu}}{F}R_{\nu\phantom{\mu}\varrho\sigma}^{\phantom{\nu}\mu}\frac{y^{\sigma}}{F}\,.
\end{equation}
Note that this construction does not correspond to the Ricci tensor of Riemannian geometry. In particular it is sometimes referred to
as the predecessor of flag curvature, which is a generalization of sectional curvature in Finsler geometry. For the special case here
$R^{\mu}_{\phantom{\mu}\varrho}$ are the components of a $(4\times 4)$ matrix. The trace of this matrix is taken to obtain the generalization
of the Ricci scalar in Finsler geometry: $\mathit{Ric}\equiv R^{\varrho}_{\phantom{\varrho}\varrho}$. Since the explicit expressions for
$R^{\mu}_{\phantom{\mu}\varrho}$ and $\textit{Ric}$ are complicated and not illuminating, they will not be stated explicitly.

The flag curvature in Finsler geometry is computed similarly to the sectional curvature in Riemannian geometry. The latter is defined in a
tangent space at a point $x$ of the manifold where two arbitrary, linearly independent directions are needed for its computation. The
resulting quantity only depends on the plane considered, but not on the particular choice of the directions. The flag curvature in
Finsler geometry carries the same spirit where one direction is chosen to correspond to $\mathbf{y}$ and the other one, say $\mathbf{L}$,
is supposed to be orthogonal to $\mathbf{y}$. These vectors are then suitably contracted with the curvature tensor of \eqref{eq:finsler-curvature-tensor}.
Note that $\mathbf{y}$ and the vector orthogonal to it can be considered to span a flag where $\mathbf{y}$ is assumed to point along the
flag pole. This explains the name for the curvature. For an $n$-dimensional Finsler manifold $R$ is the sum of $n-1$ flag curvatures. It only
depends on $r$ and $\mathbf{y}$, but not on the direction $\mathbf{L}$ chosen orthogonal to $\mathbf{y}$.

Although $R_{\mu\varrho}$ of \eqref{eq:predecessor-flag-curvature} is not understood to be the generalization of the Ricci tensor in
Finsler geometry, it is still possible to define the latter. The definition (cf.~Eq.~(7.6.4) in \cite{Bao:2000}) involves both the Finsler
structure $F$ and the Finslerian version of the Ricci scalar $\textit{Ric}$:
\begin{equation}
\label{eq:ricci-tensor-from-predecessor}
\mathit{Ric}_{\mu\nu}\equiv \frac{1}{2}\frac{\partial^2(F^2\textit{Ric})}{\partial y^{\mu}\partial y^{\nu}}\,.
\end{equation}
For Finsler metrics that are Riemannian, i.e., for the isotropic metric considered in \eqref{eq:finsler-metric-isotropic} it also holds that
$\mathit{Ric}_{\mu\nu}=R_{\mu\phantom{\alpha}\alpha\nu}^{\phantom{\mu}\alpha}$. Hence in our case the Finslerian definition of the Ricci
tensor corresponds to the Riemannian expression, computing an appropriate trace of the curvature tensor. The expression of
\eqref{eq:ricci-tensor-from-predecessor} can be used to obtain the Ricci scalar in Finsler geometry by contracting the Ricci tensor with
two vectors~$y^{\mu}/F$ (cf.~(7.6.5) in \cite{Bao:2000}):
\begin{equation}
\label{eq:ricci-scalar-finsler}
\mathit{Ric}\equiv \mathit{Ric}_{\mu\nu}\frac{y^{\mu}}{F}\frac{y^{\nu}}{F}\,.
\end{equation}
It can be shown in general that the latter corresponds to $R^{\varrho}_{\phantom{\varrho}\varrho}$ that is
obtained from tracing \eqref{eq:predecessor-flag-curvature}. This object is distinguished from the Ricci scalar $\mathcal{Ric}$ in a Riemannian
setting, which follows from tracing the Ricci curvature tensor $\mathcal{Ric}_{\mu\nu}$, cf. \eqref{eq:ricci-scalar-conventional}. Note that the
quantity of \eqref{eq:ricci-scalar-finsler} is the direct Finslerian equivalent of the Ricci scalar. Since the Finsler metric considered is
Riemannian, $\mathcal{Ric}$ only involves dependences on $r$, whereas $\textit{Ric}$ depends on $y^{\mu}$ as well.
In general and especially here $\mathcal{Ric}\neq \textit{Ric}$.

At this stage there are several possibilities of defining the Einstein tensor $G_{\mu\nu}$ in a Finsler framework using different combinations
of $\mathcal{Ric}_{\mu\nu}$, $\mathcal{Ric}$, $R_{\mu\nu}$, $\mathit{Ric}_{\mu\nu}$, and $\mathit{Ric}$. The following have been tried:
\begin{subequations}
\label{eq:propositions-einstein-tensor}
\begin{align}
(G^{\mu}_{\phantom{\mu}\nu})^{(1)}&\equiv \mathcal{Ric}^{\mu}_{\phantom{\mu}\nu}-\frac{1}{2}\delta^{\mu}_{\phantom{\mu}\nu}\textit{Ric}\,, \\[2ex]
(G^{\mu}_{\phantom{\mu}\nu})^{(2)}&\equiv R^{\mu}_{\phantom{\mu}\nu}-\frac{1}{2}\delta^{\mu}_{\phantom{\mu}\nu}\textit{Ric}\,, \\[2ex]
(G^{\mu}_{\phantom{\mu}\nu})^{(3)}&\equiv R^{\mu}_{\phantom{\mu}\nu}-\frac{1}{2}\delta^{\mu}_{\phantom{\mu}\nu}\mathcal{Ric}\,.
\end{align}
\end{subequations}
A reasonable test of whether one of these choices is suitable, requires computing their covariant derivatives, i.e., $D_{\mu}(G^{\mu}_{\phantom{\mu}\nu})^{(i)}$
for $i=1\dots 3$. The wishful result would be a nonvanishing covariant derivative bearing resemblance to the modified covariant
conservation law of the energy-momentum tensor in \eqref{eq:modified-energy-momentum-conservation-isotropic}. This makes
sense when we assume that the modified Einstein tensor (in a Finslerian framework) is linked to the energy-momentum tensor
in an explicitly Lorentz-violating theory. The corresponding covariant derivative to be used involves both the nonminimal connection
$N^{\mu}_{\phantom{\mu}\nu}$ and the Chern connection $\Gamma^{\mu}_{\phantom{\mu}\nu\varrho}$ being equal to the
Christoffel symbols $\gamma^{\mu}_{\phantom{\mu}\nu\varrho}$ in this case:
\begin{equation}
\label{eq:covariant-derivative-finsler-einstein-tensor}
D_{\mu}(G^{\mu}_{\phantom{\mu}\nu})^{(i)}=\frac{\partial(G^{\mu}_{\phantom{\mu}\nu})^{(i)}}{\partial x^{\mu}}-N^{\beta}_{\phantom{\beta}\mu}\frac{(G^{\mu}_{\phantom{\mu}\nu})^{(i)}}{\partial y^{\beta}}+\Gamma^{\mu}_{\phantom{\mu}\alpha\mu}(G^{\alpha}_{\phantom{\alpha}\nu})^{(i)}-\Gamma^{\alpha}_{\phantom{\alpha}\nu\mu}(G^{\mu}_{\phantom{\mu}\alpha})^{(i)}\,.
\end{equation}
Starting from the Finsler metric of \eqref{eq:finsler-metric-isotropic} there have been up to three derivatives with respect to the coordinates
involved, which is why in \eqref{eq:covariant-derivative-finsler-einstein-tensor} the third derivative of the refractive index appears in general.
The more of the higher derivatives of a Taylor expansion of $n(r)$ are taken into account, the smaller are the structures in changes of $n(r)$
to be resolved. Therefore relying on the geometric-optics approximation it is reasonable to consider only the first-order change of $n(r)$
incorporated in its first derivative and to neglect the higher-order derivatives, which describe small-scale changes of $n(r)$. Within this
approximation it makes sense to set $n(r)=1$, since modifications lead to higher-order contributions.

Furthermore the Finsler structure that the isotropic case was identified with is three-dimensional, cf.~\secref{sec:isotropic-case}, and it
involves spatial velocity components only. Hence $y^t$ can be considered as auxiliary and will be set to zero at the end. With this physical input the
covariant derivative of each Einstein tensor proposed in \eqref{eq:propositions-einstein-tensor} can be computed. The final result for
the third possibility looks rather promising:
\begin{equation}
D_{\mu}(G^{\mu}_{\phantom{\mu}\nu})^{(3)}|_{y^t=0}=\frac{1}{(y^r)^2+r^2[(y^{\theta})^2+(y^{\phi})^2\sin^2\theta]}\frac{n'}{r^2}\begin{pmatrix}
0 \\
(y^r)^2 \\
y^ry^{\theta}r^2 \\
y^ry^{\phi}r^2\sin^2\theta \\
\end{pmatrix}_{\nu}+\dots\,,
\end{equation}
where terms of $\mathcal{O}(n'',n''',n'^2,n'^3)$ have been neglected. Using the inverse metric $g^{\mu\nu}$ of
\eqref{eq:finsler-metric-isotropic-inverse} the second index can be raised. Besides we identify the spatial components of $\mathbf{y}$ with
the spatial components of the physical velocity, i.e., $y^r=u^r$, $y^{\theta}=u^{\theta}$, and $y^{\phi}=u^{\phi}$ where the spatial flat metric
in spherical polar coordinates is given by $(r_{ij})=\mathrm{diag}(1,r^2,r^2\sin^2\theta)$. This leads to the final result
\begin{subequations}
\begin{align}
D_{\mu}(G^{\mu\nu})^{(3)}|_{y^t=0}&=-\frac{n'}{\mathbf{u}^2r^2}\begin{pmatrix}
0 \\
(u^r)^2 \\
u^ru^{\theta} \\
u^ru^{\phi} \\
\end{pmatrix}^{\nu}+\dots\,, \\[2ex]
D_{\mu}(G^{\mu i})^{(3)}|_{y^t=0}&=-\frac{n'}{r^2}\widehat{u}^r\widehat{u}^i+\dots\,,
\end{align}
\end{subequations}
with the normalized three-velocity vector $\widehat{\mathbf{u}}=\mathbf{u}/|\mathbf{u}|$. Comparing the obtained result to
\eqref{eq:modified-energy-momentum-conservation-isotropic} reveals that the structure of both expressions is very similar.
The difference is a global prefactor of the form $r^2\sqrt{\mathbf{u}^2}$. The dimensionful factor of $r^2$ is not surprising.
Both the Riemann curvature tensor and the (modified) Einstein tensor involve two derivatives, which is why their mass dimensions
is $-2$. However the energy-momentum tensor is based on the ``Lagrangian'' of a classical light ray, \eqref{eq:lagrangian-classical-light-ray-isotropic},
which is a dimensionless quantity. The discrepancy in mass dimensions is compensated by the only dimensionful length scale
available, which is $r$. It seems that an alternative definition of the Einstein tensor in the framework of Finsler geometry can
compensate for the modified energy-momentum conservation law when explicit Lorentz violation is considered. This result is
interesting and deserves further study, e.g., whether it holds for anisotropic theories as well.


\section{Conclusions and outlook}
\label{sec:conclusion}

In this work classical-ray analogues to the photon sector of the minimal Standard-Model Extension were discussed. It was shown that
a nonvanishing photon mass allows for deriving classical point-particle Lagrangians in analogy to the fermion sector. However the
standard method used for the fermion sector does not work any more in case the photon mass vanishes. The reason is that a light ray
does not have as many degrees of freedom as a massive particle.

Instead, for the photon sector an alternative technique had to be employed which allowed to derive a Lagrangian-type function
for a classical ray directly from the modified photon dispersion relation. This was carried out for several interesting cases
of the minimal, {\em CPT}-even photon sector, which is characterized by dimensionless controlling coefficients. Subsequently it was
shown that the results obtained are consistent with the eikonal equation approach that describes the geometric-optics limit of an
electromagnetic wave. Mathematically the Lagrangian-type functions can be interpreted as Finsler structures. In contrast to the
fermion sector they only involve the spatial velocity components and they are closely linked to an effective refractive index of the
Lorentz-violating vacuum.

It has been known long since that there is a connection between the geodesic equations for a light ray in a gravitational
background and the eikonal equations. This link is warranted for weak gravitational fields at least, e.g., in the solar system.
It was crucial to set up a phenomenological description of light rays subject to Lorentz violation in a weak gravitational
field. This description made it possible to obtain sensitivities on the isotropic controlling coefficient $\widetilde{\kappa}_{\mathrm{tr}}$
that could be probed by the space missions GAIA and LATOR employing measurements of light deflection at massive bodies.
The upshot is that the planned mission LATOR may have a sensitivity on $|\widetilde{\kappa}_{\mathrm{tr}}|$ in the order of
magnitude of $10^{-16}$ where the running mission GAIA can reach $10^{-14}$. The difference in sensitivity originates from
the different precision of measuring angles for both missions.

The final part of the paper was dedicated to investigating the properties of the (isotropic) curved spacetime, which the
phenomenological studies were based on, from a Finslerian point of view. It was demonstrated that in the classical limit
(neglecting higher spacetime derivatives of the refractive index) an Einstein tensor can be defined that is not subject to
the usual Bianchi identities in Riemannian geometry. Therefore its covariant derivative is nonzero and it has a form that
is related to the modified conservation law of the energy-momentum tensor based on the classical Lagrangian-type function
studied in this context. Hence it seems that Finsler geometry provides new geometrical degrees of freedom that can serve
as a kind of ``buffer'' to allow for a momentum transfer whenever the momentum of the light ray changes. These geometrical
degrees of freedom take the role of the Nambu-Goldstone modes appearing when spontaneous Lorentz violation is considered.

To summarize, the current article provides a technique in treating Lorentz-violating photons in a curved background in
a geometric-optics approximation. As an outlook it will be interesting to apply the setup to anisotropic frameworks, first to
obtain sensitivities on related controlling coefficients and second to study the properties of the underlying Finsler geometry.

\section{Acknowledgments}

It is a pleasure to thank V.~A.~Kosteleck\'{y} for suggesting this line of research and for having fruitful discussions.
This work was performed with financial support from the \textit{Deutsche Akademie der Naturforscher Leopoldina} within
Grant No. LPDS 2012-17.

\newpage
\begin{appendix}
\numberwithin{equation}{section}

\section{Classical Lagrangians for massive Lorentz-violating photons}
\label{sec:lagrangians-massive-photons}
\setcounter{equation}{0}

The first part of the appendix shall briefly demonstrate how to derive the classical Lagrange functions in \secref{eq:lagrange-function-massive-modmax}
from the set of equations (\ref{eq:dispersion-relation-general}), (\ref{eq:group-velocity-correspondence}), and (\ref{eq:lagrange-function}).
The demonstration will be performed for the nonbirefringent, anisotropic case of the {\em CPT}-even sector and for a particular choice of
the {\em CPT}-odd framework. The calculation is easier for the {\em CPT}-even theory, which is why it will be studied first.

\subsection{{\em CPT}-even minimal photon sector}
\label{sec:lagrangians-cpt-even-massive-photons}

The base is \eqref{eq:kappas-anisotropic-nonbirefringent-case} where for convenience we set $(3/2)\widetilde{\kappa}_{e-}^{11}\equiv\kappa$.
For the remaining {\em CPT}-even cases the procedure works analogously. First of all the modified dispersion relation for a massive photon subject to this particular
Lorentz-violating framework reads
\begin{equation}
(1+\kappa)\left(k_0^2-k_1^2-k_2^2\right)-(1-\kappa)k_3^2=m_{\upgamma}^2\,.
\end{equation}
To obtain the group velocity components it is often reasonable not to solve the dispersion relation to obtain $k_0$ directly, but to differentiate
it implicitly with respect to the spatial momentum components:
\begin{subequations}
\begin{align}
2(1+\kappa)k_0\frac{\partial k_0}{\partial k_1}-2(1+\kappa)k_1=0 &\Leftrightarrow \frac{\partial k_0}{\partial k_1}=\frac{k_1}{k_0}\,, \\[2ex]
2(1+\kappa)k_0\frac{\partial k_0}{\partial k_2}-2(1+\kappa)k_2=0 &\Leftrightarrow \frac{\partial k_0}{\partial k_2}=\frac{k_2}{k_0}\,, \\[2ex]
2(1+\kappa)k_0\frac{\partial k_0}{\partial k_3}-2(1-\kappa)k_3=0 &\Leftrightarrow \frac{\partial k_0}{\partial k_3}=\frac{1-\kappa}{1+\kappa}\frac{k_3}{k_0}\,.
\end{align}
\end{subequations}
For the particular case studied, \eqref{eq:group-velocity-correspondence} leads to the following three equations:
\begin{equation}
\frac{k_1}{k_0}=-\frac{u^1}{u^0}\,,\quad \frac{k_2}{k_0}=-\frac{u^2}{u^0}\,,\quad \frac{1-\kappa}{1+\kappa}\frac{k_3}{k_0}=-\frac{u^3}{u^0}\,.
\end{equation}
Evidently only the third one is modified by Lorentz violation mirroring the spatial anisotropy. These relations can be solved directly to
express the spatial momentum components via $k_0$:
\begin{equation}
\label{eq:spatial-momentum-components-via-k0}
k_1=-\frac{k_0u^1}{u^0}\,,\quad k_2=-\frac{k_0u^2}{u^0}\,,\quad k_3=-\frac{1+\kappa}{1-\kappa}\frac{k_0u^3}{u^0}\,.
\end{equation}
We can now use \eqref{eq:lagrange-function} and express the spatial momentum components by taking into account the previously obtained
results of \eqref{eq:spatial-momentum-components-via-k0}:
\begin{equation}
L=-(k_0u^0+k_1u^1+k_2u^2+k_3u^3)=\frac{k_0}{u^0}\left[-(u^0)^2+(u^1)^2+(u^2)^2+\frac{1+\kappa}{1-\kappa}(u^3)^2\right]\,.
\end{equation}
The latter is solved with respect to $k_0$ giving an expression comprising the (unknown) Lagrange function and the
four-velocity components:
\begin{equation}
\label{eq:k0-via-velocity-components}
k_0=-L\frac{(1-\kappa)u^0}{(1-\kappa)\left[(u^0)^2-(u^1)^2-(u^2)^2\right]-(1+\kappa)(u^3)^2}\,.
\end{equation}
Now all four-momentum components in the dispersion relation can be eliminated via
\eqref{eq:spatial-momentum-components-via-k0} and a subsequent insertion of \eqref{eq:k0-via-velocity-components}:
\begin{subequations}
\begin{align}
0&=\frac{1+\kappa}{1-\kappa}\frac{k_0^2}{(u^0)^2}\left\{(1-\kappa)\left[(u^0)^2-(u^1)^2-(u^2)^2\right]-(1+\kappa)(u^3)^2\right\}-m_{\upgamma}^2\,, \\[2ex]
0&=L^2\frac{(1-\kappa)(1+\kappa)}{(1-\kappa)\left[(u^0)^2-(u^1)^2-(u^2)^2\right]-(1+\kappa)(u^3)^2}-m_{\upgamma}^2\,.
\end{align}
\end{subequations}
The final equation comprises a polynomial of the Lagrangian whose coefficients depend on four-velocity components only.
The polynomial must be solved to give $L$:
\begin{equation}
L^{\pm}=\pm m_{\upgamma}\sqrt{\frac{1}{1+\kappa}\left[(u^0)^2-(u^1)^2-(u^2)^2\right]-\frac{1}{1-\kappa}(u^3)^2}\,.
\end{equation}
The result corresponds to \eqref{eq:massive-lagrangian-anisotropic}. The procedure shown is typically applied to derive classical Lagrangians.
Four of the five equations are employed to eliminate all four-momentum components and to obtain a polynomial equation in $L$ that only comprises the
four-velocity. The latter is then solved with respect to $L$ finally.

\subsection{{\em CPT}-odd minimal photon sector}
\label{sec:lagrangians-cpt-odd-massive-photons}

Due to observer Lorentz invariance without a loss of generality $(k_{AF})^{\kappa}=(0,0,0,1)^{\kappa}$ will be chosen for the spacelike case.
The modified dispersion relation involves an isotropic contribution and a second term that does not comprise the momentum component
parallel to the preferred spacetime direction:
\begin{equation}
\label{eq:dispersion-relation-cpt-odd}
(k_0^2-\mathbf{k}^2)^2-4m_{\scriptscriptstyle{\mathrm{CS}}}^2(k_0^2-k_1^2-k_2^2)=0\,.
\end{equation}
The group velocity components are obtained by implicit differentiation of \eqref{eq:dispersion-relation-cpt-odd} with respect to the
spatial momentum components:
\begin{subequations}
\begin{align}
0&=4(k_0^2-\mathbf{k}^2)\left[k_0\frac{\mathrm{d}k_0}{\mathrm{d}k_1}-k_1\right]-8m_{\scriptscriptstyle{\mathrm{CS}}}^2\left(k_0\frac{\mathrm{d}k_0}{\mathrm{d}k_1}-k_1\right) \notag \\
&=4(k_0^2-\mathbf{k}^2-2m_{\scriptscriptstyle{\mathrm{CS}}}^2)\left[k_0\frac{\mathrm{d}k_0}{\mathrm{d}k_1}-k_1\right]\,, \\[2ex]
0&=4(k_0^2-\mathbf{k}^2-2m_{\scriptscriptstyle{\mathrm{CS}}}^2)\left[k_0\frac{\mathrm{d}k_0}{\mathrm{d}k_2}-k_2\right]\,, \\[2ex]
0&=4(k_0^2-\mathbf{k}^2)\left[k_0\frac{\mathrm{d}k_0}{\mathrm{d}k_3}-k_3\right]-8m_{\scriptscriptstyle{\mathrm{CS}}}^2k_0\frac{\mathrm{d}k_0}{\mathrm{d}k_3} \notag \\
&=4(k_0^2-\mathbf{k}^2-2m_{\scriptscriptstyle{\mathrm{CS}}}^2)k_0\frac{\mathrm{d}k_0}{\mathrm{d}k_3}-4(k_0^2-\mathbf{k}^2)k_3\,.
\end{align}
\end{subequations}
Since the preferred spacetime direction points along the third axis of the coordinate system, the first and second group velocity
components remain standard where only the third one is modified:
\begin{equation}
\frac{\mathrm{d}k_0}{\mathrm{d}k_1}=\frac{k_1}{k_0}\,,\quad \frac{\mathrm{d}k_0}{\mathrm{d}k_2}=\frac{k_2}{k_0}\,,\quad \frac{\mathrm{d}k_0}{\mathrm{d}k_3}=\frac{k_3(k_0^2-\mathbf{k}^2)}{k_0(k_0^2-\mathbf{k}^2-2m_{\scriptscriptstyle{\mathrm{CS}}}^2)}\,.
\end{equation}
Therefore \eqref{eq:group-velocity-correspondence} results in
\begin{equation}
\label{eq:group-velocity-correspondence-cpt-odd}
\frac{k_1}{k_0}=-\frac{u^1}{u^0}\,,\quad \frac{k_2}{k_0}=-\frac{u^2}{u^0}\,,\quad \frac{k_3(k_0^2-\mathbf{k}^2)}{k_0(k_0^2-\mathbf{k}^2-2m_{\scriptscriptstyle{\mathrm{CS}}}^2)}=-\frac{u^3}{u^0}\,.
\end{equation}
The first two of these relationships allow for writing $k_1$ and $k_2$ in terms of $k_0$. However the third equation would lead to
a cumbersome third-order polynomial to be solved, which is not a reasonable step to take. It is better to insert the first two of
\eqref{eq:group-velocity-correspondence-cpt-odd} into \eqref{eq:lagrange-function} and to solve the latter with respect to $k_3$.
Then it is possible to express $k_3$ via $k_0$ only:
\begin{equation}
k_3=\frac{1}{u^0u^3}\left\{k_0\left[(u^1)^2+(u^2)^2-(u^0)^2\right]-Lu^0\right\}\,.
\end{equation}
Now we can express all spatial momentum components via $k_0$. Hence we can eliminate all of them in \eqref{eq:dispersion-relation-cpt-odd}
to obtain an equation that only involves $k_0$. This can be solved to write $k_0$ in terms of four-velocity components and the Lagrangian
where one of the solutions reads
\begin{subequations}
\label{eq:k0-solution-cpt-odd}
\begin{align}
k_0&=-u^0\frac{L\sqrt{u_{\bot}^2}+m_{\scriptscriptstyle{\mathrm{CS}}}(u^3)^2+|u^3|\sqrt{L^2+2m_{\scriptscriptstyle{\mathrm{CS}}}\sqrt{u_{\bot}^2}L+m_{\scriptscriptstyle{\mathrm{CS}}}^2(u^3)^2}}{u^2\sqrt{u_{\bot}^2}}\,, \\[2ex]
(u_{\bot}^{\mu})&=(u^0,u^1,u^2,0)^T\,.
\end{align}
\end{subequations}
Here $u^0>0$ has been assumed for simplicity. The last step is to eliminate all four-momentum components in the third of
\eqref{eq:group-velocity-correspondence-cpt-odd} to obtain a polynomial equation for $L$:
\begin{equation}
L^2+2m_{\scriptscriptstyle{\mathrm{CS}}}\sqrt{u_{\bot}^2}L+m_{\scriptscriptstyle{\mathrm{CS}}}^2(u^3)^2=0\,,
\end{equation}
which leads to the Lagrange functions
\begin{equation}
L^{\pm}=m_{\scriptscriptstyle{\mathrm{CS}}}\left[\pm\sqrt{u^2}-\sqrt{u_{\bot}^2}\right]\,.
\end{equation}
Reinstating the preferred spacetime direction, it is possible to write the latter in the form of \eqref{eq:lagrange-function-mcs-theory}.
Using the other solution of $k_0$ similar to \eqref{eq:k0-solution-cpt-odd} the Lagrangians with the opposite signs are obtained.
A computation for $u^0<0$ leads to analogous results. Due to observer Lorentz invariance the form of the Lagrangian stays
the same for general spacelike $k_{AF}$.

\section{Light deflection in Schwarzschild spacetimes}
\label{sec:light-deflection-schwarzschild}

In \cite{Betschart:2008yi} it was found that a constant refractive index $n\neq 1$ due to Lorentz violation leads to a change in light deflection. This result
is in contrast to what we obtain from Bouguer's formula in \secref{sec:gravitational-backgrounds-isotropic-case}. A rough explanation is that Bouguer's
formula relies on the eikonal equation, which is equivalent to the null geodesic equations only for a weak gravitational field. However the latter reference
is based on a Schwarzschild metric,
\begin{equation}
\mathrm{d}\tau^2=\left(1-\frac{2GM}{r}\right)\mathrm{d}t^2-\left(1-\frac{2GM}{r}\right)^{-1}\mathrm{d}r^2-r^2(\mathrm{d}\theta^2+\sin^2\theta\mathrm{d}\phi^2)\,,
\end{equation}
which in this form is not generally isotropic. To get a more profound understanding, consider the geodesic equations for a photon in a generally isotropic spacetime
of \eqref{eq:line-interval} with $A=A(r)$. The Christoffel symbols are computed in Riemmanian geometry according to
\eqref{eq:christoffel-symbols} and the geodesic equations read
\begin{equation}
\frac{\mathrm{d}x^{\mu}}{\mathrm{d}\lambda}+\gamma^{\mu}_{\phantom{\mu}\nu\varrho}\frac{\mathrm{d}x^{\nu}}{\mathrm{d}\lambda}\frac{\mathrm{d}x^{\varrho}}{\mathrm{d}\lambda}=0\,,\quad (x^{\mu})=(t,r,\theta,\phi)^T\,.
\end{equation}
In what follows, differentiation with respect to the curve parameter $\lambda$ and with respect to $r$, respectively, will be denoted by a dot and a prime.
The geodesic equations can then be cast into the following form:
\begin{subequations}
\label{eq:geodesic-equations-spherically-symmetric-spacetime}
\begin{align}
\label{eq:equation-t}
0&=\ddot{t}-\frac{A'}{A}\dot{r}\dot{t}\,, \displaybreak[0]\\[2ex]
\label{eq:equation-r}
0&=\ddot{r}+\frac{A'}{2A}\dot{r}^2-\frac{A'}{2A^3}\dot{t}^2-\left(1+\frac{A'}{2A}r\right)r\dot{\theta}^2-\left(1+\frac{A'}{2A}r\right)r\sin^2(\theta)\dot{\phi}^2\,, \displaybreak[0]\\[2ex]
\label{eq:equation-theta}
0&=\ddot{\theta}+\left(\frac{2}{r}+\frac{A'}{A}\right)\dot{r}\dot{\theta}-\sin(\theta)\cos(\theta)\dot{\phi}^2=0\,, \displaybreak[0]\\[2ex]
\label{eq:equation-phi}
0&=\ddot{\phi}+\left(\frac{2}{r}+\frac{A'}{A}\right)\dot{r}\dot{\phi}+2\cot(\theta)\dot{\theta}\dot{\phi}=0\,, \displaybreak[0]\\[2ex]
\label{eq:equation-null}
0&=\frac{1}{A}\dot{t}^2-A\dot{r}^2-Ar^2\left[\dot{\theta}^2+\sin^2(\theta)\dot{\phi}^2\right]\,,
\end{align}
\end{subequations}
where the fifth of those is the condition for a null-trajectory. They correspond to the equations stated in \cite{Wu:1988} in case that $A$ is a
function of the radial coordinate $r$ only. Now the right-hand side of \eqref{eq:equation-t} can be written as the derivative of a
conserved quantity that is denoted as $K_0$ in \cite{Wu:1988}:
\begin{equation}
\label{eq:conserved-quantity-1}
0=A\frac{\mathrm{d}}{\mathrm{d}\lambda}\left(\frac{\dot{t}}{A}\right) \Rightarrow K_0=\frac{\dot{t}}{A}\,,\quad \dot{t}=K_0A\,.
\end{equation}
With the choice of $\theta=\pi/2$ \eqref{eq:equation-theta} is fulfilled automatically. Using the previous results, \eqref{eq:equation-phi}
can be expressed as the time-derivative of another conserved quantity $K_1$:
\begin{equation}
\label{eq:angular-momentum-conservation}
0=\frac{1}{Ar^2}\frac{\mathrm{d}}{\mathrm{d}\lambda}(Ar^2\dot{\phi}) \Rightarrow K_1=Ar^2\dot{\phi}\,,\quad \dot{\phi}=\frac{K_1}{Ar^2}\,.
\end{equation}
Looking at \eqref{eq:conserved-quantities-killing} we see that both $K_0$ and $K_1$ correspond to the conserved quantities that are obtained
via the Killing vectors, cf.~\appref{sec:killing-vectors-generally-isotropic}. From now on the trajectory shall be parameterized with respect
to proper time: $\lambda=\tau$. Since $K_0$ is then linked to infinitesimal time translations, it is reasonable
to identify it with the total photon energy $E$. Furthermore $K_1$ is connected to infinitesimal changes in the angle $\phi$, which is why it
corresponds to the angular momentum $L$. When these conserved quantities are compared to Eqs.~(5.7a,b) in \cite{Betschart:2008yi}
we see that the energy is the same, but the angular momentum differs by an additional factor of $A$. Finally \eqref{eq:equation-r} can be written as follows:
\begin{equation}
\label{eq:conserved-quantity-3}
0=\frac{1}{2A\dot{r}}\frac{\mathrm{d}}{\mathrm{d}\tau}\left(A\dot{r}^2-E^2A+\frac{L^2}{r^2A}\right) \Rightarrow K_2=A\dot{r}^2-E^2A+\frac{L^2}{r^2A}\,.
\end{equation}
Therefore the latter comprises even another conserved quantity $K_2$. Setting $K_2=0$ is in accordance with the null-trajectory condition of
\eqref{eq:equation-null}. Taking into account
that $\dot{r}=(\mathrm{d}r/\mathrm{d}\phi)\dot{\phi}$ where $\dot{\phi}$ is again expressed by the conserved angular momentum, it is possible
to solve \eqref{eq:conserved-quantity-3} with respect to $\mathrm{d}\phi/\mathrm{d}r$:
\begin{equation}
\label{eq:change-in-angle}
\frac{\mathrm{d}\phi}{\mathrm{d}r}=\frac{L}{r\sqrt{E^2A(r)^2r^2-L^2}}\,,\quad \phi(r)=L\int_{r_0}^{\infty} \frac{\mathrm{d}r}{r\sqrt{E^2A(r)^2r^2-L^2}}\,.
\end{equation}
Comparing to \eqref{eq:integration-deflection-angle} we see that $C=L/E$, i.e., the constant appearing in Bouguer's formula can be
understood as the ratio of angular momentum and total energy.

Now there are some differences between the final result of \eqref{eq:change-in-angle} and the corresponding Eq.~(5.9) in
\cite{Betschart:2008yi}. In the latter paper a black-hole gravitational background is considered in Schwarzschild coordinates. This line
interval does not have the form of a generally isotropic metric given in \eqref{eq:line-interval-isotropic-case}. In fact, there are isotropic
coordinates allowing us to write the Schwarzschild solution in the form (see, e.g., page 93 of \cite{Eddington:1923}):
\begin{subequations}
\begin{align}
\varrho&=\frac{1}{2}\left(r-GM+\sqrt{r(r-2GM)}\right)\,, \\[2ex]
\mathrm{d}\tau^2&=\left(\frac{1-GM/(2\varrho)}{1+GM/(2\varrho)}\right)^2\,\mathrm{d}t^2-\left(1+\frac{GM}{2\varrho}\right)^4\left[\mathrm{d}\varrho^2+\varrho^2\,\mathrm{d}\theta^2+\varrho^2\sin^2\theta\,\mathrm{d}\phi^2\right]\,.
\end{align}
\end{subequations}
Using this set of coordinates the equation encoding angular momentum conservation and the change in the angle $\phi$ with respect to the
new radial coordinate $\rho$ read as follows:
\begin{subequations}
\begin{align}
\label{eq:angular-momentum-conservation-isotropic}
L&=g_{\rho\rho}\rho^2\dot{\phi}\,, \\[2ex]
\frac{\mathrm{d}\phi}{\mathrm{d}\rho}&=\frac{L}{\rho}\frac{\sqrt{g_{\rho\rho}g_{tt}}}{\sqrt{E^2g_{\rho\rho}^2\rho^2-L^2g_{\rho\rho}g_{tt}}}=\frac{L}{\rho}\frac{1}{\sqrt{E^2g_{\rho\rho}^2\rho^2-L^2}}+\mathcal{O}\left(\frac{GM}{2\rho}\right)^2\,.
\end{align}
\end{subequations}
The first corresponds to \eqref{eq:angular-momentum-conservation} and the second to \eqref{eq:change-in-angle} neglecting second-order gravity
effects. Multiplying the modified line interval of their Eq.~(4.12) by the constant $\sqrt{1+\epsilon}$ leads to
\begin{equation}
\mathrm{d}\widetilde{\tau}^2=\frac{1}{\sqrt{1+\epsilon}}\left(1-\frac{2GM}{r}\right)\mathrm{d}t^2-\sqrt{1+\epsilon}\left[\frac{1}{1-2GM/r}\mathrm{d}r^2+r^2\,\mathrm{d}\Omega^2\right]\,.
\end{equation}
Since photons move on null-trajectories, $\mathrm{d}\widetilde{\tau}^2=0$ anyhow, which is why a multiplication of the line element by a constant
should not change the physics.
In this case the Lorentz-violating contribution governed by a position-independent $\epsilon$ drops out of $\mathrm{d}\phi/\mathrm{d}\rho$
when taking into account \eqref{eq:angular-momentum-conservation-isotropic}. Therefore in the isotropic coordinates the particular Lorentz-violating
contribution of their case 3 produces second-order gravity effects associated to Lorentz violation. Far away from the back-hole event horizon
there are no novel physical effects and this corresponds to the outcome of the eikonal approach.

\subsection{Killing vectors of a spherically symmetric spacetime}
\label{sec:killing-vectors-generally-isotropic}

In the current paragraph the Killing vectors for a spherically symmetric spacetime, cf.~\eqref{eq:line-interval} with $A(r,\theta,\phi)=A(r)$, will be
listed. The Killing vectors $\xi_{\mu}$ describe infinitesimal
isometries for a spacetime and they are linked to underlying symmetries and conserved quantities. In general they are obtained from a set
of partial differential equations called the Killing equations:
\begin{equation}
D_{\alpha}\xi_{\beta}+D_{\beta}\xi_{\alpha}=0\,,\quad D_{\nu}\xi_{\lambda}=\partial_{\nu}\xi_{\lambda}-\gamma^{\mu}_{\phantom{\mu}\nu\lambda}\xi_{\mu}\,,
\end{equation}
with the covariant derivative $D_{\alpha}$ and the Christoffel symbols $\gamma^{\mu}_{\phantom{\mu}\nu\lambda}$. The latter can be directly
extracted from \eqref{eq:geodesic-equations-spherically-symmetric-spacetime}. For the spherically symmetric spacetime it is possible to solve
the Killing equations analytically. To do so, it is
reasonable to make a certain \textit{Ansatz}, e.g., one with vanishing spatial components of $\xi_{\mu}$. This simplifies the set of equations
dramatically where several are immediately fulfilled automatically. They are then solved successively to obtain four Killing
vectors. Since the metric is isotropic, it is reasonable to make an \textit{Ansatz} for $\xi_{\mu}$ that only involves a nonvanishing timelike
component that does not depend on time itself:
\begin{equation}
(\xi_{\mu})=\begin{pmatrix}
\xi_0(r,\theta,\phi) \\
0 \\
0 \\
0 \\
\end{pmatrix}\,.
\end{equation}
In this case the following three differential equations must to be solved:
\begin{equation}
\xi_0\frac{A'}{A}+\frac{\partial\xi_0}{\partial r}=0\,,\quad \frac{\partial\xi_0}{\partial\theta}=0\,,\quad \frac{\partial\xi_0}{\partial\phi}=0\,.
\end{equation}
The remaining ones are fulfilled automatically.
The latter two tell us immediately that $\xi_0$ neither depends on $\theta$ nor $\phi$. Therefore the first differential equation is an ordinary
one that can be solved directly by integration:
\begin{equation}
\frac{\xi_0'}{\xi_0}=-\frac{A'}{A} \Rightarrow \ln|\xi_0|=-\ln|c_0A| \Rightarrow \xi_0(r)=\frac{\widetilde{c}_0}{A(r)}\,,\quad c_0,\,\widetilde{c}_0\in \mathbb{R}\,.
\end{equation}
A similar approach leads to the remaining Killing vectors. In total one obtains
\begin{subequations}
\begin{align}
\xi_{\mu}^{(1)}&=\begin{pmatrix}
1/A \\
0 \\
0 \\
0 \\
\end{pmatrix}\,,\quad \xi_{\mu}^{(2)}=r^2A\begin{pmatrix}
0 \\
0 \\
\sin\phi \\
\sin\theta\cos\theta\cos\phi \\
\end{pmatrix}\,, \displaybreak[0]\\[2ex]
\xi_{\mu}^{(3)}&=r^2A\begin{pmatrix}
0 \\
0 \\
\cos\phi \\
-\sin\theta\cos\theta\sin\phi \\
\end{pmatrix}\,,\quad \xi_{\mu}^{(4)}=r^2A\begin{pmatrix}
0 \\
0 \\
0 \\
\sin^2\theta \\
\end{pmatrix}\,.
\end{align}
\end{subequations}
Suitable contractions of the Killing vectors with $(\dot{x}^{\mu})=(\dot{t},\dot{r},\dot{\theta},\dot{\phi})^T$ (and additional linear combinations)
lead to conserved quantities. The first conserved quantity follows from a contraction with the first Killing vector:
\begin{subequations}
\label{eq:conserved-quantities-killing}
\begin{equation}
\label{eq:conserved-quantities-killing-1}
\xi_{\mu}^{(1)}\dot{x}^{\mu}=\frac{\dot{t}}{A}=\mathrm{const.}
\end{equation}
The latter corresponds to the result obtained in \eqref{eq:conserved-quantities-killing-1} and it is related to energy conservation. The second
conserved quantity involves the remaining Killing vectors where it is understood to be evaluated at $\theta=\pi/2$:
\begin{equation}
\label{eq:conserved-quantities-killing-2}
\sqrt{\sum_{i=2}^4 \left.(\xi_{\mu}^{(i)}\dot{x}^{\mu})^2\right|_{\theta=\pi/2}}=Ar^2\dot{\phi}=\mathrm{const.}
\end{equation}
\end{subequations}
This conserved quantity is the same as what was obtained in \eqref{eq:angular-momentum-conservation} and it means angular momentum
conservation. Hence the Killing vectors $\xi^{(i)}$ for $i=2\dots 4$ are related to rotational symmetry of the spherically symmetric spacetime.

\section{Eikonal equation for inhomogeneous and anisotropic media}
\label{sec:eikonal-equation-inhomogeneous-anisotropic}

The current section serves with providing some general results on the physics of the eikonal equation, which are used in
\secref{sec:gravitational-backgrounds} extensively.
In general, the eikonal equation provides a set of three coupled nonlinear differential equations. In what follows a refractive index
bearing a dependence on the radial distance $r$ and an angle $\phi$ is assumed (cf., e.g., \eqref{eq:anisotropic-angle}). The photon trajectory
shall be parameterized by the angle $\phi$, i.e., $\mathbf{r}(\phi)=r(\phi)\widehat{\mathbf{e}}_r(\phi)$. Its first and second derivative read
\begin{subequations}
\begin{equation}
\label{eq:photon-trajectory-derivatives}
\mathbf{r}'=\dot{r}\widehat{\mathbf{e}}_r+r\widehat{\mathbf{e}}_{\phi}\,,\quad \mathbf{r}''=(\ddot{r}-r)\widehat{\mathbf{e}}_r+2\dot{r}\widehat{\mathbf{e}}_{\phi}\,.
\end{equation}
The arc length depends on $\phi$ and we obtain a set of useful relationships:
\begin{align}
\label{eq:eikonal-equation-identities}
s(\phi)&=\int^{\phi} \mathrm{d}\phi'\,|\mathbf{r}'|=\int^{\phi} \mathrm{d}\phi'\,\sqrt{r^2+\dot{r}^2}\,,\quad \frac{\mathrm{d}s}{\mathrm{d}\phi}=\sqrt{r^2+\dot{r}^2}\,, \\[2ex]
\left(\frac{\mathrm{d}s}{\mathrm{d}\phi}\right)^{-2}&=\frac{1}{r^2+\dot{r}^2}\,,\quad \frac{\mathrm{d}}{\mathrm{d}\phi}\left(\frac{\mathrm{d}s}{\mathrm{d}\phi}\right)^{-1}=\frac{\mathrm{d}}{\mathrm{d}\phi}\left(\frac{1}{\sqrt{r^2+\dot{r}^2}}\right)=-\frac{(r+\ddot{r})\dot{r}}{(r^2+\dot{r}^2)^{3/2}}\,, \\[2ex]
\frac{\mathrm{d}\mathbf{r}}{\mathrm{d}s}&=\frac{\mathrm{d}\mathbf{r}}{\mathrm{d}\phi}\left(\frac{\mathrm{d}s}{\mathrm{d}\phi}\right)^{-1}\,.
\end{align}
\end{subequations}
Now the derivative on the left-hand side of the eikonal equation can be computed. Instead of differentiating with respect to the arc
length we have to calculate derivatives with respect to $\phi$, which leads to three terms:
\begin{align}
\frac{\mathrm{d}}{\mathrm{d}s}\left[n\frac{\mathrm{d}\mathbf{r}}{\mathrm{d}\phi}\left(\frac{\mathrm{d}s}{\mathrm{d}\phi}\right)^{-1}\right]&=\frac{\mathrm{d}n}{\mathrm{d}s}\frac{\mathrm{d}\mathbf{r}}{\mathrm{d}\phi}\left(\frac{\mathrm{d}s}{\mathrm{d}\phi}\right)^{-1}+n\frac{\mathrm{d}^2\mathbf{r}}{\mathrm{d}\phi^2}\left(\frac{\mathrm{d}s}{\mathrm{d}\phi}\right)^{-2}+\frac{\mathrm{d}}{\mathrm{d}s}\left(\frac{\mathrm{d}s}{\mathrm{d}\phi}\right)^{-1}n\frac{\mathrm{d}\mathbf{r}}{\mathrm{d}\phi} \notag \displaybreak[0]\\
&=\frac{\mathrm{d}n}{\mathrm{d}\phi}\frac{\mathrm{d}\mathbf{r}}{\mathrm{d}\phi}\left(\frac{\mathrm{d}s}{\mathrm{d}\phi}\right)^{-2}+n\frac{\mathrm{d}^2\mathbf{r}}{\mathrm{d}\phi^2}\left(\frac{\mathrm{d}s}{\mathrm{d}\phi}\right)^{-2} \notag \displaybreak[0]\\
&\phantom{{}={}}+\frac{\mathrm{d}}{\mathrm{d}\phi}\left[\left(\frac{\mathrm{d}s}{\mathrm{d}\phi}\right)^{-1}\right]\left(\frac{\mathrm{d}s}{\mathrm{d}\phi}\right)^{-1}n\frac{\mathrm{d}\mathbf{r}}{\mathrm{d}\phi}\,.
\end{align}
Now employing the derivatives of \eqref{eq:photon-trajectory-derivatives} and the identities given in \eqref{eq:eikonal-equation-identities}
the eikonal equation can be expressed in terms of the radial coordinate $r$, the angle $\phi$, and the basis vectors:
\begin{align}
\widehat{\mathbf{e}}_r\frac{\partial n}{\partial r}+\frac{1}{r}\frac{\partial n}{\partial\phi}\widehat{\mathbf{e}}_{\phi}&=\frac{1}{r^2+\dot{r}^2}\left\{\frac{\mathrm{d}n}{\mathrm{d}\phi}(\dot{r}\widehat{\mathbf{e}}_r+r\widehat{\mathbf{e}}_{\phi})+n\left[(\ddot{r}-r)\widehat{\mathbf{e}}_r+2\dot{r}\widehat{\mathbf{e}}_{\phi}\right]\right\} \notag \\
&\phantom{{}={}}-n(\dot{r}\widehat{\mathbf{e}}_r+r\widehat{\mathbf{e}}_{\phi})\frac{(r+\ddot{r})\dot{r}}{(r^2+\dot{r}^2)^2}
\end{align}
Sorting terms associated to $\widehat{\mathbf{e}}_r$ and $\widehat{\mathbf{e}}_{\phi}$, respectively, results in a system of two differential equations:
\begin{subequations}
\begin{align}
\frac{\partial n}{\partial r}&=\frac{1}{r^2+\dot{r}^2}\left[\frac{\mathrm{d}n}{\mathrm{d}\phi}\dot{r}+n(\ddot{r}-r)\right]-n\frac{(r+\ddot{r})\dot{r}^2}{(r^2+\dot{r}^2)^2}\,, \\[2ex]
\frac{1}{r}\frac{\partial n}{\partial\phi}&=\frac{1}{r^2+\dot{r}^2}\left[\frac{\mathrm{d}n}{\mathrm{d}\phi}r+2n\dot{r}\right]-n\frac{(r+\ddot{r})r\dot{r}}{(r^2+\dot{r}^2)^2}\,.
\end{align}
\end{subequations}
Multiplying the second with $\dot{r}/r$ and subtracting it from the first eliminates various terms, which simplifies the equation drastically:
\begin{subequations}
\begin{align}
\frac{\partial n}{\partial r}-\frac{\dot{r}}{r^2}\frac{\partial n}{\partial\phi}&=\frac{n}{r^2+\dot{r}^2}\left(\ddot{r}-\frac{2\dot{r}^2}{r}-r\right)\,, \\[2ex]
\label{eq:eikonal-equation-main-result}
0&=(r^2+\dot{r}^2)\left[r\frac{\partial n}{\partial r}-\frac{\dot{r}}{r}\frac{\partial n}{\partial\phi}\right]+n(r^2+2\dot{r}^2-r\ddot{r})\,.
\end{align}
\end{subequations}
This is the final result that we are interested in and that shall be used for practical purposes. However multiplying the latter with $r\dot{r}/(r^2+\dot{r}^2)^{3/2}$,
it can be written in a form that allows for a deeper physical understanding:
\begin{equation}
\frac{\mathrm{d}}{\mathrm{d}\phi}(nr\sin\alpha)-\sqrt{r^2+\dot{r}^2}\frac{\partial n}{\partial\phi}=0\,,\quad \sin\alpha=\frac{r}{\sqrt{r^2+\dot{r}^2}}\,.
\end{equation}
If the refractive index only depends on the radial coordinate the second term on the left-hand side vanishes, which then leads us directly
to the formula of Bouguer. Physically this result means angular momentum conservation. For a refractive index that additionally depends
on the angle $\phi$ angular momentum is not a conserved quantity any more. Instead, there is a driving term that modifies the angular
momentum. The change is bigger the larger the velocity is and the stronger the refractive index changes with the angle.

\end{appendix}

\newpage


\end{document}